\documentclass[aps,prd,superscriptaddress,amsmath,twocolumn,10pt]{revtex4}

\usepackage{color,hangcaption,hhline,psfrag,rotating,amssymb}
\usepackage[hang,nooneline]{subfigure}
\usepackage{dcolumn}
\usepackage{verbatim}

\newcommand{\nn}{\nonumber}
\newcommand{\be}{\begin{equation}}
\newcommand{\ee}{\end{equation}}
\newcommand{\bea}{\begin{eqnarray}}
\newcommand{\eea}{\end{eqnarray}}

\newcommand{\Tr}{\mathrm{Tr}}

\begin{document}
\title{The Upsilon spectrum and the determination of the lattice spacing from lattice QCD including charm quarks in the sea}

\author{R.~J.~Dowdall}
\email[]{Rachel.Dowdall@glasgow.ac.uk}
\affiliation{SUPA, School of Physics and Astronomy, University of Glasgow, Glasgow, G12 8QQ, UK}
\author{B.~Colquhoun}
\affiliation{SUPA, School of Physics and Astronomy, University of Glasgow, Glasgow, G12 8QQ, UK}
\author{J.~O.~Daldrop}
\thanks{Current address: Helmholtz Institut f\"{u}r Strahlen und Kernphysik and Bethe Centre for Theoretical Physics, Universit\"{a}t Bonn, 53115 Bonn, Germany}
\affiliation{SUPA, School of Physics and Astronomy, University of Glasgow, Glasgow, G12 8QQ, UK}
\author{C.~T.~H.~Davies}
\email[]{c.davies@physics.gla.ac.uk}
\affiliation{SUPA, School of Physics and Astronomy, University of Glasgow, Glasgow, G12 8QQ, UK}
\author{I.~D.~Kendall}
\affiliation{SUPA, School of Physics and Astronomy, University of Glasgow, Glasgow, G12 8QQ, UK}
\author{E.~Follana}
\affiliation{Departamento de F\'{i}sica Te\'{o}rica, Universidad de Zaragoza,
Cl. Pedro Cerbuna 12, E-50009 Zaragoza, Spain}
\author{T.~C.~Hammant}
\affiliation{DAMTP, University of Cambridge, Wilberforce Road, Cambridge CB3 0WA, UK}
\author{R.~R.~Horgan}
\affiliation{DAMTP, University of Cambridge, Wilberforce Road, Cambridge CB3 0WA, UK}
\author{G.~P.~Lepage}
\affiliation{Laboratory of Elementary-Particle Physics, Cornell University, Ithaca, New York 14853, USA}
\author{C.~J.~Monahan}
\thanks{Current address: Department of Physics, College of William and Mary, Williamsburg, VA 23187, USA}
\affiliation{DAMTP, University of Cambridge, Wilberforce Road, Cambridge CB3 0WA, UK}
\author{E.~H.~M\"{u}ller}
\affiliation{SUPA, School of Physics, University of Edinburgh, King's Buildings, Edinburgh, EH9 3JZ, UK}

\collaboration{HPQCD collaboration}
\homepage{http://www.physics.gla.ac.uk/HPQCD}
\noaffiliation

\date{\today}

\begin{abstract}
We give results for the Upsilon spectrum from lattice QCD using an improved 
version of the NRQCD action for $b$ quarks which includes radiative corrections 
to kinetic terms at $\mathcal{O}(v^4)$ in the velocity expansion. 
We also include for the 
first time the effect of up, down, strange and charm quarks in 
the sea using `second generation' gluon field 
configurations from the MILC collaboration. 
Using the $\Upsilon$ $2S-1S$ splitting to determine the 
lattice spacing, we are able to obtain
the $1P-1\overline{S}$ splitting to 1.4\% and the $3S-1S$ 
splitting to 2.4\%.  Our improved result for  
$M(\Upsilon) -M(\eta_b)$ is 70(9) MeV 
and we predict $M(\Upsilon^{\prime})-M(\eta_b^{\prime})$ = 35(3) MeV. 
We also calculate $\pi,K$ and $\eta_s$ correlators using the Highly Improved 
Staggered Quark action and perform a chiral and continuum extrapolation 
to give values for $M_{\eta_s}$ (0.6893(12) GeV) 
and $f_{\eta_s}$ (0.1819(5) GeV) 
that allow us to tune the 
strange quark mass as well as providing an independent and 
consistent determination of 
the lattice spacing. Combining the NRQCD and HISQ analyses gives 
$m_b/m_s$ = 54.7(2.5) and a value for the heavy quark 
potential parameter of $r_1 = $ 0.3209(26) fm. 
\end{abstract}


\maketitle

%
\section{Introduction}
\label{sec:intro}
%
Lattice QCD calculations have developed rapidly both in 
accuracy and in scope in the last few years. 
This growth has built on the first 
demonstration that numerical simulations including $u$, $d$ and 
$s$ quarks in the sea with light enough $u/d$ quarks
give results in agreement with experiment 
for simple 
`gold-plated' quantities across the full 
range of hadron physics~\cite{ourlattexpt}. 
Errors at the level of a few \% make this highly non-trivial. 
A key element of those calculations was the determination 
of the $\Upsilon$ spectrum because there are many 
gold-plated states below threshold for strong Zweig-allowed decay.
In addition radial and orbital excitation energies are 
very insensitive to quark masses (including that of the $b$ itself) 
making them useful for determining the lattice spacing, $a$, without 
a complicated tuning process. A further incentive for lattice 
$\Upsilon$ studies is the 
importance of testing $b$ quark physics from lattice QCD 
so that the same action can be used for results 
in $B$ physics required, 
in conjunction with experiment, 
for the determination of elements of the Cabibbo-Kobayashi-Maskawa matrix.
Here we give new results for the $\Upsilon$ spectrum improving 
on those earlier results in several ways to keep pace with 
improvements in other areas of lattice QCD. We have improved 
statistical errors, improved 
the NRQCD action and we are also now 
using `second generation' gluon 
field configurations that include charm quarks in the sea. 

The $b$ quarks in these first calculations that included
the full effect of sea quarks~\cite{Gray:2005ur, alanphd} were 
implemented using lattice Nonrelativistic QCD 
(NRQCD) with an action accurate through $v^4$ in the velocity expansion 
for the $b$ quark~\cite{Lepage:1992tx}. 
The coefficients of the $v^4$ terms were 
matched to full QCD at tree level, 
having removed the most significant source of radiative corrections, 
that of tadpole diagrams generated in lattice QCD from the form 
of the lattice gluon field, by the use of 
`tadpole-improvement'~\cite{Lepage:1992xa}. 
The gluon field configurations used were generated 
by the MILC collaboration~\cite{Bazavov:2009bb} using a Symanzik-improved gluon 
action in which radiative corrections at $\mathcal{O}(\alpha_sa^2)$ 
were included except for radiative corrections 
from quark loops~\cite{Hao:2007iz} 
($\mathcal{O}(n_f\alpha_sa^2)$ where $n_f$ is the number 
of sea quark flavors), which were omitted. Configurations at 
three different values of the lattice spacing were available: 
`supercoarse' ($a \approx$ 0.18fm); `coarse'  ($a \approx$ 0.12fm) 
and `fine' ($a \approx$ 0.09fm). $u/d$ and $s$ sea 
quarks were included using the improved staggered (asqtad) 
action~\cite{Orginos:1998ue, Orginos:1999cr, Lepage:1998vj} 
which is numerically relatively fast. A range of $u/d$ masses 
(taken to be the same)
were used ranging down to a ratio with the $s$ sea quark mass of 
around 0.2. 
The key mass splittings in the bottomonium spectrum 
studied were those between the 
ground $S$-wave states and the first radially excited 
$S$-wave states, the $2S-1S$ splitting, and that between 
the first $P$-wave states and the ground $S$-wave states, 
the $1P-1S$ splitting.   
The statistical errors from the lattice 
calculation for these splittings were 1--2\% (i.e. 5-10 MeV) 
and systematic errors were estimated to be similar to this 
or smaller, depending on the lattice spacing.  
Within these
errors, agreement with experiment was confirmed. 

More recently the $\Upsilon$ spectrum has been calculated 
using the same NRQCD action on gluon configurations at a 
`coarse' ($a \approx$ 0.11 fm) and a `fine' ($a \approx$ 0.09 fm) 
lattice spacing generated 
by the RBC/UKQCD collaboration using the Iwasaki gluon action 
and 2+1 flavors of sea quarks implemented with the domain 
wall formalism~\cite{Meinel:2009rd, Meinel:2010pv}. 
Results in close agreement and with similar errors 
to those found  
on the MILC configurations are obtained, confirming the 
independence of the results from the sea quark formalism. 

The systematic errors in the calculation of the 
$\Upsilon$ $2S-1S$ and $1P-1S$ 
splittings were studied in some detail in~\cite{Gray:2005ur}. 
Sources of error there were missing 
radiative corrections to the $v^4$ terms in 
the lattice NRQCD action (beyond tadpole-improvement), 
as well as radiative corrections to 
discretisation correction terms and from higher 
order ($v^6$) missing relativistic corrections. In addition 
systematic errors from the missing radiative corrections to the 
improvement terms in the gluon action were estimated.   
These errors were typically each of order 1\% in the $1P-1S$ splitting 
on the fine lattices and about half that for the $2S-1S$ splitting 
because of some cancellation between $1S$ and $2S$ states. 
Errors were similar for the radiative and relativistic 
errors on coarser lattices but of course the 
discretisation errors were larger. 

Subsequent to this, we have made estimates of the 
effect of missing $c$ quarks in the sea~\cite{newfds, Gregory:2010gm}. 
These have negligible 
effect on mesons apart from bottomonium, where internal 
momenta can be large enough to generate $c$ quarks from the 
vacuum. We found the shift in the ground-state $S$ wave masses 
might be of $\mathcal{O}({\rm 5 MeV})$~\cite{Gregory:2010gm} (it 
is spin-independent) with approximately 
half the shift 
for $2S$ states because of a smaller `wave-function at the origin'
and no shift for $1P$ states. This would give rise to systematic 
errors of 0.5\% for the $2S-1S$ splitting and 1\% for the $1P-1S$ 
splitting, similar to the systematic errors from other effects 
quoted above. 

The conclusion from these results is that the errors 
in bottomonium masses and radial and orbital mass 
splittings have been pinned down 
and tested from this NRQCD action at the level of 5-10 MeV.  
There is also a contribution to systematic errors at the 
same level coming from the gluon field configurations. 
The NRQCD systematic errors also feed in to the calculation 
of $B$, $B_s$ and $B_c$ meson masses using NRQCD $b$ quarks. 
The state-of-the-art calculation for the masses of these mesons has 
$\mathcal{O}$(10 MeV) errors dominated by systematic errors 
from this NRQCD action~\cite{Gregory:2010gm, Gregory:2009hq}. 

In the last five years, however, other 
lattice QCD calculations have become 
increasingly accurate. 
For example the mass of the $D_s$ meson was recently calculated 
by HPQCD with combined statistical and systematic errors 
of 3 MeV and its decay constant calculated to 1\%~\cite{newfds}. 
These errors are at the level where we must 
allow for missing electromagnetism from lattice QCD. 

There have been several contributions to this progress. 
Advances in computational speed have meant better 
statistical errors from calculating many more meson correlators 
on larger samples of configurations. 
It has also been possible to generate lattices with smaller 
lattice spacing, so that the $D_s$ calculation includes 
`superfine' ($a \approx$ 0.06 fm) and `ultrafine' ($a \approx$ 0.045 fm) 
lattices~\cite{Bazavov:2009bb}. Significant improvements 
have been made to relativistic quark actions too. For example, 
the $D_s$ meson mass calculation used the  
Highly Improved Staggered Quark (HISQ) action for both
valence quarks. The HISQ action~\cite{Follana:2006rc} has 
smaller discretisation errors than the asqtad action 
by about a factor of 3 and can 
be used for quarks as heavy as charm on 
lattices with a lattice spacing of 0.1fm or smaller. 
This has revolutionised charm physics calculations~\cite{oldfds} 
in lattice QCD 
and is having an impact also on calculations 
for mesons containing a $b$ quark 
through a combination of an extrapolations in the 
mass of the heavy HISQ quark acting as the `$b$' to 
the physical point for the real $b$ quark, combined 
with extrapolations to the continuum ($a \rightarrow 0$) 
limit from results at many values of $a$~\cite{McNeile:2011ng}. 
The heavy HISQ calculations are computationally much more expensive than 
those using NRQCD and this currently limits 
their utility. 
The results for $B_s$ and $B_c$ meson masses have 
comparable errors to the existing NRQCD results,  
but are dominated by statistical and $a \rightarrow 0$ 
extrapolation uncertainities. They then provide a complementary 
way of testing $b$ physics to that of NRQCD and it is 
clear that combining the strengths of both methods will be 
optimal in future.  

Meanwhile the MILC collaboration have moved on to the 
production of `second generation' gluon field configurations 
which have a number of improvements over the 
earlier ensembles~\cite{Bazavov:2010ru}. They include a more highly improved 
gluon action~\cite{Hart:2008sq}, HISQ quarks in the sea with the addition of 
$c$ quarks as well as $u$, $d$ and $s$ and with lighter 
$u$ and $d$ masses than before. 

The availability of these configurations along with the 
incentives discussed above to improve errors in 
$\Upsilon$ and $B$ physics using NRQCD $b$ quarks 
has meant that we have begun a new programme of 
improved NRQCD calculations. Here we present the first 
results, giving the radial and orbital splittings in 
the $\Upsilon$ spectrum, tuning the lattice $b$ quark 
mass and determining the lattice spacing from the $(2S-1S)$
splitting. As well as using 
the second generation gluon field configurations we 
have improved the NRQCD action by adding radiative corrections 
to the $v^4$ kinetic terms including discretisation errors. 
We also have improved statistics and improved methods for 
tuning the $b$ quark mass. This has meant that we can 
test the effect of radiative corrections to the $v^4$ 
kinetic terms on the meson dispersion relation. Using 
both perturbative and nonperturbative methods for 
determining the radiative corrections to spin-dependent 
terms we are able to improve the determination of 
the $\Upsilon$ hyperfine splitting.  

A useful complementary method for determining 
the lattice spacing was developed in~\cite{Davies:2009tsa}. It uses 
the fictitious $s\overline{s}$ pseudoscalar particle 
known as the $\eta_s$. This particle does not exist in 
the real world because of mixing with light quarks 
to form the $\eta$ and 
$\eta^{\prime}$ but on the lattice this can be prevented. 
The mass and decay constant of 
the $\eta_s$ can be determined accurately in a lattice 
QCD calculation using the Highly Improved Staggered 
Quark (HISQ) action and their physical values fixed from 
$M_{\pi}$, $M_{K}$, $f_{\pi}$ and $f_K$ from a simultaneous 
chiral and continuum extrapolation.  
Here we update the results of~\cite{Davies:2009tsa} 
for these 2+1+1 configurations 
and use these also to give a determination of the lattice spacing. 

The two different methods for determining the lattice 
spacing can be combined through the use of a third quantity, 
$r_1$~\cite{Aubin:2004wf}, which can be derived accurately from determination 
of the heavy quark potential~\cite{Sommer:1993ce}. $r_1/a$ values are provided 
for these configurations by the MILC collaboration~\cite{doug}. 
$r_1/a$ provides a good determinant of the relative lattice 
spacing between different sets of gluon configurations but 
its physical value must be determined from other quantities. 
From the separate determination of the lattice 
spacing from the two methods above we have two sets 
of results for $r_1$ in fm as a function of lattice spacing. 
From this we are able to test that the two methods 
give the same result in the continuum and chiral limits 
(which they do) and provide a physical value of 
$r_1$ that could be used, in the absence of either 
of the other methods, to determine the lattice spacing 
on other ensembles with 2+1+1 flavors of sea quarks. 

We also combine results for 
tuned $b$ quark masses in NRQCD and tuned 
$s$ quark masses from HISQ 
along with one-loop renormalisation constants
to give a value for $m_b/m_s$ for comparison 
to other results obtained purely from 
the HISQ action.  

The layout of the paper is as follows. 
Section~\ref{sec:secgen} discusses the second-generation gluon 
field ensembles giving more details of the improvements 
present there. Section~\ref{sec:ups} describes the improvements to the  
NRQCD calculations and results for the $\Upsilon$ spectrum. 
Section~\ref{sec:etas} discusses the $\pi$, $K$, $\eta_s$ analysis
on these same configurations and the additional information 
that provides to determine the lattice spacing.  
This is tied together via the determination of the 
heavy quark potential parameter, $r_1$, in section~\ref{sec:r1}
and $m_b/m_s$ in section~\ref{sec:mbms}. 
Section~\ref{sec:conclude} provides our conclusions. 

%
\section{Second generation 2+1+1 gluon field ensembles}
\label{sec:secgen}
%
%
%
\begin{table}
\caption{
Details of the MILC gluon field ensembles used in this paper. 
$\beta=10/g^2$ is the $SU(3)$ gauge coupling and $L/a$ and $T/a$ 
are the number of lattice spacings in the space and 
time directions for each lattice. 
$am_{l},am_{s}$ and $am_c$ are the light (up and down taken to 
have the same mass), strange and charm sea quark masses in lattice units.
$r_1/a$ is the static-quark potential parameter in 
lattice units determined by 
the MILC collaboration~\cite{Bazavov:2010ru, doug}. Note that 
this has not been `smoothed'. 
The ensembles 1 and 2 will be referred to in the text as ``very coarse'', 3 and 4 as ``coarse'' and 5 as ``fine.'' 
}
\label{tab:params}
\begin{ruledtabular}
\begin{tabular}{llllllll}
Set & $\beta$ & $r_1/a$ & $am_{l}$ & $am_{s}$ & $am_c$ & $L/a \times T/a$ \\
\hline
1 & 5.80 & 2.041(10) & 0.013   & 0.065  & 0.838 & 16$\times$48 \\
2 & 5.80 & 2.0621(45)& 0.0064  & 0.064  & 0.828 & 24$\times$48 \\
\hline
3 & 6.00 & 2.574(5)  & 0.0102  & 0.0509 & 0.635 & 24$\times$64 \\
4 & 6.00 & 2.623(11) & 0.00507 & 0.0507 & 0.628 & 32$\times$64 \\
\hline
5 & 6.30 & 3.549(13)  & 0.0074  & 0.037  & 0.440 & 32$\times$96 \\
\end{tabular}
\end{ruledtabular}
\end{table}

The gauge configurations used in this calculation are listed in 
Table~\ref{tab:params}~\cite{Bazavov:2010ru}. 
These were generated by the MILC collaboration using a 
tadpole-improved L\"{u}scher-Weisz gauge action with coefficients 
corrected perturbatively through $\mathcal{O}(\alpha_s)$ 
including pieces proportional to $n_f$, the number 
of quark flavors in the sea~\cite{Hart:2008sq}
(see Appendix \ref{appendix:gauge_action}).
The gauge action is then improved completely through $\mathcal{O}(\alpha_sa^2)$, 
unlike the earlier asqtad configurations.
Sea quarks are included with the 
HISQ action \cite{Follana:2006rc} which also 
has smaller discretisation errors compared to the 
asqtad action (see the discussion in section~\ref{sec:etas}). 
The configurations include a sea
charm quark in addition to up, down and strange. 
These configurations are then said to have 2+1+1 flavors in 
the sea, since the $u$ and $d$ quarks are taken to have the 
same mass, which is heavier than average $u/d$ mass in the real world, and the 
$s$ and $c$ masses are tuned as closely as possible to 
their correct values at that lattice spacing. The tuning of the 
sea $s$ quark mass is much more accurately done -- to better 
than 5\% -- than on 
the previous asqtad configurations. 
This means that the $u/d$ quark mass (denoted $m_l$ here) can 
be more accurately calibrated in terms of the $s$ quark 
mass for chiral extrapolations. 
Here we use a ratio of $m_{l}/m_s$ as 
low as one tenth (see Table~\ref{tab:params}) 
whereas in our previous work on the asqtad configurations our 
most chiral ensemble had a ratio of the $m_{l, sea}/m_{s,physical}$ 
of one quarter. 
This means 
that we have a much smaller chiral extrapolation to do to reach 
the physical $u/d$ mass (where $m_l=m_s/27$~\cite{Bazavov:2009bb}) than before. 

The sea quarks are included with the standard method of 
incorporating the determinant of the quark matrix raised 
to the one quarter power for each flavor, in order to 
implement the correct counting for sea staggered quarks. 
The algorithm used for including the sea quarks has 
now been improved by MILC to the exact RHMC algorithm~\cite{Bazavov:2010ru}
i.e. all errors in the time step for the updating 
algorithm have been removed.  

The configurations are separated by 5 trajectories 
in the time units 
of the updating algorithm for the very coarse and 
coarse ensembles and by 6 trajectories for the 
fine ensemble. In subsections~\ref{subsec:smear} and 
~\ref{subsec:etasfits} we will study the autocorrelations 
in our meson correlators to show how independent 
the configurations are for different observables. 

The $r_1/a$ values given in Table~\ref{tab:params} are 
determined by the MILC collaboration after extraction 
of the potential between two infinitely heavy (static) 
quarks at separation $r/a$ in lattice units. 
$r_1/a$ is defined~\cite{Aubin:2004wf} as the point 
where the force $F(r)$ derived 
from the derivative of the potential satisfies  
\begin{equation}
r^2F(r) = 1.
\label{eq:r1def}
\end{equation}
The values of $r_1/a$ for these ensembles have been 
chosen to match approximately those of the previous 
results including 2+1 flavors of asqtad quarks and 
can be used to determine the lattice spacing if 
the physical value for $r_1$ is known. 
Using the $r_1$ value determined previously 
on configurations with 2+1 flavors of sea quarks, 
this means that the lattice spacing values will be
approximately 0.15 fm, 0.12fm and 0.09fm. 
The physical spatial size of the lattices then 
exceeds 2.5 fm and is as high as 3.8 fm on 
the ensembles that correspond to $m_l/m_s = 0.1$. 
In section~\ref{sec:r1} we will derive a physical 
value for $r_1$ based on the results from 
sections~\ref{sec:ups} and~\ref{sec:etas} to calibrate 
more accurately the lattice spacing values for these 
configurations. 

%
\section{The Upsilon Spectrum}
\label{sec:ups}
%

%
\subsection{The NRQCD action}
\label{subsec:nrqcd}
%

The spectrum of bottomonium mesons is extracted by computing 
appropriate correlators 
constructed from $b$-quark propagators on the gluon field 
ensembles listed in Table~\ref{tab:params}. 
We make use of NRQCD, an effective field theory that gives an expansion 
of the Dirac action in powers of the heavy quark velocity, $v$. 
This is discretised onto a space-time lattice as lattice NRQCD
~\cite{Thacker:1990bm,Lepage:1992tx} 
and is a good formalism to use for $b$ quarks 
since they are known to be very nonrelativistic 
inside their bound states ($v^2 \approx 0.1$). 
As used on the lattice NRQCD has the 
advantage that propagators can be generated 
using a simple time evolution equation 
rather than having to invert the Dirac matrix.
The quark and antiquark fields are separated in this formalism 
as 2-component spinors.  

The NRQCD Hamiltonian we use is given by:
 \begin{eqnarray}
 aH &=& aH_0 + a\delta H; \nonumber \\
 aH_0 &=& - \frac{\Delta^{(2)}}{2 am_b}, \nonumber \\
a\delta H
&=& - c_1 \frac{(\Delta^{(2)})^2}{8( am_b)^3}
            + c_2 \frac{i}{8(am_b)^2}\left(\bf{\nabla}\cdot\tilde{\bf{E}}\right. -
\left.\tilde{\bf{E}}\cdot\bf{\nabla}\right) \nonumber \\
& & - c_3 \frac{1}{8(am_b)^2} \bf{\sigma}\cdot\left(\tilde{\bf{\nabla}}\times\tilde{\bf{E}}\right. -
\left.\tilde{\bf{E}}\times\tilde{\bf{\nabla}}\right) \nonumber \\
 & & - c_4 \frac{1}{2 am_b}\,{\bf{\sigma}}\cdot\tilde{\bf{B}}  
  + c_5 \frac{\Delta^{(4)}}{24 am_b} \nonumber \\
 & & -  c_6 \frac{(\Delta^{(2)})^2}{16n(am_b)^2} .
\label{eq:deltaH}
\end{eqnarray}
Here $\nabla$ is the symmetric lattice derivative and $\Delta^{(2)}$ and 
$\Delta^{(4)}$ the lattice discretization of the continuum $\sum_iD_i^2$ and 
$\sum_iD_i^4$ respectively. $am_b$ is the bare $b$ quark mass. 
$\bf \tilde{E}$ and $\bf \tilde{B}$ are the chromoelectric 
and chromomagnetic fields calculated from an improved clover term~\cite{Gray:2005ur}.
The $\bf \tilde{B}$ and $\bf \tilde{E}$ are made anti-hermitian 
but not explicitly traceless, to match the perturbative calculations 
done using this action.  

In terms of the velocity expansion $H_0$ is $\mathcal{O}(v^2)$ and 
$\delta H$ is $\mathcal{O}(v^4)$, including discretisation corrections. 
$H_0$ contains the bare quark 
mass parameter which is nonperturbatively tuned to the correct 
value for the $b$ quark as discussed below in subsection~\ref{subsec:tune}. 
The terms in $\delta H$ have coefficients $c_i$ whose values are 
fixed from matching lattice NRQCD to full QCD. This matching takes account 
of high momentum modes that differ between NRQCD and full QCD and 
so it can be done perturbatively, giving the $c_i$ the 
expansion $1 + c^{(1)}_i\alpha_s + \mathcal{O}(\alpha_s^2)$. 
In previous calculations~\cite{Gray:2005ur} we used the tree level value of 1 for 
all the $c_i$, after tadpole-improving the gluon fields. 
This means dividing all the gluon fields, $U_{\mu}(x)$ by a tadpole-parameter,
$u_0$, before constructing covariant derivatives or 
${\bf E}$ and ${\bf B}$ fields for the Hamiltonian above.  
The $u_0$ parameter corrects for tadpole diagrams that 
arise in a universal way from the way in which the 
lattice gluon field is constructed. 
For $u_0$ we took the mean trace of the gluon field 
in Landau gauge, $u_{0L}$. 
With tadpole-improvement 
in place we expect the radiative corrections to the 
$c_i$ coefficients to be of normal size 
i.e. $\mathcal{O}(1)$~\cite{Morningstar:1994qe}; 
without this they can be rather large.  

Here, on top of tadpole-improvement with $u_{0L}$, we use 
$\mathcal{O}(\alpha_s)$ corrected coefficients for the kinetic terms, 
i.e. $c_1$, $c_5$ and $c_6$, so improving on the NRQCD action 
used previously, and significantly reducing the systematic 
errors in the tuning of the $b$ quark mass and in the determination of the radial 
and orbital mass splittings.   
The calculation of the $c^{(1)}_i$ for $i=1,5,6$ is discussed in 
Appendix~\ref{appendix:cicalc}~\cite{eikephd}. 
Table~\ref{tab:wilsonparams} gives the values for $c_1$, $c_5$ and 
$c_6$ that we use on the very coarse, coarse and fine lattices as 
a result. 
As expected, after tadpole-improvement, 
the coefficients $c^{(1)}_{1,5,6}$ are not large and 
they are well-behaved as a function of the $b$ quark mass.
In subsection~\ref{subsec:tune} we test these coefficients 
through a precision study of the dispersion relation for 
$\Upsilon$ and $\eta_b$ mesons. 

The other coefficients in the NRQCD action are $c_2$, $c_3$ and 
$c_4$. $c_3$ and $c_4$ multiply spin-dependent terms that 
give rise respectively to spin-orbit and spin-spin fine 
structure in the spectrum. Most of the splittings we will discuss
here are `spin-averaged' to remove the effect of these terms and 
so we will generally set $c_3$ and $c_4$ to their tree level 
values of 1. However, in section~\ref{subsub:hyp} we will 
discuss the hyperfine splitting ($M(\Upsilon) - M(\eta_b)$) and show results for 
both perturbatively improved and nonperturbatively determined $c_4$. 
The calculation of the appropriate $c^{(1)}_4$~\cite{Hammant:2011bt} is discussed
in Appendix~\ref{appendix:cicalc}, and the nonperturbative 
determination of $c_4$ and $c_3$ in Appendix~\ref{appendix:cinonpert}. 
The nonperturbative studies 
indicate that the value of $c_3$ is very close 
to 1 for this NRQCD action. 
$c_2$ multiplies a spin-independent term, the Darwin term, 
which can affect spin-independent splittings such as radial and 
orbital excitation energies. Because the Darwin term is field-dependent 
we do not expect it to have such a large effect as kinetic terms, 
and therefore do not expect radiative corrections to $c_2$ to be as 
important as for $c_1$, $c_5$ and $c_6$. However, in subsection~\ref{subsec:tune}
we will investigate the effect of changing $c_2$ so that we 
can estimate concretely the systematic error from not knowing 
its $\mathcal{O}(\alpha_s)$ correction.  

Given the NRQCD action above, the time evolution of 
the heavy quark propagator is given by:
\begin{eqnarray}
G({\bf x},t+1) &=& 
	\left( 1-\frac{a\delta H}{2}\right)\left(1-\frac{aH_0}{2n}\right)^nU^{\dag}_{t}(x) 
	\nonumber \\
        & & \times \left(1-\frac{aH_0}{2n}\right)^n\left(1-\frac{a\delta H}{2}\right) G(\vec{x},t) 
\label{eq:evol}
\end{eqnarray}
with starting condition:
\begin{equation}
G({\bf x},0) = \phi({\bf x})\mathtt{1}.
\label{eq:nrqcdstart}
\end{equation}
The smearing function $\phi({\bf x})$ is used to improve the projection onto a particular state in the spectrum. 
Including a variety of smearing functions is essential to 
obtain accurate results for the splittings between the low 
lying excited states. Full details of the smearing functions 
used will be given in subsection~\ref{subsec:smear}.
The $1$ in equation~\ref{eq:nrqcdstart} is the unit matrix in 
color and (2-component) spin space. 
The parameter $n$ has no physical significance, but is included 
for improved numerical stability of 
high momentum modes that do not contribute to bound states~\cite{Lepage:1992tx}. 
In~\cite{Gray:2005ur} it was demonstrated that radial and orbital mass 
splittings were the same within the statistical errors available there 
for $n=2$ and $n=4$ on coarse lattices. 
The minimum value of $n$ for stability increases as the $b$ quark mass 
in lattice units falls on finer lattices. 
Rather than varying 
$n$ as we change the quark mass, here we use $n=4$ throughout 
which is the value appropriate to the fine lattices.
At zero spatial momentum the anti-quark 
propagator is the complex conjugate of the 
quark propagator for a source of the kind given in 
equation~\ref{eq:nrqcdstart}. 

\begin{table}
\caption{ The coefficients $c_1$, $c_5$ and $c_6$ used 
in the NRQCD Hamiltonian of equation~\ref{eq:deltaH}
on the very coarse (sets 1 and 2), coarse (sets 3 and 4) 
and fine (set 5) ensembles. 
Other coefficients had values 1 except for calculations 
in which we specifically changed their values to test
the effect, as described in the text. 
}
\label{tab:wilsonparams}
\begin{ruledtabular}
\begin{tabular}{llll}
Set & $c_1$ & $c_5$ & $c_6$ \\
\hline
very coarse & 1.36 & 1.21 & 1.36 \\
coarse & 1.31 & 1.16 & 1.31 \\
fine & 1.21 & 1.12 & 1.21 \\
\end{tabular}
\end{ruledtabular}
\end{table}

\begin{table}
\caption{ Parameters used in the NRQCD action for our calculations 
that included a full $5\times5$ matrix of correlators. Other parameters 
have been used in subsidiary test calculations as described in the text. 
$am_b$ is the 
bare $b$ quark mass and $u_{0L}$ the Landau link tadpole-improvement 
factor used in the NRQCD action. The different number of digits 
given in the $u_{0L}$ column reflect the precision with 
which it was determined. $n_{{\rm cfg}}$ gives the number 
of configurations used in each ensemble and $n_t$ is the number 
of starting time sources per configuration. $T_p$ is the time 
length of each propagator in lattice units. $a_{sm}$ is 
the parameter for the smearing function described in 
subsection~\ref{subsec:smear}.  	
}
\label{tab:upsparams}
\begin{ruledtabular}
\begin{tabular}{lllllll}
Set & $am_b$ & $u_{0L}$ & $n_{{\rm cfg}}$ & $n_t$ &  $T_p$ & $a_{sm}$  \\
\hline
1 & 3.42 & 0.8195 & 1021 & 16 & 40 & 0.79 \\
2 & 3.39 & 0.82015 & 1000 & 16 & 40 & 0.80 \\
\hline
3 & 2.66 & 0.834  & 1053 & 16 & 40 & 1.0 \\ 
4 & 2.62 & 0.8349 & 1000 & 16 & 40 & 1.0 \\ 
\hline
5 & 1.91 & 0.8525 & 874  & 16 & 48 & 1.37 \\ 
\end{tabular}
\end{ruledtabular}
\end{table}

Details of various parameters used in our calculation are listed 
in table \ref{tab:upsparams}. Tuning of the bare $b$ quark mass 
will be discussed in subsection~\ref{subsec:tune}. 
The tadpole parameters $u_{0L}$ were calculated by fixing a 
subset of each ensemble to lattice Landau gauge using a Fourier-accelerated 
steepest descents algorithm~\cite{Davies:1987vs} to maximise the average 
trace link ($\sum_{\mu=1,4;x}\Tr U_{\mu}(x)$), which value, normalised, 
then becomes $u_{0L}$. 
The whole ensemble was then fixed 
to Coulomb gauge by using the same algorithm to maximise 
the spatial trace link ($\sum_{i=1,3;x} \Tr U_{i}(x)$) 
to allow us to use `wave-function' smearing 
operators, with parameter $a_{sm}$ as described in subsection~\ref{subsec:smear}. 
Propagators were calculated from 16 time sources on 
each configuration to minimise statistical errors. 
Because in NRQCD we operate a simple time evolution we 
can choose the time length of each propagator. This we 
take to be greater than or equal to half the time extent of the 
lattice as detailed in Table~\ref{tab:upsparams}. 

%
\subsection{Smearing functions and multiexponential fits}
\label{subsec:smear}
%

Quark propagators are generated using three different smearing functions 
which we label as local, ground state and excited state. 
They are chosen to improve the projection onto different 
radially excited states and previous experience has shown 
that `hydrogen-like' wavefunctions work well~\cite{Gray:2005ur}.
\begin{eqnarray}
\phi_l(r) &=& \delta_{r,0} 
\nn \\
\phi_{gs}(r) &=& \exp(-r/a_{sm})
\nn \\
\phi_{es}(r) &=& (2a_{sm} - r)\exp(-r/a_{sm}).
\end{eqnarray}
$a_{sm}$ is the smearing radius and is chosen to be approximately the same in physical units for each ensemble. Values are given in Table \ref{tab:upsparams}.
Since a different smearing function can be applied 
separately to the quark and anti-quark we can make five different 
combinations as detailed in Table~\ref{tab:smearing}. 

\begin{table}[h]
\caption{Smearing combinations used for either 
the source or the sink in the construction of $S$-wave correlators. }
\label{tab:smearing}
\begin{center}
\begin{tabular}{l|c|c}
Name & quark  & anti-quark \\
  & smearing  & smearing\\
\hline
l & $\phi_l $   & $\phi_l$\\
g & $\phi_{gs}$ & $\phi_l$ \\
e & $\phi_{es}$ & $\phi_l$ \\
G & $\phi_{gs}$ & $\phi_{gs}$\\
E & $\phi_{es}$ & $\phi_{es}$
\end{tabular}
\end{center}
\end{table}

A different smearing can also be applied at the source 
and the sink making correlator cominations labelled by e.g. lg,le,gG. 
The different smearing combinations 
allow the construction of up to a $5\times5$ matrix of 
correlators for the $S$-wave states that can be fit simultaneously.
The cross-correlators provide further useful information 
beyond that in the diagonal terms 
that can be used in the fitting to extract the excited states more precisely. 
The correlators with quantum numbers of $^3S_1$ or 
$^1S_0$ are distinguished by the insertion of either 
a ${\bf \sigma}$ or a ${\bf 1}$ in spin space at source 
and sink~\cite{Davies:1994mp}.  

To make $P$-wave states we use only the l and g smearings above 
and apply a symmetric difference operator, $\Delta$ to the smeared 
source to give a $P$-`wavefunction'. This propagator is 
combined with that from a $\delta$ function source and 
a derivative applied at the sink to make a $P$-wave meson correlator. 
The complete set of combinations of $\sigma$ matrices with 
derivatives that are needed for the $P$-wave states is 
given in~\cite{Davies:1994mp}. On the lattice the 5-dimensional 
spin 2 representation is split into $E$ and $T_2$ representations 
of the lattice rotational group and we fit these representations 
separately since differences in mass between them can arise 
from discretisation errors on the lattice.  

For the $S$-wave states, statistical errors were improved further 
by using random wall sources in combination with the smearings discussed 
above. The delta function quark source is replaced with a (pseudo-)random colour vector $\eta_a(\vec{x}) \in U(1)$ at each spatial point of the initial time slice.
When the meson correlator is constructed, 
the white noise property $\langle\eta_a(\vec{x}) \eta^\dagger_b(\vec{y}) \rangle = \delta_{ab}\delta(\vec{x}-\vec{y}) $ 
ensures that the random noise cancels at all points except those 
where the initial spatial sites are the same.
This can be combined with the smearing functions by distributing the random 
number associated with the centre of each smearing function along with 
the smearing function. Then once again the white noise property will mean 
that the resultant correlator averages over the initial time source the 
effect of having a smeared source at every point~\cite{kendallphd}. 
Previous studies have found a significant improvement in the precision of the Upsilon ground state energy using random wall sources \cite{Davies:2009tsa}. The improvement 
is less clear for excited states and therefore we did not use this 
technique for the $P$-wave states. 

\begin{figure*}
\includegraphics[width=0.3\hsize]{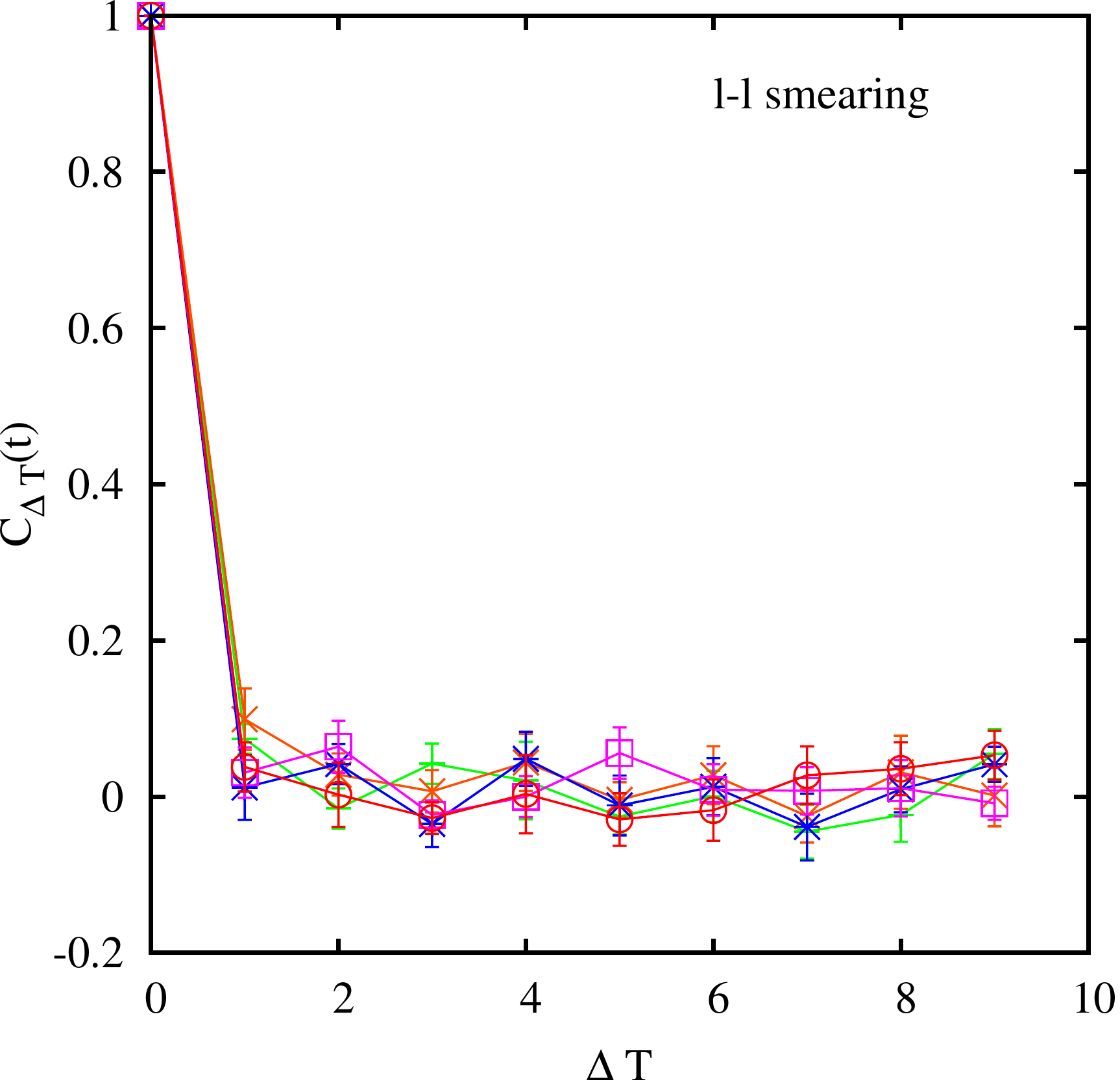}
\includegraphics[width=0.3\hsize]{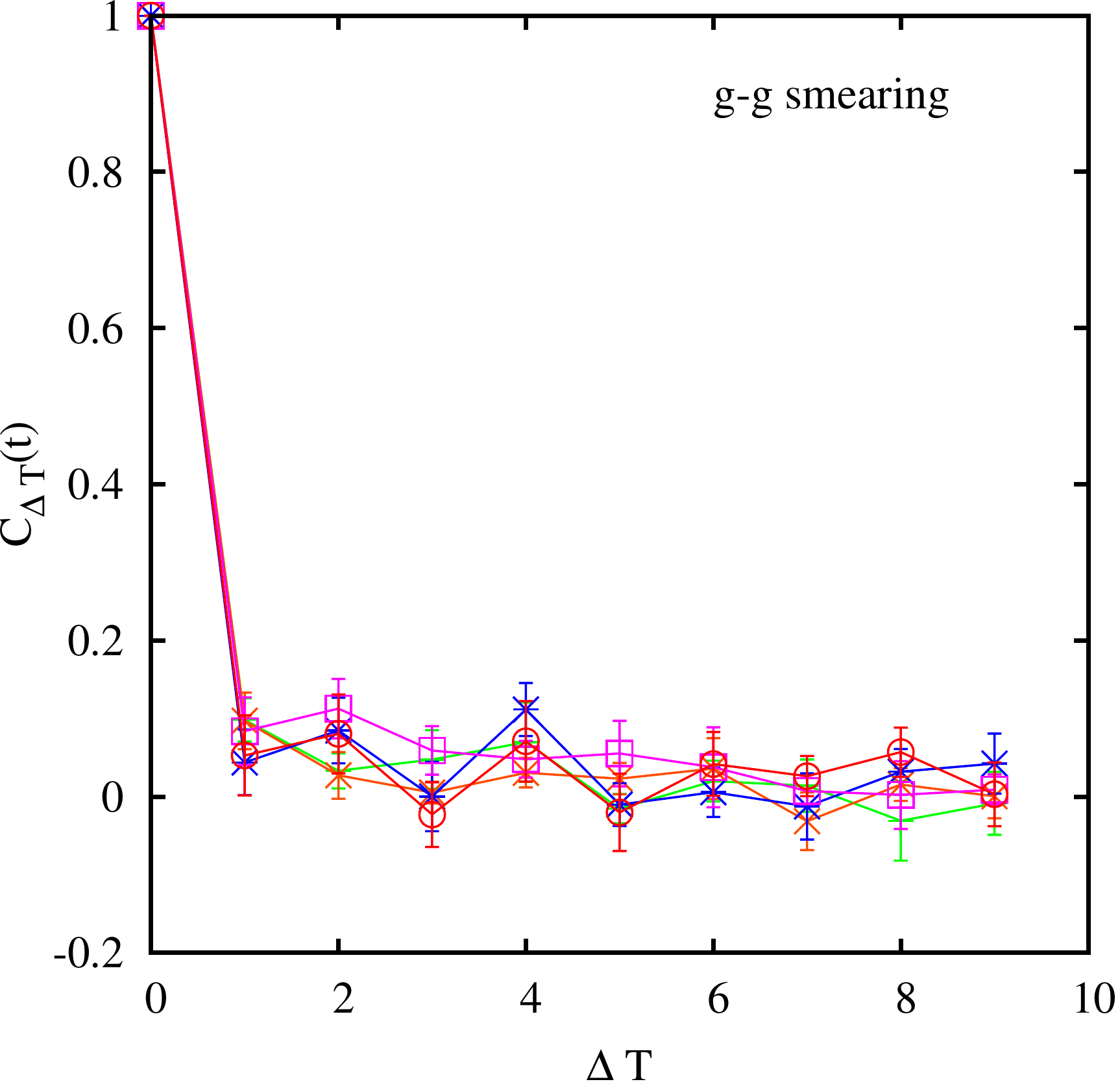}
\includegraphics[width=0.3\hsize]{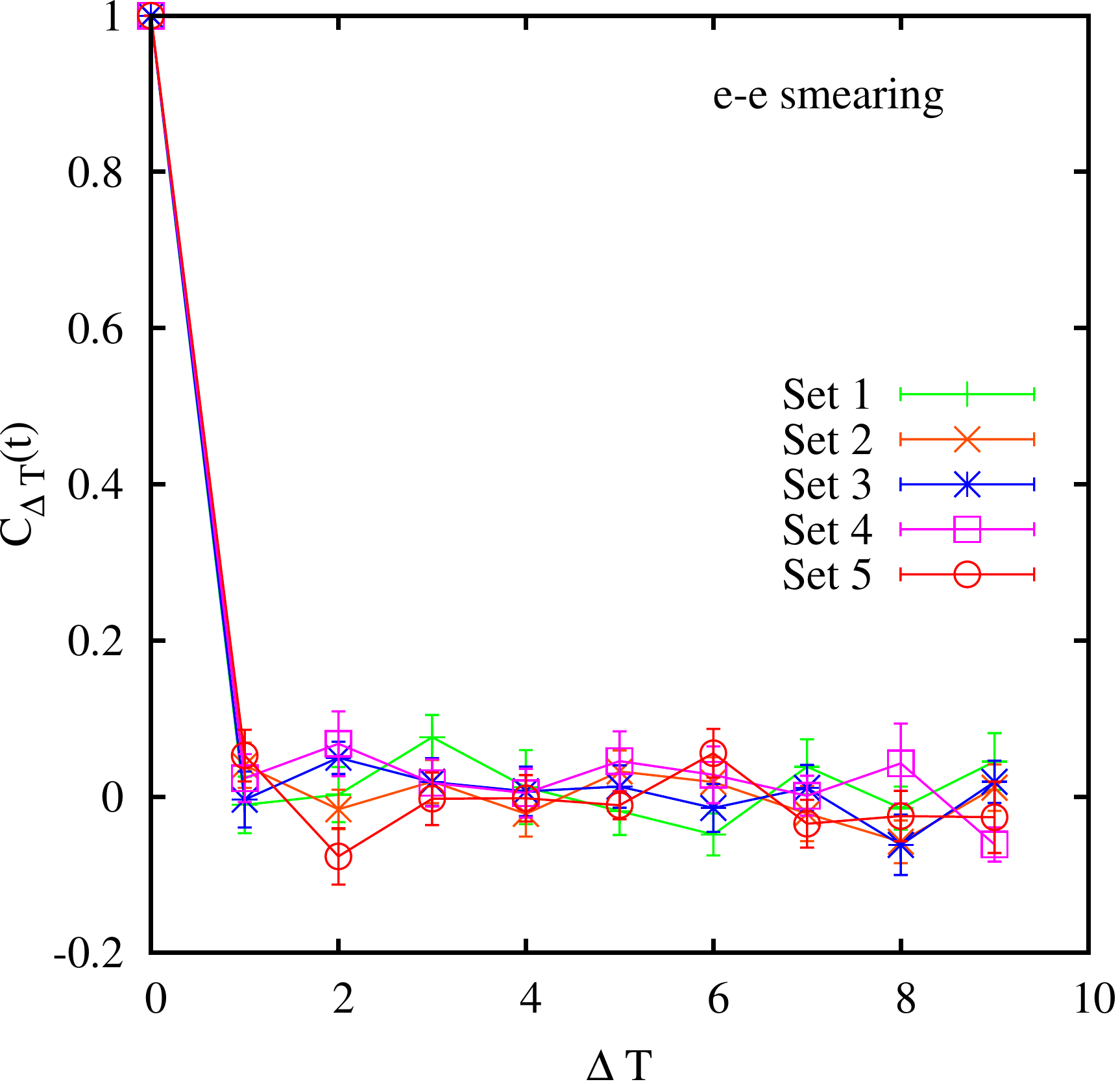}
\caption{Autocorrelation function $C_{\Delta T}$ for $\Upsilon$ correlators 
made from different smearing combinations, from left to right: ll, gg and 
ee. Different symbols are given to different ensembles according 
to the key on the right in the ee plot (color online). 
The correlators are evaluated at lattice time separation $t/a=4$ on 
very coarse lattices (sets 1 and 2), $t/a=5$ on coarse lattices (sets 3 and 4) 
and $t/a=8$ on fine lattices (set 5). This corresponds to a $t$ value 
where the gg correlators have reached the ground-state plateau and 
the ee correlators have a short plateau corresponding approximately 
to the first excited state mass.  
$\Delta T$ gives the separation at which the autocorrelation 
is measured in units of numbers in the ordered ensemble list. 
 }
\label{fig:ups-autocorr}
\end{figure*}

Propagators were calculated from 16 time sources on 
each configuration but to avoid correlations between time sources,
the correlators were binned over all sources on the same configuration.
Autocorrelations between results on successive configurations in an 
ensemble were studied by calculating the autocorrelation function 
$C_{\Delta T}$~\cite{Gattringer:2010zz}:
\begin{equation}
C_{\Delta T} = \frac{\langle x_ix_{i+\Delta T}\rangle - \langle x_i \rangle\langle x_{i+\Delta T}\rangle}{\langle x_i^2\rangle -\langle x_i\rangle^2}.
\label{eq:autocorr}
\end{equation}
Here $x_i$ represents a correlator on a given ensemble, $i$. $x_{i + \Delta T}$ 
is the correlator on an ensemble separated by ${\Delta T}$ from $i$ in the ordered
ensemble i.e. $\Delta T = 1$ corresponds to neighbouring configurations in 
the ensemble. The ensembles have been generated taking into account the fact 
that autocorrelations increase on finer lattices. Thus neighbouring configurations 
are 5 trajectories apart for very coarse and coarse ensembles but 
6 trajectories apart for the fine ensemble~\cite{Bazavov:2010ru}. 
$C_{\Delta T}$ is plotted against $\Delta T$ in Figure~\ref{fig:ups-autocorr} 
for the case where $x$ is an $\Upsilon$ correlator measured with a 
time separation on the lattice of approximately 0.6 fm. This value 
was chosen to correspond to a point where correlators were 
dominated by the ground-state. The picture is qualitatively 
the same for different time separations, however. $C_{\Delta T}$ 
drops to zero very rapidly, within the separation $\Delta T=1$.
We therefore do not have to worry about autocorrelations 
between configurations but can treat them all as statistically 
independent. 

\begin{figure}[t]
\includegraphics[width=0.9\hsize]{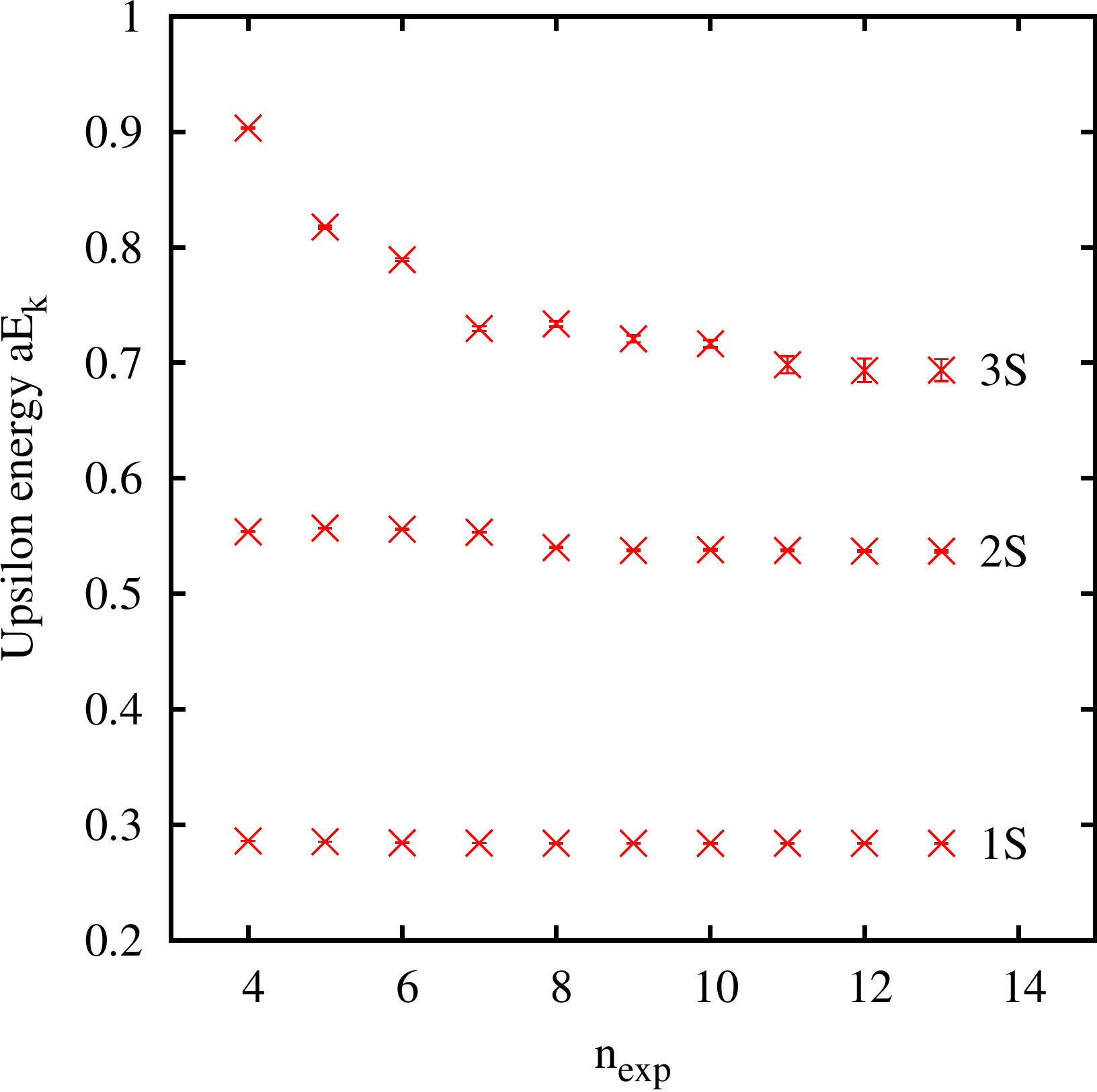}
\caption{Energies in lattice units of the low lying $\Upsilon$ states for the fine ensemble, 
set 5, from the full $5\times5$ lgeGE fit plotted 
against the number of exponentials, $n_{\rm{exp}}$, included in the fit.}
\label{fig:error_vs_nexp}
\end{figure}

Bayesian fitting is used to extract the spectrum from the 
correlators~\cite{gplbayes}. 
The fit function
\begin{equation}
G_{\mathrm{meson}}(n_{sc},n_{sk};t) = \sum_{k=1}^{n_{\rm{exp}}}a(n_{sc},k)a^*(n_{sk},k)e^{-E_kt}
\label{eq:fit}
\end{equation}
is used, where $aE_k$ is the energy of the $(k-1)$th radial 
excitation in lattice units and $a(n_{sc/sk},k)$ are the corresponding 
amplitudes labelled by the smearing used at the source and 
sink of the correlator, i.e. $sc,sk\in \{l,g,e,G,E \}$.
We fit the full range of $t$ values for the correlator
from 1 to $T_p$, where $T_p$ values are given for $S$-wave 
fits in Table~\ref{tab:upsparams} and $T_p=20$ for $P$-waves. 
The number of terms, $n_{\mathrm{exp}}$, in the fit is varied, however, 
and Bayesian model selection criteria are applied to determine 
which fit is used. In practice, this means adding 
additional terms to the fit until the results and the errors stabilise. 
An example is given in Figure~\ref{fig:error_vs_nexp}.

The Bayesian approach allows the inclusion of prior data into the fitting procedure. The $\chi^2$ test function is amended to 
\begin{equation}
 \chi^2_{\mathrm{aug}} = \chi^2 + \chi^2_{\mathrm{prior}}
\end{equation}
and the function $\chi^2_{\mathrm{aug}}$ is minimised. 
By Bayes' theorem this corresponds to maximising the posterior probability 
p(parameters$|$data) as opposed to a standard $\chi^2$ test 
which maximises only the likelihood function p(data$|$parameters).
$\chi^2_{\mathrm{prior}}$ is taken to be
\begin{equation}
 \chi^2_{\mathrm{prior}}=
\sum_k
\frac{(p_k - \tilde{p}_k)^2 }{\tilde{\sigma}_{p_k}^2}
\end{equation}
for each fit parameter $p_k$. 
This assumes that the prior probablility density function 
for each parameter is a Gaussian with central value $p_k$ and 
width $\tilde{\sigma}_{p_k}$. The fit parameters are: the 
amplitudes, which are taken to have a prior of $0.1 \pm 1.0$; 
the ground state energies $\ln(E_0)$ which are estimated from 
an effective mass plot and given a suitably wide width; and 
the splittings $\ln(E_{n+1} - E_n)$ which prior information 
tells us should be of the order 500 MeV with a width of 250 MeV. 
Taking the fit parameters to be the logarithms of the energy 
splittings ensures that the ordering of the states is respected.

$\chi^2_{\rm{aug}}$ is minimised using the singular value 
decomposition (SVD) method. In the larger matrix fits, the 
correlation matrix can become ill-conditioned and it can be 
necessary to introduce a cutoff, $w_{\mathrm{cut}}$, on the 
lowest eigenvalues of the correlation matrix in order to 
fit the data. A variation of this method is used in which, 
instead of setting eigenvalues below $w_{\mathrm{cut}}w_{\mathrm{max}}$ 
to zero, they are set to $w_{\mathrm{max}}$ times $w_{\mathrm{cut}}$. 
This is a less severe truncation of the correlation matrix and 
it improves the fits in some cases. $w_{\mathrm{cut}}$  was 
typically taken to be $10^{-4}$ for the $5\times5$ matrix fits.

\begin{figure}
\includegraphics[width=0.9\hsize]{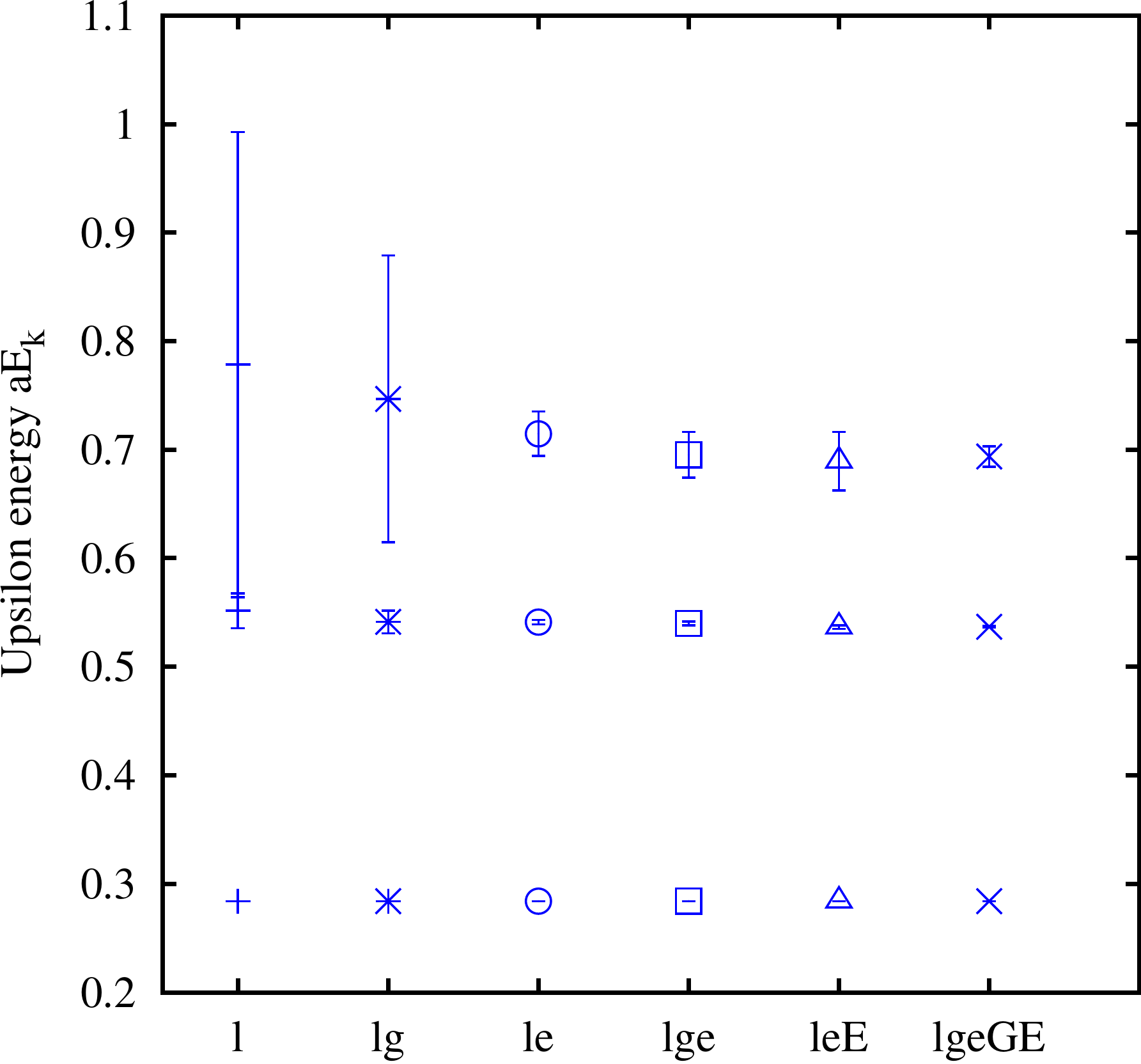}
\caption{Comparison of the effect of different smearing combinations 
for extraction of the energies of the ground and first two radially 
excited $\Upsilon$ states.
The energy in lattice units from the fine ensemble is shown for a $1\times1$ l fit (plus), $2\times2$ fits lg (star) and le (circle), $3\times3$ fits lge (square) and leE (triangle), and the $5\times5$ fit containing all sources lgeGE (cross).
 }
\label{fig:smearing_comparison}
\end{figure}

In order to determine whether the inclusion of five 
different smearing operators actually leads to 
improved results, the energies of the low lying $\Upsilon$ states 
are plotted in Figure~\ref{fig:smearing_comparison} for a 
variety of different matrix fits from the fine ensemble. 
The effect on the precision of the ground state is negligible 
but the full $5\times5$ fit has significantly smaller errors for 
the first two excited states.

Because NRQCD is a nonrelativistic effective theory, there is an 
energy offset. Thus the energies obtained from correlators at zero 
momentum do not correspond to meson masses. Energy differences 
do correspond to mass differences, however and so, for example, 
the mass difference between the $\Upsilon^{\prime}$ and 
the $\Upsilon$ (in lattice units) is given simply 
by $aE_2-aE_1$ from equation~\ref{eq:fit}. To obtain absolute mass 
values requires the study of correlators for mesons at 
nonzero spatial momentum  as discussed in Sec.~\ref{subsec:tune}. 

%
\subsection{NRQCD systematics in tuning the $b$ quark mass}
\label{subsec:tune}
%

In this calculation the parameters of QCD that need to be determined 
are the $b$ quark mass and $\Lambda_{\rm QCD}$. In practice this 
translates into the fact that we
need to tune the $b$ quark mass parameter in the lattice NRQCD 
Hamiltonian until we obtain the correct value for one calibration 
hadron mass and we need to determine the lattice spacing  
from another calibration hadron mass. After that is done all other hadron 
masses are determined with no further tuning. 
The two calibration hadrons should be chosen with rather different 
properties. The mass chosen to fix the $b$ quark mass should ideally 
be very sensitive to that value; the mass chosen to determine the 
lattice spacing should be as independent of the $b$ quark mass 
as possible to avoid a complicated iterative tuning process. 
To determine the lattice spacing we choose the radial excitation 
energy of the $\Upsilon$, i.e. $M(\Upsilon^{\prime}) - M(\Upsilon)$. 
This is known from experiment to be very insensitive to the 
heavy quark mass since it changes by only 4\% between the $b$ and 
the equivalent quantities for the $c$ quark, which has a mass a factor of 
4.5 smaller. The determination of the lattice spacing from this 
quantity will be discussed in section~\ref{subsec:results}. 
Here we focus on the tuning of the $b$ quark mass and in particular 
on the effect of the improvements to the NRQCD action which 
we have implemented here for the first time. 

As discussed in section~\ref{subsec:smear} the fitted energy from a 
zero momentum hadron correlator made from NRQCD propagators
is not the hadron's mass because there is an energy offset. 
Instead we must determine the `kinetic mass' from the 
energy-momentum dispersion relation: 
\begin{equation}
 aM_{\mbox{\tiny Kin}} = 
\frac{a^2P^2 - (a\Delta E)^2}
{2a\Delta E},
\label{eq:mkin}
\end{equation}
where $a\Delta E$ is the energy difference between the meson 
with momentum $Pa$ in lattice units and the meson at rest. 
Equation~\ref{eq:mkin} assumes a fully relativistic dispersion 
relation, i.e. 
\begin{equation}
aE(P) = aE(0) + \sqrt{a^2P^2 + a^2M_{\mbox{\tiny Kin}}^2}.
\label{eq:disp}
\end{equation}
Systematic errors will then be present in the kinetic 
mass for lattice NRQCD both because the action is only accurate 
to a specific order 
in the expansion in $v^2/c^2$ and from lattice discretisation 
errors. Here we study both of these effects. 
First it is worth briefly recapitulating a discussion from the literature 
(see, for example,~\cite{freeland}) on how the kinetic mass is built up in a 
nonrelativistic approach as successive orders in $v^2/c^2$ 
are added to the nonrelativistic expansion, 
because it provides a useful
handle on systematic errors. 

By definition the mass of a meson is given by the sum of the masses of 
its constituent quarks plus the binding energy. The binding energy 
has contributions from the internal kinetic energy, i.e. the motion 
of the constituent quarks relative to the centre of mass, and 
from the interaction energy. If we write the meson dispersion 
relation in the standard nonrelativistic expansion as:
\begin{equation}
E({\bf P}) = M_1 + \frac{{\bf P}^2}{2M_2} + \ldots
\label{eq:dispmeson}
\end{equation}
then $M_1$ is known as the static mass and $M_2$ is the kinetic mass, equal 
to $M_{\mbox{\tiny Kin}}$ in equation~\ref{eq:mkin} up to relativistic corrections. 
It should be possible to construct the correct meson mass from 
both $M_1$ and $M_2$ i.e. the binding energy contribution needs 
to feed correctly into both of them. 

To see how this works in outline it is sufficient to study two 
free particles. The total energy of the two particle system 
is the sum of the masses, $m_{i}$, 
plus the kinetic energies, ${\bf q}_i^2/2m_{i}$ for each particle. 
In the center of mass frame (${\bf P}=0$) this is simply $m_{1}+m_{2}$ 
plus the internal kinetic energy. As is well-known, the internal kinetic 
energy can be written to leading 
nonrelativistic order as 
${\bf p}^2/2\mu$ where ${\bf p}$ is the momentum of either particle 
in this frame and $\mu$ is the reduced mass 
($1/\mu = 1/m_{1} + 1/m_{2})$. 
Thus $M_1$ takes 
the expected form for this two particle system. 
To study $M_2$ we must include the 
motion of the centre of mass and expand the sum of the two particle kinetic 
energies to $\mathcal{O}({\bf P}^2)$. For $M_2$ to have the 
correct form including the leading piece of the internal 
kinetic energy we need $E(P)$ to take the form    
\begin{eqnarray}
E(P) &=& m_{q1} + m_{q2} + \frac{{\bf p}^2}{2\mu} +\ldots\\ \nonumber 
+&& \frac{{\bf P}^2}{2(m_{q1}+m_{q2})}\left( 1 - \frac{{\bf p}^2}{2\mu(m_{q1}+m_{q2})} +\ldots\right)
\label{eq:dispmeson2}
\end{eqnarray}
i.e. we need to locate a ${\bf P}^2{\bf p}^2$ term in the sum of the 
individual particle kinetic energies. This requires the individual 
kinetic energies to be expanded beyond leading order in the nonrelativistic 
expansion to include terms at fourth order in the momentum. 
Thus $M_2$ will have the correct form to leading order in the internal 
kinetic energy if the individual kinetic energy terms are correct 
through next-to-leading-order in momentum. 
In an interacting theory we also need the interaction terms to be 
correct through $\mathcal{O}(v^4)$ to have the binding energy 
correctly included in the kinetic mass. 

These issues are discussed in some detail in~\cite{freeland} for heavy quarks using the 
clover action since there are important differences in discretisation 
errors there between 
choosing $M_1$ or $M_2$ as the appropriate meson mass against which 
to tune the quark mass. 
In NRQCD we must use $M_2$ ($M_{\mbox {\tiny Kin}}$). The quark Hamiltonian given 
in equation~\ref{eq:deltaH} has no quark mass term, so to reconstruct the 
meson mass from $M_1$ would require adding back in the 
zero of energy. This is perturbatively calculable but we wish 
the tune the quark mass fully nonperturbatively. 
$M_2$ on the other hand acquires its quark mass pieces from the 
quark kinetic energy terms and so has no zero of energy 
problem. As discussed above, $M_2$ will also 
correctly include the internal kinetic energy if the $v^4$ 
relativistic corrections to the kinetic energy 
are included in the quark 
Hamiltonian, as they are in equation~\ref{eq:deltaH}.   
Indeed we are now including the radiative corrections to 
the $v^4$ kinetic terms through adjustments to $c_1$, $c_5$ 
and $c_6$, and we will show below the effect that this has.  

\begin{figure}
\includegraphics[width=0.9\hsize]{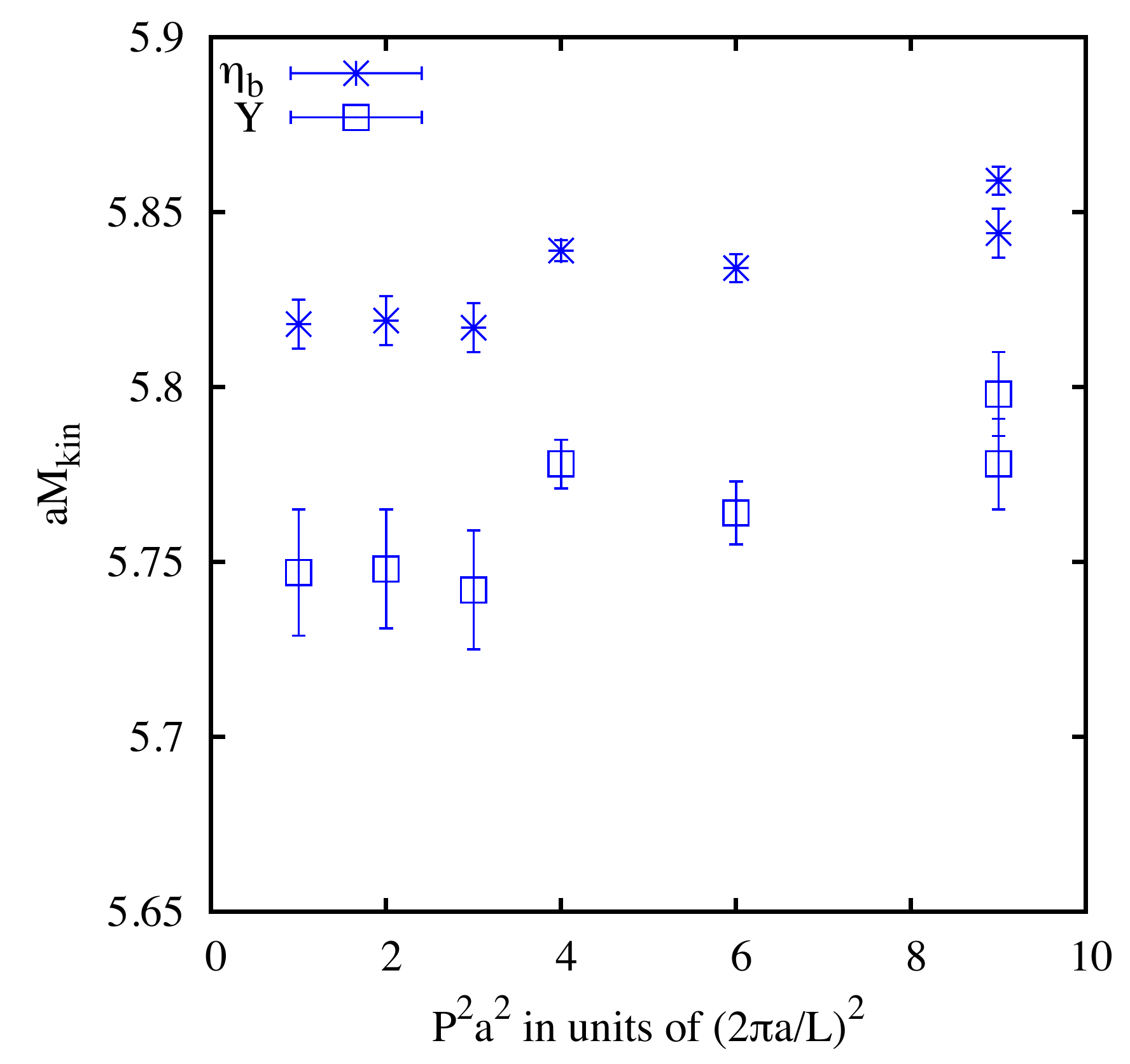}
\caption{ Kinetic mass values in lattice units obtained
on the coarse ensemble, set 3, for the $am_b$ and
$c_i$ values given in Tables~\ref{tab:wilsonparams} and~\ref{tab:upsparams}.
Kinetic mass values are given separately for the 
$\Upsilon$ and $\eta_b$ and plotted
against the square of the lattice momentum in 
units of $2\pi a/L$. The two results at $x$-axis value 
of 9 correspond to momenta with indices $(3,0,0)$ and 
$(2,1,1)$. The higher one is (3,0,0).  
 }
\label{fig:mkinupsetab}
\end{figure}

\begin{figure}
\includegraphics[width=0.9\hsize]{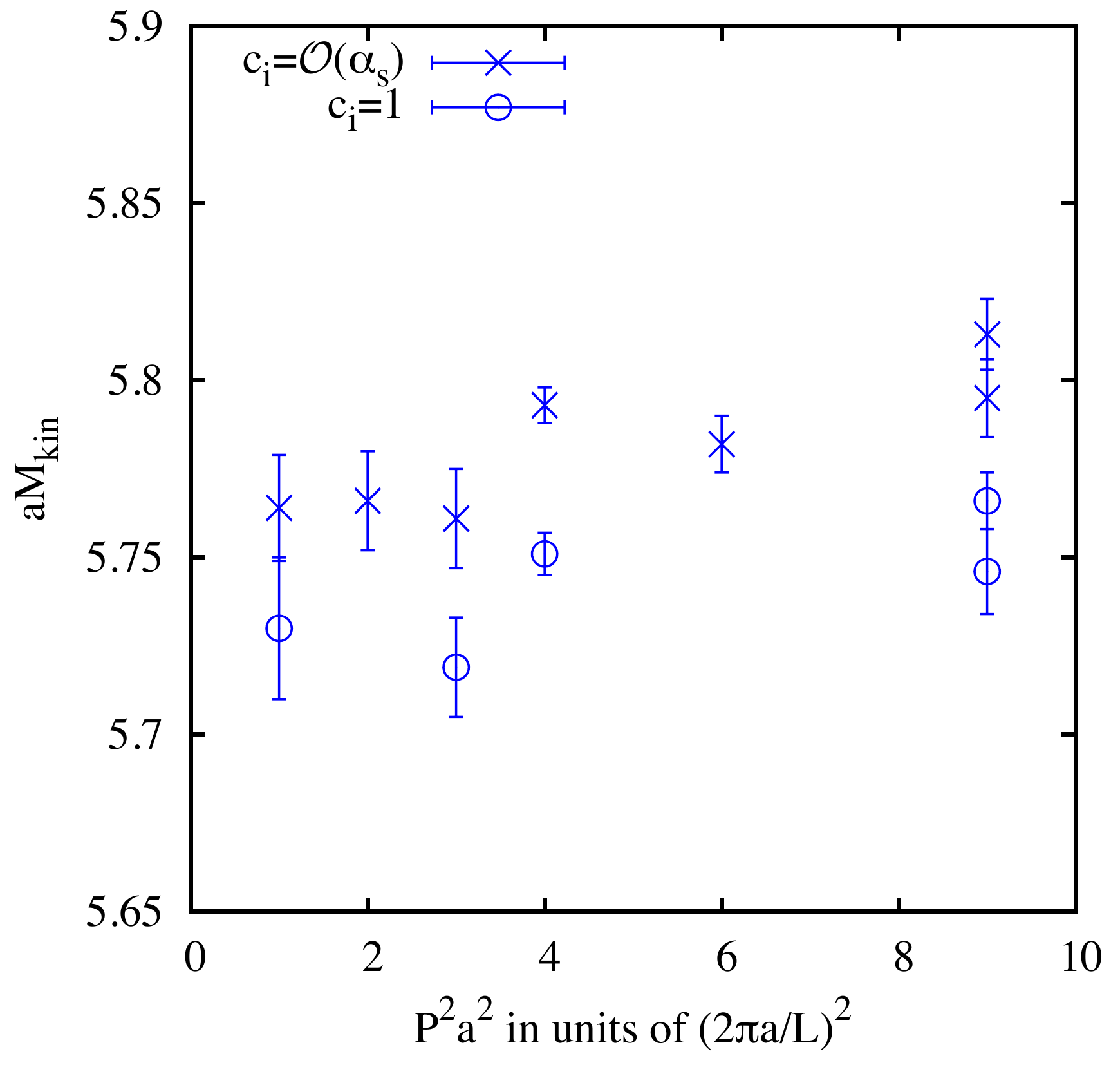}
\caption{ Spin-averaged values for the 
kinetic mass in lattice units obtained
on the coarse ensemble, set 3, for $am_b =2.66$ (as in 
Table~\ref{tab:upsparams}). 
Results for the
$c_i$ values given in Table~\ref{tab:wilsonparams} 
are compared to the results for $c_i=1$.
The kinetic mass is plotted 
against the square of the lattice momentum in 
units of $2\pi a/L$. 
 }
\label{fig:mkinci}
\end{figure}

\begin{figure}
\includegraphics[width=0.9\hsize]{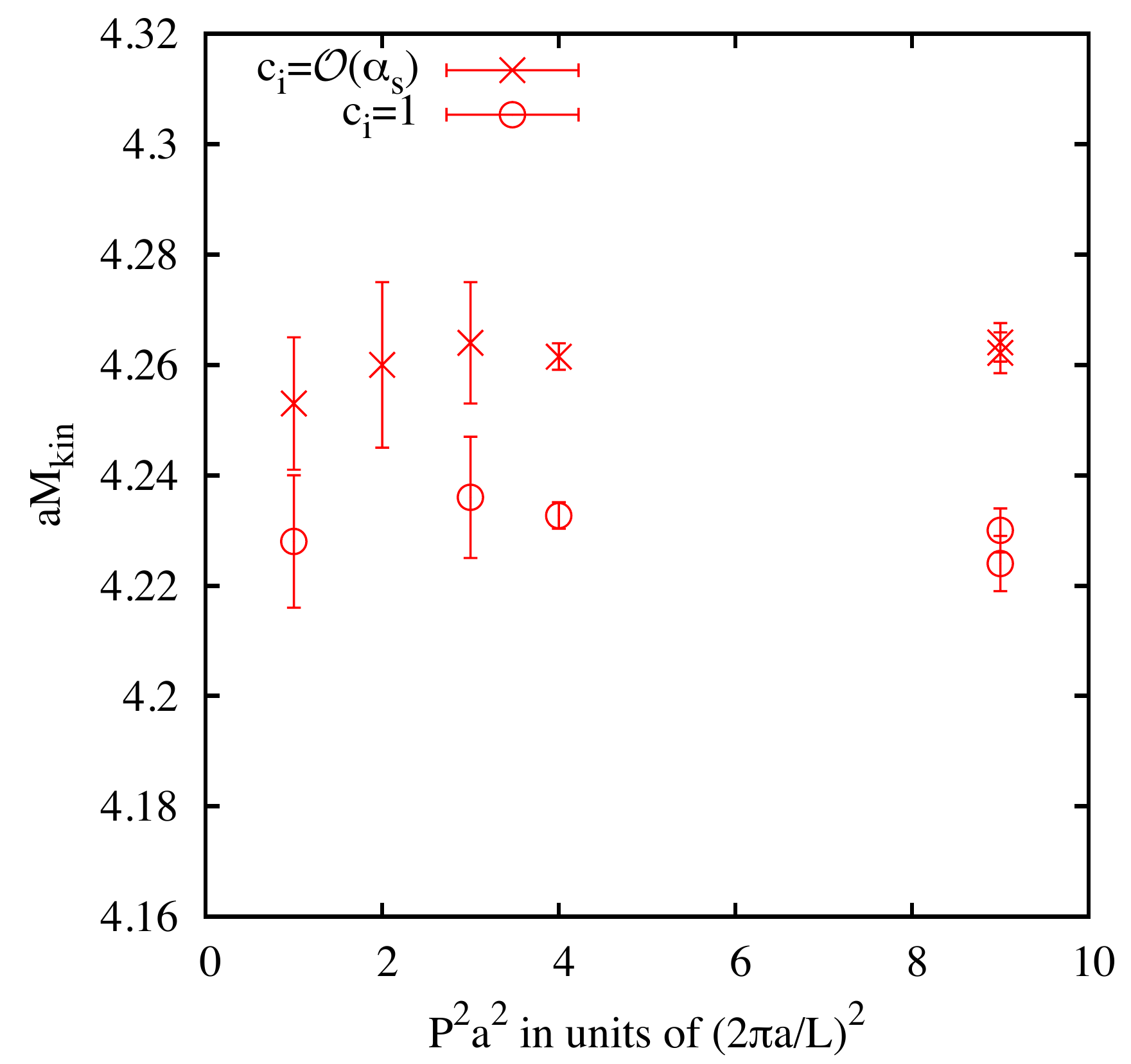}
\caption{ Spin-averaged values for the 
kinetic mass in lattice units obtained
on the fine ensemble, set 5, for the $am_b$ and
$c_i$ values given in Tables~\ref{tab:wilsonparams} and~\ref{tab:upsparams}, 
compared to the results for $c_i=1$.
The kinetic mass is plotted 
against the square of the lattice momentum in 
units of $2\pi a/L$. 
 }
\label{fig:mkinfine}
\end{figure}

We can determine the kinetic mass very precisely by use of 
propagators made starting with a
 random wall source patterned by an $\exp(i{\bf p}\cdot{{\bf x}})$
factor to give the quark momentum~\cite{kendallphd}. 
We use only a $\delta(x)$ smearing 
function
for these calculations so they are very fast, but we must 
evolve both a quark and an antiquark propagator because the complex 
conjugate of a quark propagator of momentum ${\bf p}$ is an 
antiquark of momentum $-{{\bf p}}$. Typically we take quark and 
antiquark momenta to be equal so that the meson 
momentum, when they are combined, is ${\bf P}=2{\bf p}$.   

We fit the meson correlator of momentum ${\bf P}$ simultaneously 
with the meson correlator at rest so that the energy difference 
$a\Delta E$ between the ground state energies 
can be determined directly by the fit taking the 
correlations into account. In this way we obtain $a\Delta E$ 
values with errors typically in the 5th decimal place. 
To avoid cluttering the main body of the text, the 
detailed tables of values for $\Upsilon$ and $\eta_b$ energies 
as a function of momentum
and $aM_{\mbox{\tiny kin}}$ are collected 
in Appendix~\ref{appendix:mkin}. 
Propagators were calculated 
for the full number of configurations given for each ensemble in 
Table~\ref{tab:upsparams}, but in some cases we used 
fewer time sources per configuration than is given there. 

We can then plot out the kinetic mass for 
a range of meson momenta to study systematic effects in 
equation~\ref{eq:mkin} which would show up as a disagreement 
between kinetic masses obtained from different momentum 
values. Previous calculations saw no significant differences in 
kinetic mass values for momenta up to 
$P^2a^2$ = 9 with errors of around 1\%~\cite{Gray:2005ur}.  
This is equivalent to a test, as a function of momentum, 
of the constancy of the `speed of light'. 
Here we are able to achieve errors down to 0.1\%, 
depending on the momentum. 
Then systematic variations of $aM_{\mbox{\tiny Kin}}$ 
with momentum can be seen at the 0.5\% level. 
 
$aM_{\mbox{\tiny Kin}}$ values for $\Upsilon$ and $\eta_b$ mesons
on the coarse lattices, set 3, are plotted in Figure~\ref{fig:mkinupsetab} 
and show several features. One is that there is a systematic 
difference between the values of $aM_{\mbox{\tiny Kin}}$ for 
on-axis (those in one lattice direction only) 
and off-axis momenta. This was hinted at in~\cite{Gray:2005ur} 
but the errors were too large for it to be clear. The on-axis kinetic masses 
are higher, and this reflects a breaking of rotational invariance 
on the lattice which is a discretisation error. It is particularly 
obvious for the momenta with components along the spatial 
directions labelled by integers (3,0,0) and (2,2,1), both 
of which have $P^2a^2 = 9(2\pi a/L)^2$. The difference is 
tiny but visible. 
We will return to this point below. 

Another feature of Figure~\ref{fig:mkinupsetab} 
is that 
the kinetic mass for the $\eta_b$ is above that of the $\Upsilon$ 
which is the opposite way round to the energy difference at 
zero momentum and to experiment. A similar but somewhat smaller 
effect is seen on the fine lattices. The discussion 
above on the way in which the meson kinetic mass is 
built up order by order in the nonrelativistic expansion 
shows how this has happened. It results from the fact that 
the $\sigma \cdot B$ term that gives rise to the hyperfine 
splitting is only included at leading order in our 
NRQCD action, equation~\ref{eq:deltaH}. Relativistic corrections 
to this term would be needed for it to feed correctly 
into the kinetic mass, $M_2$. The effect of the $\sigma\cdot B$ term 
splitting is correctly incorporated in 
the meson energy at zero momentum ($M_1$), however, 
and it is from differences in $M_1$ for $\Upsilon$ and $\eta_b$ that 
we determine the hyperfine splitting (see subsection~\ref{subsub:hyp}). 
This small but non-zero systematic error in $M_2$  
is simply removed by working instead with the 
spin-averaged kinetic mass of the $\Upsilon$ and $\eta_b$: 
\begin{equation}
\overline{M}_{\mbox{\tiny Kin}}(1S) = \frac{(3M_{\mbox{\tiny Kin}}(\Upsilon) +M_{\mbox{\tiny Kin}}(\eta_b) )}{4} 
\label{eq:spinav}
\end{equation}
and using this to fix the $b$ quark mass. 

The above arguments also allow insight into the effect of radiative 
corrections to the $v^4$ kinetic terms in the NRQCD Hamiltonian 
that we include here for the first time. Changing the coefficient 
of the $p^4/8m_b^3$ term, $c_1$, from 1 to $1+\mathcal{O}(\alpha_s)$ 
will modify the amount of the internal kinetic energy that is 
incorporated into the meson kinetic mass, effectively correcting 
for an $\mathcal{O}(\alpha_s)$ mismatch between this contribution 
to $M_1$ and $M_2$ from binding energy effects.
The effect of this radiative correction is seen clearly in 
Figure~\ref{fig:mkinci} where we compare the spin-averaged 
kinetic mass with all $c_i$ set to 1 to that from having 
the radiatively improved coefficients given in Table~\ref{tab:wilsonparams}.  
The difference would be expected to be $\mathcal{O}(\alpha_s \times B)$ 
where $B$ is the binding energy of $\mathcal{O}$(500 MeV). This could in principle 
be as large as 150--200 MeV. From Figure~\ref{fig:mkinci} we see 
that the effect is somewhat smaller than this on the coarse 
ensemble set 3 -- a shift of kinetic mass
of 0.05 in lattice units corresponds to around 80 MeV on 
these lattices. The shift is clearly visible, however. 
The radiative correction acts to increase the kinetic mass 
for a given bare $b$ quark mass. This is because $c_1 > 1$ and 
the binding energy is positive. Thus the correctly tuned quark mass will 
be lower (by the same percentage shift as that 
for the kinetic mass) when radiative corrections are included. 
A similar shift is observed on the fine lattices as shown 
in Figure~\ref{fig:mkinfine}. 

Remaining systematic errors from higher order radiative corrections 
to $v^4$ terms in the NRQCD action will be suppressed by a further 
power of $\alpha_s$ beyond the shift seen here. We therefore expect 
the remaining error in the kinetic mass from this 
source to be $\mathcal{O}$(0.3\%). Systematic errors 
from missing higher order, $v^6$, terms at tree level in the 
NRQCD action are a factor of $v^2$, or 10\%, smaller than the 
size of the effect of $v^4$ terms, and therefore of similar size 
to missing $\alpha_s^2v^4$ terms. They will also have the
effect of correcting for momentum-dependence in $M_{\mbox {\tiny Kin}}$.  
From Figure~\ref{fig:mkinci} we can see that there is a 
sign of an upward drift 
of $M_{\mbox {\tiny Kin}}$ with momentum but the 
effect is smaller than the shift of $M_{\mbox {\tiny Kin}}$ with 
the radiative correction to the $c_i$ coefficients. 

We now return to the issue of discretisation errors 
in the kinetic mass. These arise from the replacement of 
time and space derivatives in the NRQCD action 
with finite differences on the lattice. 
The terms with coefficients $c_5$ and $c_6$ contain 
$a^2v^4$ and $av^4$ correction terms to remove these 
errors. With the inclusion of radiative corrections 
to $c_5$ and $c_6$, the remaining errors are at 
$\mathcal{O}(\alpha_s^2 a^2v^4)$ in this calculation. 
The term with coefficient $c_5$, i.e. the term proportional 
to $\Delta^{(4)}$ is of interest because this 
is rotationally non-invariant. 
The signal for a lack of continuum rotational invariance in our results is 
a disagreement between the kinetic mass for on-axis 
momenta, that typically have a high value for $P_i^4$, 
and off-axis momenta. This was seen in Figure~\ref{fig:mkinupsetab} 
for the coarse lattices. Less variation is evident 
on the fine lattices (Figure~\ref{fig:mkinfine}), as expected
for a discretisation effect. 

\begin{figure}
\includegraphics[width=0.9\hsize]{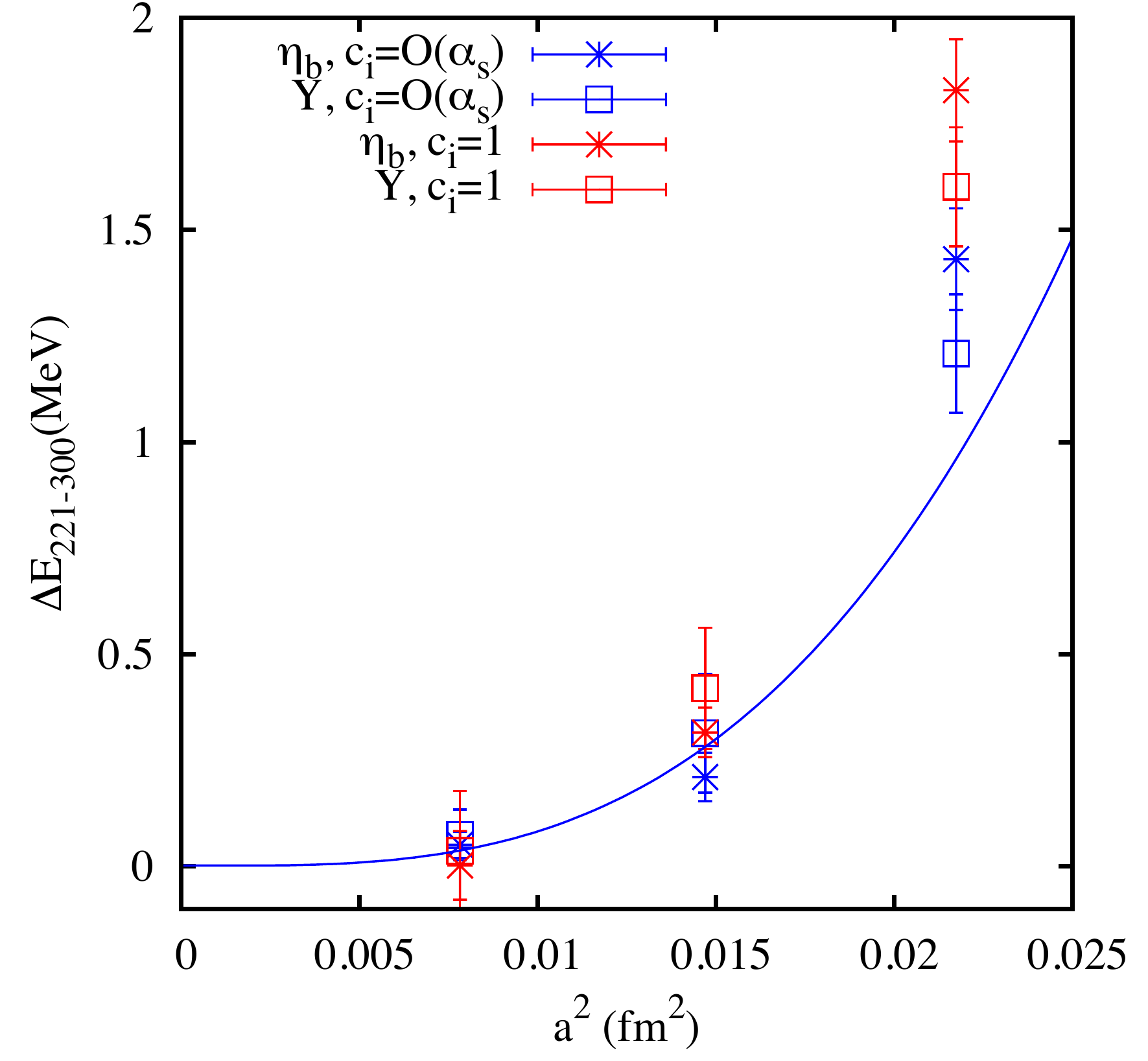}
\caption{ The energy difference in MeV between
mesons with momentum (3,0,0) and (2,2,1) 
in units of $2\pi a/L$ 
on the lattice plotted against the 
square of the lattice spacing in fm.   
Results are shown for the case $c_{1,5,6}=1$ 
as well as for $c_{1,5,6}$ taking their 
$\alpha_s$-improved values. 
An example fit curve with $a^4$ and $a^6$ 
dependence is shown through the $\Upsilon$ 
data for $c_{1,5,6}$ 
$\alpha_s$-improved. 
 }
\label{fig:esplit}
\end{figure}

To make clearer the way in which the 
rotationally noninvariant discretisation 
errors depend on the lattice spacing 
Figure~\ref{fig:esplit} 
plots the energy difference 
in physical units between mesons with 
momentum (3,0,0) and (2,2,1) as a function of 
$a^2$ using results from all three values of the 
lattice spacing. 
$P^2a^2= 9(2\pi a/L)^2$ corresponds to 
approximately the same physical momentum 
at all three lattice spacing values, 
so the results should be a good test of how 
rotational invariance is restored as $a \rightarrow 0$. 
In fact the energy difference is tiny on all except the 
very coarse lattices, where it reaches 1 MeV. 
The case in which the $c_{1,5,6}$ coefficients
are set to their tree level values of 1 is plotted 
as well as the case with the $c_{1,5,6}$ coefficients 
taking the radiatively improved values that we 
have used for the rest of our calculation 
here. The radiatively improved 
values give very slightly smaller energy splittings, 
since they have improved the $a^2$ 
contribution to this error by one order in $\alpha_s$ 
to $\alpha_s^2 a^2 v^4$.  
The energy difference between 
mesons with momentum (3,0,0) and (2,2,1) also has contributions 
at $\mathcal{O}(a^4v^6)$, however, and both the effect 
of radiative improvement and the shape of the 
curve in Figure~\ref{fig:esplit} tend to imply that 
these $a^4$ terms dominate over any remaining $a^2$ terms. 

Rotationally invariant discretisation errors 
would give rise to a kinetic mass that 
varied with $P^2$. This is the same effect as that 
of relativistic 
errors, because the correcting operators are the same.
Discretisation errors require an $a$-dependent coefficient 
to correct them. 
However, as discussed above under relativistic corrections, 
there is no sign in our results of such errors to better
than 0.5\%. 

%
%
\begin{table*}[t]
\centerline{
\begin{tabular}{llllllll}
\hline
\hline
Set  & $am_b$ & $c_{1,5,6}$ & $c_2$ & $c_4$ & $aM_{\mbox{\tiny Kin}}(\Upsilon)$ & $aM_{\mbox{\tiny Kin}}(\eta_b)$ & $a\overline{ M}_{\mbox{\tiny Kin}}(1S) $ \\
\hline
1 & 3.42 & $\alpha_s$ & 1    & 1    & 7.269(18) & 7.405(10)  & 7.303(15)  \\
1 & 3.42 & $\alpha_s$ & 1    & 1.22    & 7.271(22) & 7.472(10)  & 7.321(18)  \\
\hline
2 & 3.39 & $\alpha_s$ & 1    & 1    & 7.228(10) & 7.345(4)  & 7.257(8)   \\
2 & 3.42 & $\alpha_s$ & 1    & 1    & 7.310(14) & 7.423(7)  & 7.338(13)  \\
\hline
3 & 2.66 & 1          & 1    & 1    & 5.703(17) & 5.767(7)  & 5.719(14) \\
3 & 2.66 & $\alpha_s$ & 1    & 1    & 5.742(17) & 5.817(7)  & 5.761(14) \\
3 & 2.66 & $\alpha_s$ & 1.25 & 1    & 5.748(8)  & 5.823(4)  & 5.766(7)  \\ 
3 & 2.66 & $\alpha_s$ & 1    & 1.25 & 5.767(10) & 5.889(4)  & 5.798(8)  \\ 
\hline
4 & 2.62 & $\alpha_s$ & 1    & 1    & 5.706(9)  & 5.761(4)  & 5.719(7)   \\
4 & 2.66 & $\alpha_s$ & 1    & 1    & 5.778(11) & 5.833(5)  & 5.792(10)  \\
\hline
5 & 1.91 & 1 & 1    & 1    & 4.230(13) & 4.252(6) & 4.236(11)  \\
5 & 1.91 & $\alpha_s$ & 1    & 1    & 4.256(14) & 4.287(6) & 4.264(11)  \\
5 & 2.0  & $\alpha_s$ & 1    & 1    & 4.431(11) & 4.466(5) & 4.439(10)  \\
\hline
\hline
\end{tabular}
}
\caption{Summary of the kinetic masses obtained on different ensembles for a variety 
of parameter values. 
We use the energy difference between lattice momentum zero and momentum 
$a{\bf p} = (1,1,1)$ in units of $2\pi a/L$. 
The column $c_{1,5,6}$ denotes whether the $\mathcal{O}(\alpha_s)$ 
improved coefficients were used in the action and the 
columns $c_2,c_4$ indicate additional values of those coefficients 
that were run on coarse set 3 and very coarse set 1 to estimate systematic errors. 
}
\label{tab:kinmassallsets}
\end{table*}

\begin{figure}
\includegraphics[width=0.9\hsize]{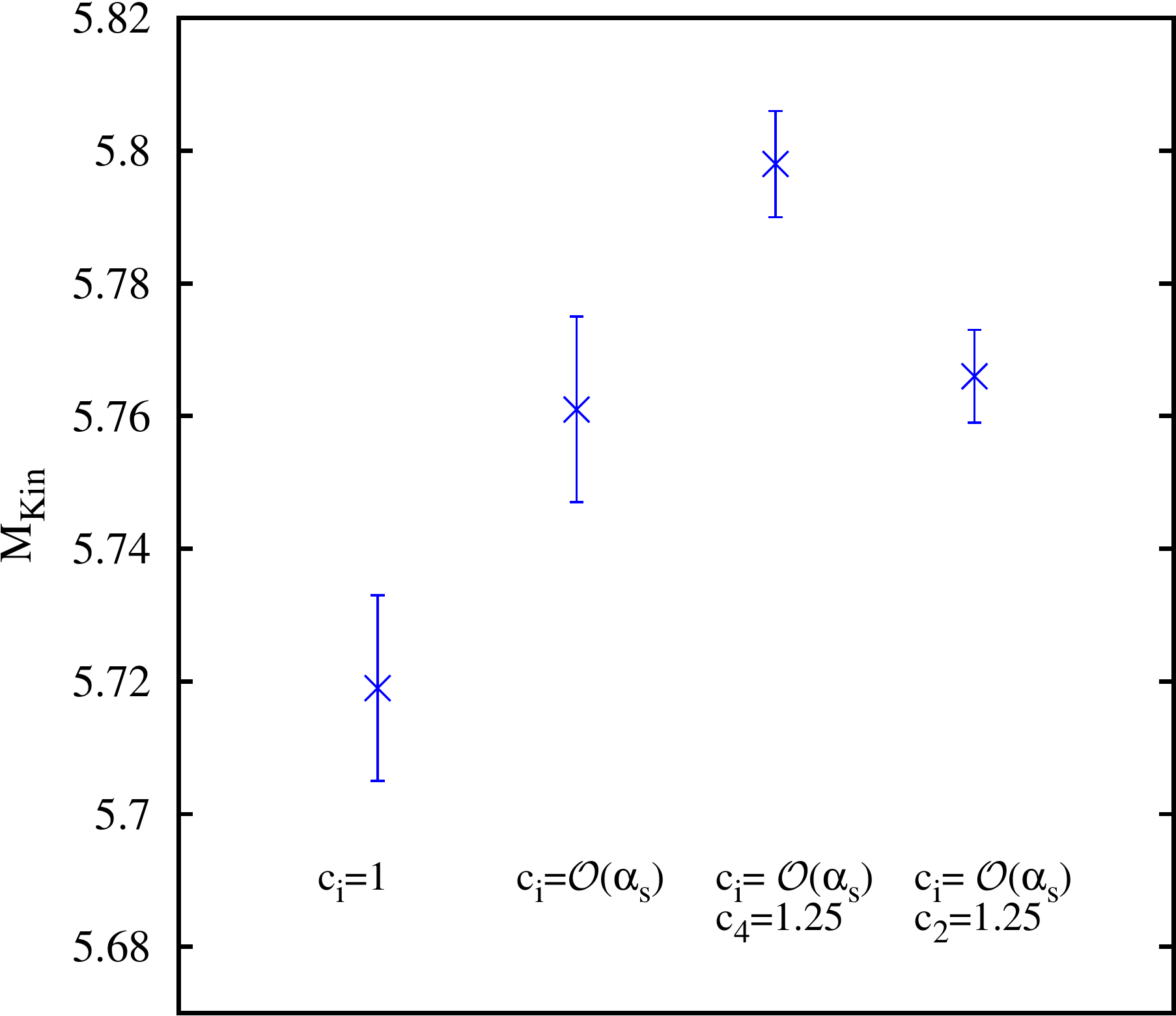}
\caption{ Comparison of values obtained for the kinetic 
mass from a variety of different parameter values 
on coarse set 3. 
 }
\label{fig:mkincompare}
\end{figure}

The conclusion from this subsection is that, to 
minimise systematic errors, we should tune 
the $b$ quark mass by calculating 
the spin-averaged kinetic mass $\overline{M}_{\mbox{\tiny Kin}}(1S)$
and matching that to experiment. We do this from 
the comparison of meson energies at zero momentum 
and the `maximally off-axis' momentum (1,1,1) to 
minimise discretisation errors. 
Table~\ref{tab:kinmassallsets} gives results for this kinetic 
mass on all ensembles for the given values of the $b$ 
quark mass and coefficients, $c_i$. To convert these results to physical 
units we need a value for the lattice spacing to 
be determined in subsection~\ref{subsec:results}. 
Table~\ref{tab:kinmassallsets} gives statistical/fitting 
errors on the values. As discussed above, remaining 
systematic errors from missing radiative, relativistic 
and discretisation errors amount to a total of 0.5\%. 
We are able to pin down the size of these systematic 
errors by using the improved methods described here to 
study the dispersion at this level of detail. 

Figure~\ref{fig:mkincompare} compares the results 
for the spin-averaged kinetic mass on the coarse ensemble, set 3 
for a variety of different choices for the coefficients 
in the NRQCD action to show the size of variations 
in the kinetic mass. 
The figure shows that we can see the difference between 
taking tree-level values for $c_{1,5,6}$ and radiatively 
improved values. 
Changing $c_2$ (the coefficient of the Darwin 
term) has very little effect. 
The effect of changing $c_4$ (the coefficient of the 
${\bf \sigma}\cdot{\bf B}$ term which should be 
spin-averaged away at leading order in this kinetic mass) 
is also not large. Another check of this is given 
in Table~\ref{tab:kinmassallsets} on set 1. 

The experimental result for the $\Upsilon$ mass is 
9.4603(3) GeV and that of the $\eta_b$, 9.391(3) GeV,~\cite{pdg}  giving 
a spin-average of 9.443(1) GeV. 
The real world includes effects that are missing from our 
lattice calculation, however, and so we must correct for 
this. Electromagnetism affects the $\Upsilon$ and $\eta_b$ 
approximately equally and, from a potential model we 
estimate that it reduces their masses by ~1.6 MeV~\cite{Gregory:2010gm}. 
In addition the $\eta_b$ can annihilate to gluons and 
we estimate that this effect also reduces its mass by 
~2.4 MeV, taking the same value as that estimated 
for the $\eta_c$~\cite{Follana:2006rc}. 
The `experimental' mass that we should compare 
our results to is then increased from above to 9.445(2) 
GeV where we allow for a 100\% error in our estimate of the shifts in the 
masses~\footnote{Note that in our previous 
work~\cite{Gregory:2010gm} we had to allow for a mass shift 
from missing $c$ quarks in the sea. That is no longer 
necessary here.}.

%
\subsection{NRQCD systematics in radial and orbital splittings}
\label{subsec:systs}
%

\begin{table*}[t]
\centerline{\begin{tabular}{cccccc}
\hline
\hline
Correction & relativistic & radiative & radiative & 4-quark & {\bf Total} \\
& & kinetic & Darwin & spin-independent & relativistic  \\
& & & & & + radiative \\
\hline
Form 
& $\delta p^6/(m_b)^5 $
& $\alpha_s^2\delta p^4/4(m_b)^3 $
& $4\pi\alpha_s^2\psi(0)^2/(3m_b^2)$
& $\alpha_s^2\psi(0)^2/m_b^2$
&
\\
\hline
Est. \%age in $2S-1S$ 
& & & & 
\\
very coarse
& 0.5 
& 0.2
& 0.4
& 0.4
&  0.8
\\
coarse
& 0.5 
& 0.15 
& 0.3
& 0.3
& 0.7
\\
fine
& 0.5
& 0.1
& 0.2
& 0.2
& 0.6
\\
\hline
Est. \%age in $1P-1S$ 
& & & &
\\
very coarse
& 1.0
&  0.7
&  0.9
& 0.9
&  1.8
\\
coarse
& 1.0
& 0.5
& 0.7
& 0.7
& 1.5
\\
fine
& 1.0
& 0.3
& 0.4
& 0.4
& 1.2
\\
\hline
\hline
\end{tabular}}
\caption{An estimate of systematic errors in the $2S-1S$ and $1P-1S$ 
splittings in the $\Upsilon$ in our lattice QCD calculation arising 
from missing higher order relativistic and radiative corrections 
to the NRQCD action that we use (equation~\ref{eq:deltaH}).
}
\label{tab:systrelrad}
\end{table*}

\begin{table*}[t]
\centerline{\begin{tabular}{ccccc}
\hline
\hline
Correction & discretisation in & discretisation in & discretisation in & {\bf Total} \\
& NRQCD action (i) & NRQCD action (ii) & gluon action & discretisation \\
\hline
Form 
& $\alpha_s^2a\delta p^4/8n(m_b)^2 $
& $\alpha_s^2a^2\delta p_i^4/12m_b $
& $4\pi\alpha_s^2a^2\psi(0)^2/15$
&
\\
\hline
Est. \%age in $2S-1S$ 
& & & & 
\\
very coarse
& 0.2
& 0.4
& 0.3
& 0.5
\\
coarse
& 0.1
& 0.2
& 0.15
& 0.3
\\
fine
& 0.05
& 0.06
& 0.05
& 0.1
\\
\hline
Est. \%age in $1P-1S$ 
& & & &
\\
very coarse
& 0.7
& 2.0
& 1.0
& 2.3
\\
coarse
& 0.4
& 1.0
& 0.5
& 1.2
\\
fine
& 0.2
& 0.3
& 0.1
& 0.4
\\
\hline
\hline
\end{tabular}}
\caption{An estimate of systematic errors in the $2S-1S$ and $1P-1S$ 
splittings in the $\Upsilon$ in our lattice QCD calculation arising 
from discretisation errors in the NRQCD and gluon actions. }
\label{tab:systdisc}
\end{table*}

Here we discuss the remaining sources of systematic 
error in our calculation of the 
radial and orbital excitation energies.
These systematic errors will feed subsequently into the 
determination of the lattice spacing from the $\Upsilon$
$2S-1S$ splitting. 

Radial and orbital excitation energies arise at leading 
order from the time derivative and $H_0$ in the NRQCD 
action (equation~\ref{eq:deltaH}). The relativistic corrections 
at $v^4$ in $\delta H$ thus provide relative $\mathcal{O}(v^2) \approx 10\%$ 
corrections to these splittings. 
Missing radiative corrections 
to the $v^4$ terms dominated the 
errors in earlier calculations~\cite{Gray:2005ur,Meinel:2009rd}, 
since $\alpha_sv^2 \approx 2--3\%$ is larger than $v^4 \approx 1\%$ 
from missing higher order 
relativistic corrections.  
We now include for the first time the radiative 
corrections to most of the $v^4$ terms in $\delta H$.
The remaining errors are then largely 
at relative $\mathcal{O}(\alpha_s^2v^2)$, i.e. less than 1\%. 

Table~\ref{tab:systrelrad} lists the remaining systematic errors from 
spin-independent terms in the $2S-1S$ and 
$1P-1S$ splittings in more detail following~\cite{Gray:2005ur}. The errors were 
determined using a potential model to make estimates 
of the energy shifts in each of the $1S$, $2S$ and $1P$ 
states. For example, radiative corrections at 
$\mathcal{O}(\alpha_s^2)$ to the $p^4/(8m_b^3)$ term 
in the NRQCD action give shifts of size 
$\alpha_s^2<p^4>/4m_b^3$ where $<p^4>$ is the expectation 
value of $p^4$ in that state. 

The effects of the Darwin term term appear at $\mathcal{O}(\alpha_sv^4)$ 
since we have not included a radiative correction to $c_2$. 
However, since this term vanishes in the free theory it is already 
suppressed by an additional power of $\alpha_s$. Its effects 
are proportional to the square of the wavefunction at the origin 
so it does not affect $P$-wave states.  
A very similar term arises from missing spin-independent 4-quark 
opertors. The spin-dependent 4-quark operators are discussed 
in Appendix~\ref{appendix:cicalc} along with the coefficients 
they have in order to match NRQCD to QCD. The spin-independent 
ones arise from the same diagrams and the calculation of their 
coefficients is in progress. Here we take an error from missing 
these 4-quark operators which is of the same size as the error from 
radiative corrections to the Darwin term.  

Note that errors cancel to a significant 
extent between the $2S$ and $1S$ states because of 
their similarities~\cite{Gray:2005ur}. This is the 
reason for focussing on the $2S-1S$ splitting to 
determine the lattice spacing, because it has the 
smallest systematic error.  

We see from Table~\ref{tab:systrelrad} that the 
largest remaining systematic error 
is now that from missing 
$v^6$ terms. The key kinetic term at $v^6$ that would 
appear in a higher order NRQCD action is 
$-(\Delta^{(2)})^3/(16(am_b)^5)$ at tree level. 
This term is proportional to $+(v^2)^3$
and so, if it dominates 
the $v^6$ errors, they will have the same sign at every value 
of the lattice spacing. Including this $v^6$ 
term would act in the direction of reducing 
both the $2S-1S$ and $1P-1S$ splitting but the $1P-1S$ 
splitting would be reduced the most. 

Table~\ref{tab:systdisc} similarly quantifies remaining 
systematic errors from missing $\alpha_s^2$ radiative corrections 
to the discretisation correction terms 
with coefficients $c_5$ and $c_6$. These are significantly reduced over 
our earlier calculations~\cite{Gray:2005ur} now that the 
$\alpha_s$ radiative corrections are included. In addition the gluon action 
is now improved completely through $\mathcal{O}(\alpha_s^2a^2)$~\cite{Hart:2008sq} 
and this means that the discretisation errors coming from the 
gluon action are similarly reduced. 

We can estimate the size of $a^4$ errors from 
the analysis in subsection~\ref{subsec:tune} where we 
study discretisation errors in the kinetic mass. 
The energy difference between mesons of momenta 
$(3,0,0)$ and $(2,2,1)$ in units of $2\pi a/L$ can be 
taken as a measure 
of at least the rotationally noninvariant
 $a^4$ errors, as discussed there. 
The energy difference (Figure~\ref{fig:esplit}) 
is barely visible except on 
the very coarse lattices where it amounts to 1 MeV, 
or 0.2\% of the $2S-1S$ splitting. This is much less 
than the estimate of remaining $a^2$ errors in that 
case so we do not include it in Table~\ref{tab:systdisc}. 

%
\subsection{Results}
\label{subsec:results}
%

\subsubsection{Radial and orbital excitation energies}
\label{subsub:spsplits}

Our main results for the fitted energies for the ground-state and first 
two radial excitations of the $\Upsilon$ and $\eta_b$ 
are given in Table~\ref{tab:latdata}. 
The values
come from multi-exponential fits to a $5\times 5$ matrix 
of correlators for each meson as described in section~\ref{subsec:smear}.
We take 9 exponentials on sets 1, 2 and 3; 11 exponentials 
on set 4 and 12 on set 5. 
We also give the 
fitted ground-state energy for the $h_b(1P)$ state on sets 3 and 5 from a
5 exponential fit to $2 \times 2$ matrix of correlators. 
The $b$ quark masses and coefficients, $c_i$, used in the 
NRQCD action are those of 
Tables~\ref{tab:wilsonparams} and~\ref{tab:upsparams}.
Errors are very small on the ground-state $S$-wave masses 
but increase rapidly with the radial excitation number. 
The table also includes energy splittings 
in lattice units for radial and orbital excitations. 
 
%
%
\begin{table*}[t]
\centerline{
\begin{tabular}{llllll}
\hline
\hline
  & 1 & 2 & 3 & 4 & 5 \\
\hline
$aE(1 ^1S_0)$  & 0.25080(5) & 0.25361(3) & 0.26096(3) & 0.26524(2) & 0.25851(2) \\
$aE(2 ^1S_0)$  & 0.6898(16) & 0.6909(8)  & 0.6235(8)  & 0.6246(6)  & 0.5248(7)  \\
$aE(3 ^1S_0)$  & 0.975(14)  & 0.940(22)  & 0.849(9)   & 0.854(4)   & 0.677(11)  \\
$aE(1^3S_1)$   & 0.28532(6) & 0.28809(3) & 0.29245(3) & 0.29681(2) & 0.28405(2)  \\
$aE(2^3S_1)$   & 0.7078(14) & 0.7074(8)  & 0.6416(7)  & 0.6393(9)  & 0.5370(9)   \\
$aE(3^3S_1)$   & 0.988(16)  & 0.975(8)   & 0.855(11)  & 0.867(10)   & 0.693(10)   \\
\hline
$aE(2 \overline{S} - 1 \overline{S})$  & 0.4266(11) & 0.4238(7)  & 0.3525(6) & 0.3467(7) & 0.2563(7) \\
$aE(3 \overline{S} - 1 \overline{S})$  & 0.708(12)  & 0.687(8)   & 0.569(9)  & 0.575(8)  & 0.411(8)  \\
$aE(2^1S_0 - 1^1S_0)$  & 0.4390(16) & 0.4373(8)  & 0.3626(8)  & 0.3594(6) & 0.2663(7)  \\
$aE(3^1S_0 - 1^1S_0)$  & 0.724(14)  & 0.687(22)  & 0.588(9)   & 0.588(4)  & 0.418(11)  \\
$aE(2^3S_1 - 1^3S_1)$  & 0.4225(14) & 0.4193(8)  & 0.3492(7)  & 0.3425(9) & 0.2530(9)  \\
$aE(3^3S_1 - 1^3S_1)$  & 0.703(16)  & 0.687(8)   & 0.563(11)  & 0.570(10)  & 0.409(10)  \\
\hline
$R_S$	& 1.664(38) 	    & 1.638(19)   & 1.611(32) & 1.665(31)  & 1.617(40) \\
$a\Delta$  			&  0.00190(1)    &  0.00190(1)     & 0.00151(1) & 0.00151(1) & 0.00091(1)\\
\hline
$aE(1^1P_1 )$  			&  -         	    &      -     & 0.5654(23)&	-	 & 0.4833(10)\\
$aE(1^1P_1 - 1 \overline{S})$  	&  -         	    &      -     & 0.2809(22)&  -	 & 0.2056(10) \\
$R_P$  				&  -         	    &      -     & 0.808(7)  &  -	 & 0.816(5)  \\
\hline
$aE(1^3S_1 - 1^1S_0)$  & 0.03452(8) & 0.03448(4) & 0.03149(4)  & 0.03157(3) & 0.02554(3) \\
$aE(2^3S_1 - 2^1S_0)$  & 0.0180(21)   & 0.0165(11) & 0.0181(11) & 0.0147(10)  & 0.0122(11) \\
$R_H$ 
 & 0.521(62) & 0.479(33) & 0.575(35) & 0.465(34) & 0.478(45) \\
\hline
\hline
\end{tabular}
}
\caption{Radial, orbital and $S$-wave fine structure splittings in lattice units for sets 1 to 5 for the NRQCD parameters and coefficients given in Tables~\ref{tab:wilsonparams} 
and~\ref{tab:upsparams}. $c_3=c_4=1.0$. Errors are statistical/fitting only. $R_S$, $R_P$ and $R_H$ are defined in the text. }
\label{tab:latdata}
\end{table*}
\begin{table*}[t]
\centerline{
\begin{tabular}{llllll}
\hline
\hline
       & 3    & 4    & 5    \\
$c_3$  & 1.0  & 1.0  & 1.0 \\
$c_4$  & 1.25 & 1.25 & 1.10 \\
\hline
$aE(1 ^1S_0)$  & 0.20943(3) & 0.21289(2) & 0.23204(2) \\
$aE(2 ^1S_0)$  & 0.5796(6)  & 0.5777(7)  & 0.5021(12) \\
$aE(3 ^1S_0)$  & 0.788(12)  & 0.802(6)   & 0.660(12)  \\
$aE(1^3S_1)$   & 0.25628(4) & 0.25978(2) & 0.26206(3) \\
$aE(2^3S_1)$   & 0.6022(7)  & 0.5999(7)  & 0.5170(18) \\
$aE(3^3S_1)$   & 0.827(7)   & 0.821(5)   & 0.663(24)  \\
\hline
$aE(2 \overline{S} - 1 \overline{S})$   & 0.3520(6) & 0.3463(6)  & 0.2584(13)\\
$aE(3 \overline{S} - 1 \overline{S})$   & 0.573(6)  & 0.568(4)   & 0.405(18) \\
$aE(2^1S_0 - 1^1S_0)$  & 0.3702(6) & 0.3648(7) & 0.2701(12) \\
$aE(3^1S_0 - 1^1S_0)$  & 0.579(12) & 0.589(6)  & 0.428(12)  \\
$aE(2^3S_1 - 1^3S_1)$  & 0.3460(7) & 0.3401(7) & 0.2549(18) \\
$aE(3^3S_1 - 1^3S_1)$  & 0.571(7)  & 0.561(5)  & 0.401(24)  \\
$R_S$		       & 1.651(20) & 1.650(15) & 1.573(95) \\
\hline
$aE(1^1P_1 )$  				& 0.5247(22)  & 0.5253(20) & - \\
$aE(1^1P_1 - 1 \overline{S})$  		& 0.2801(22)  & 0.2773(20) & - \\
$R_P$					& 0.810(7)    & 0.815(6)   & -	\\
\hline
$aE(1^3S_1 - 1^1S_0)$  & 0.04684(5) & 0.04689(3) & 0.03003(4) \\
$aE(2^3S_1 - 2^1S_0)$  & 0.0226(9) & 0.0222(10) & 0.0149(22) \\
$R_H$  & 0.482(19) & 0.473(21) & 0.496(73) \\
\hline
\hline
\end{tabular}
}
\caption{Radial, orbital and $S$-wave fine structure splittings in lattice units for sets 
3, 4 and 5 with NRQCD coefficients and parameters as in Tables~\ref{tab:wilsonparams} 
and~\ref{tab:upsparams}. In addition $c_4$ is nonperturbatively tuned taking 
values from Table~\ref{tab:c4vals}. $c_3=1$ for all results. 
Errors are statistical/fitting only. Reduced statistics of 400 configurations 
were used for the $S$-wave states from set 5. $R_S$, $R_P$ and $R_H$ are 
defined in the text. }
\label{tab:latdatac4np}
\end{table*}

As explained earlier we can use the radial excitation 
energy, $M(\Upsilon^{\prime})-M(\Upsilon)$, to fix 
the lattice spacing, by setting 
\begin{equation}
a^{-1} ({\rm GeV}) = \frac{0.5630(9)}{aE(2^3S_1)-aE(1^3S_1)}. 
\end{equation}
0.5630(4) GeV is the experimental mass difference and we 
have increased the error to allow for a possible relative shift 
in the two masses as a result of the electromagnetic attraction 
between quark and antiquark missing in our calculation. 
As discussed earlier, a potential model estimate would 
give a shift of 1.6 MeV to the $\Upsilon$ from the electrostatic
attraction between quark and antiquark, and somewhat 
less for the $\Upsilon^{\prime}$ since typical separations 
between quark and antiquark are larger. We do not shift the 
result but allow for an error 
of 0.8 MeV. 

As long as we deal with spin-averaged splittings we do not 
have to consider errors in spin-dependent terms. 
However, for the $2S-1S$ splitting the match to experiment 
cannot be spin-averaged since no experimental information 
is available for the $\eta_b(2S)$. 
In that case we have to consider sources of systematic error 
in the hyperfine splitting 
that will induce errors in the $\Upsilon$ and $\Upsilon^{\prime}$ 
energies. 
This will discussed further in subsection~\ref{subsub:hyp}. 

The main source of error is from missing radiative corrections 
when we take the coefficient of the $\sigma \cdot B$ term, $c_4$, 
to be 1. In section~\ref{subsub:hyp} we compare results 
for $c_4 =1$ to those from $c_4$ corrected perturbatively 
through $\mathcal{O}(\alpha_s)$ and nonperturbatively, 
to give the correct $1^3P$ fine structure. Both methods 
for correcting $c_4$ give values above 1 and increase 
the lattice result for the hyperfine splitting 
(which is proportional to $c_4^2$ at leading order). 
Thus with $c_4=1$ the $\Upsilon$ energy is too low. 
Since $M(\Upsilon) = M(1\overline{S}) + (M(\Upsilon)-M(\eta_b))/4$, 
the shift from $c_4$ in the $\Upsilon$ mass is 
one quarter of the change in the hyperfine splitting. 
In section~\ref{subsub:hyp} we also 
determine the ratio of the $2S$ hyperfine 
splitting to that of the $1S$ hyperfine splitting and 
find a result close to 0.5, independent of 
$c_4$. Thus the shift from a change in $c_4$ to the 
$2S-1S$ splitting is one eighth of 
change in the 1S hyperfine splitting. 
An increase in $c_4$ above 1 reduces the 
$2S-1S$ splitting. 
From Table~\ref{tab:hypallsets} we can compare results for the 
1S hyperfine splitting for $c_4=1$ to the 
value obtained for $c_4$ improved through 
$\mathcal{O}(\alpha_s)$ for sets 1, 3 and 5. Dividing by 8
then gives shifts in lattice units that can 
be applied to correct the $2S-1S$ splitting 
on very coarse, coarse and fine lattices. 
These shifts are denoted by $a\Delta$ in 
Table~\ref{tab:latdata}, and are to be 
subtracted from the $2S-1S$ splitting 
to give the corrected lattice result. 
It can be seen that $a\Delta$ is not much 
larger than the statistical errors on the 
$2S-1S$ splitting. The statistical error in $a\Delta$ 
is negligible, but there is a systematic 
error
which is taken as $0.5 \times a\Delta$. This accounts 
for the errors in the hyperfine splitting 
from 4-quark operators, higher order radiative corrections 
to $c_4$ and relativistic corrections to the 
${\bf \sigma}\cdot{\bf B}$ term. This error 
is then included in the systematic error 
for the corrected $2S-1S$ 
splitting.  

Note that we do not expect the spin-orbit 
term with coefficient $c_3$ to have significant 
effect on the $S$-wave states. In any case our nonperturbative 
determination of $c_3$ discussed in 
Appendix~\ref{appendix:cinonpert} gives a result 
consistent with the value of 1.0 that we 
are using. Possible errors from radiative 
corrections to $c_2$ are included in our 
systematic error budget for NRQCD (Table~\ref{tab:systrelrad}). 

Table~\ref{tab:aval} gives the values of 
the lattice spacing in fm obtained from the 
$2S-1S$ splitting on each ensemble, along with their 
associated statistical/fitting error and systematic error. 
The systematic errors are combined in quadrature 
from Tables~\ref{tab:systrelrad} and~\ref{tab:systdisc} and 
from $a\Delta$ in Table~\ref{tab:latdata}. The systematic 
errors are dominated by those from missing higher order relativistic 
corrections to the NRQCD action and these will be correlated 
to some extent between ensembles. 
There is an additional overall systematic 
error of 0.2\% coming from the experimental value for the 
splitting and electromagnetic effects missing from our 
calculation. 
\begin{table}
\centerline{
\begin{tabular}{llll}
\hline
\hline
Set  & $a_{\Upsilon}$ (fm) & $a_{\eta_s}$ (fm)	& $a_{r_1/a}$ (fm)  \\
\hline
1     & 0.1474(5)(14)(2)    & 0.1546(10)(5) & 0.1569(8)(13)  \\
2     & 0.1463(3)(14)(2)    & 0.1526(6)(5) & 0.1553(3)(13)  \\
3     & 0.1219(2)(9)(2)    & 0.1234(7)(4) & 0.1244(2)(10)  \\
4     & 0.1195(3)(9)(2)   & 0.1218(5)(4) & 0.1221(5)(10)  \\
5     & 0.0884(3)(5)(1)    & 0.0899(6)(3) & 0.0902(3)(7) \\
\hline
\hline
\end{tabular}
}
\caption{Lattice spacing values in fm determined from several 
methods. The first column gives results from the 
$\Upsilon$ $2S - 1S$ splitting. The first 
error is from statistics/fitting, the second from remaining 
systematic errors from the NRQCD action (from Tables~\ref{tab:systrelrad} 
and~\ref{tab:systdisc}) and the third is a correlated
0.2\% error from experiment and electromagnetic corrections. 
The second column gives lattice spacing values from 
the decay constant of the $\eta_s$ meson as described in 
section~\ref{sec:etas}. The first error is from 
statistics/fitting and the second is a correlated 
0.3\% error from the uncertainty in the physical value of 
$f_{\eta_s}$ as discussed in section~\ref{sec:etas}. The 
third column gives lattice spacing values 
determined from $r_1/a$ values in Table~\ref{tab:params}. 
The first error is from statistics/fitting and the 
second is a correlated 0.8\% error from the 
uncertainty in the physical value of $r_1$ as discussed 
in section~\ref{sec:r1}. 
}
\label{tab:aval}
\end{table}

In Table~\ref{tab:latdatac4np} we give results for 
cases where $c_4$ is set to its nonperturbatively 
tuned value on sets 3, 4 and a test value of 1.10 on 
set 5 (the nonperturbatively tuned value is in 
fact 1.18, see Appendix~\ref{appendix:cinonpert}). 
Changing $c_4$ shifts the fitted energies 
of all the states but this is simply because 
the zero of energy has changed. As expected, 
changing $c_4$ has very little effect 
on splittings between spin-averaged $S$ wave states or 
between the ${}^1P_1$ mass and the spin-averaged 
$1S$ state. 

An important test of the results is whether, using 
these values for the lattice spacing, we get 
results in agreement with experiment for other 
mass differences i.e. whether ratios of splittings 
are correct. In our previous work on 2+1 flavor 
gluon configurations~\cite{Gray:2005ur} agreement with 
experiment was found within 3\% statistical/systematic 
errors. Here we have substantially improved errors, 
including improved statistical errors, so we 
can improve on our earlier analysis. 

\begin{figure}
\includegraphics[width=0.9\hsize]{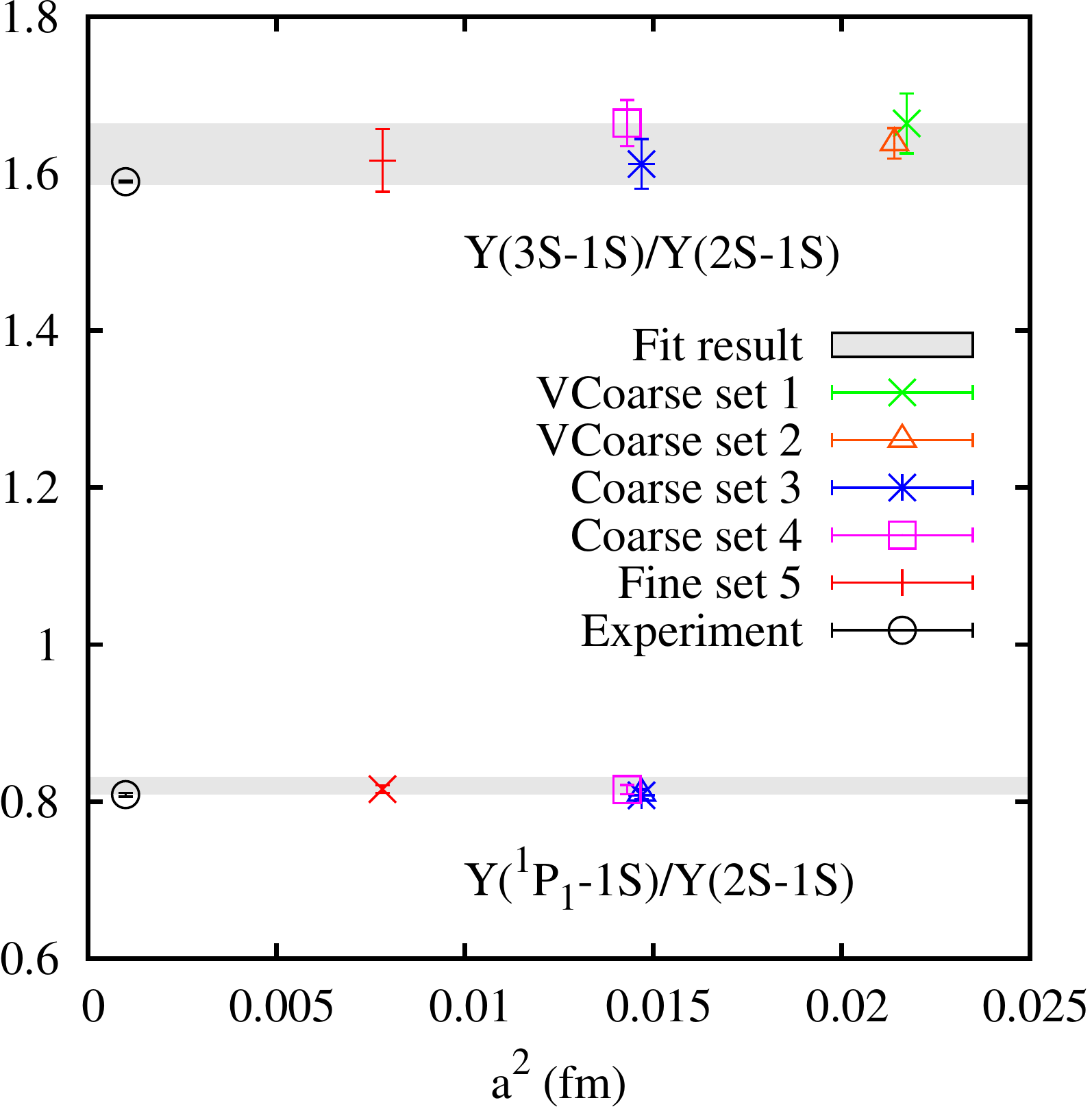}
\caption{ Results for the ratio of the $3S-1S$ and 
$1P-1S$ splittings to the $2S-1S$ in the $\Upsilon$
system plotted against the square of the lattice spacing 
determined from the $2S-1S$ splitting. 
The grey shaded bands give the physical result 
obtained from a fit to the data as described in the text. 
The black open circles slightly offset from $a=0$ are from experiment~\cite{pdg}. 
 }
\label{fig:ratios}
\end{figure}

Table~\ref{tab:latdata} gives values for the 
ratios of the $\Upsilon$ $3S-1S$ 
and $1^1P_1-1\overline{S}$ splittings 
to the $\Upsilon$ $2S-1S$ splitting from our results 
for $c_4=1$. Table~\ref{tab:latdatac4np} gives 
the same ratios for the case where $c_4$ takes 
its nonperturbatively tuned value.  
In forming the ratio $R_P = (1^1P_1-1\overline{S})/(2^3S_1-1^3S_1)$ 
for the case $c_4=1$ we correct the denominator for $c_4$ errors using 
the $a\Delta$ values in Table~\ref{tab:latdata}. 
The numerator should not be sensitive to $c_4$ 
because the $S$-state energies have been spin-averaged and 
the ${}^1P_1$ state is unaffected by $c_4$, as discussed in 
Appendix~\ref{appendix:cinonpert}.
Both the $\Upsilon$ $3S-1S$ and $2S-1S$ splittings 
will have some sensitivity to $c_4$. However, any shifts 
will 
cancel between the two splittings up to 
an amount equal to one quarter of the difference in the 
$3S$ and $2S$ hyperfine splittings. This is negligible 
compared to statistical errors in this ratio. 
The ratio $R_S = (3^3S_1-1^3S_1)/(2^3S_1-1^3S_1)$ 
in Table~\ref{tab:latdata} 
is therefore not corrected for $c_4$. 

The ratio $R_S$ from Table~\ref{tab:latdata}, where we 
have results for all five sets, is plotted against the 
square of the lattice spacing in Figure~\ref{fig:ratios}. 
We see very little dependence on lattice spacing
and on the $u/d$ sea quark mass (the $s$ and $c$ sea 
quark masses are already well-tuned to 
their physical values, see section~\ref{sec:etas}). 
Figure~\ref{fig:ratios} also shows results 
for $R_P$, combining results from Tables~\ref{tab:latdata}
and~\ref{tab:latdatac4np} since we have results for 
only 2 ensembles in each table. The results change 
very little between the ensembles, 
however, as Figure~\ref{fig:ratios} 
shows.   

To derive a physical value for each ratio 
we can use the results to fit for the dependence
on these two quantities and then determine 
the result at physical sea quark mass and 
at $a=0$ to compare to experiment. 
For the sea $u/d$ quark mass dependence
a simple polynomial in $m_l/m_s$
(values in Table~\ref{tab:params}) suffices 
because the $m_l$ values are already 
very close to the physical point with 
$m_l/m_s = 0.1$ and 0.2. 
The dependence on the lattice spacing 
is more complicated because in NRQCD we 
must allow for unphysical $a$-dependence 
coming from $am_b$-dependent radiative 
corrections to discretisation errors. 
This $am_b$-dependence is mild when 
$am_b$ is sufficiently large, as here, 
and this is seen explicitly in 
the radiative corrections that are 
included in our calculation for 
$c_5$ and $c_6$ (Table~\ref{tab:wilsonparams}). 

We therefore fit each ratio, $R$, to the 
following functional form: 
\begin{eqnarray}
R &=& R_{\mathrm{phys}}\left[ 1 \right.\\
&+&\sum_{j=1,2}c_j(a\Lambda)^{2j}(1+c_{jb}\delta x_m + c_{jbb}(\delta x_m)^2) \nonumber \\
&+& \left. 2b_l \delta x_l (1 + c_l(a\Lambda)^2) \right] . \nonumber 
\label{eq:fitxa}
\end{eqnarray}
Here $\delta x_l$ is $(am_l/am_s) - (m_l/m_s)_{\mathrm{phys}}$ 
for each ensemble. $(m_l/m_s)_{\mathrm{phys}}$ is 
taken from lattice QCD as 27.2(3)~\cite{Bazavov:2009bb}.  
The strange sea quark mass is tuned to better than 
3\% with the lattice spacing taken from the 
$\Upsilon$ $2S-1S$ splitting so we can ignore any 
effects from this mistuning since sea quark 
mass effects are so small. 
$\delta x_m$ allows for variation in the value of $am_b$ over 
the range we are using and therefore a change in the 
NRQCD radiative corrections to discretisation errors. 
We choose $\delta x_m$  to vary from -0.5 
to +0.5 over our full range of masses by setting $\delta x_m = (am_b - 2.65)/1.5$.  
$\Lambda$ sets the scale for physical $a$-dependence. 
We take it to be 500 MeV.  

The fit prior on $R_{\mathrm{phys}}$ is taken to be $0.8 \pm 0.1$ 
for $R_P$ and $1.6 \pm 0.2$ for $R_S$.   
Since tree-level $a^2$ errors have been removed in this 
calculation we take the prior on $a^2$ terms to be 
$0.0 \pm 0.3$; we take $0.0 \pm 1.0$ for higher order 
terms in $a$.  For $b_l$ we take $0.0 \pm 0.015$ allowing 
for a 3\% shift if the $u/d$ quarks were as heavy as 
strange. Previous results~\cite{Gray:2005ur} saw a 10\% shift 
in results in the quenched approximation.  

Good fits using the form in 
equation~\ref{eq:fitxa} are easily obtained for both $R_P$ and $R_S$. 
For $\chi^2/\mathrm{dof} \{\mathrm{dof}\}$ we obtain $0.2\{4\}$ and $0.4\{5\}$
for $R_P$ and $R_S$ respectively. 
The physical results we obtain are:
\begin{eqnarray}
\frac{1^1P_1 - 1\overline{S}}{(2S-1S)_{\Upsilon}} &=& 0.820(12) \nonumber \\
\frac{(3S-1S)_{\Upsilon}}{(2S-1S)_{\Upsilon}}  &=& 1.625(39) 
\end{eqnarray}
These values are plotted along with the lattice 
results in Figure~\ref{fig:ratios}. 

The complete error 
budget for the two ratios is given 
in Table~\ref{tab:ratsysts}. Most of 
the errors are obtained directly from our fit. 
The NRQCD systematic error in $R_P$ is taken from 
combining results in
Tables~\ref{tab:systrelrad} and~\ref{tab:systdisc}.
Since these errors are correlated between the 
numerator and denominator of $R_P$ we take 
the systematic error in $R_P$ to be the difference
between them. The total NRQCD systematic error 
at each lattice spacing is then included in our 
fit as a correlated error on the data. In fact 
we find no significant difference whether we 
include it as a correlated or uncorrelated error. We
obtain the error in our final result from this 
systematic error by observing the change in the final 
answer from including it or not including it. Variation 
in the NRQCD systematic errors
as a function of $am_b$ is included in our fit form 
and the error from this estimated from the variation 
of $\chi^2$ in the fit. 
We use the same approach for $R_S$ and take the 
NRQCD systematic error to be the same as for 
$R_P$. We might expect some further cancellation 
of errors within $R_S$ because of the similarity 
between the $S$-wave states. However, this is less 
true when comparing $3S$ to $1S$ than for $2S$ and 
$1S$ so we ignore that possibility to be conservative. 

We believe that errors from any mistuning of 
$m_b$ are completely negligible. $R_P$ and 
$R_S$ change experimentally very little between 
$b$ and $c$ and we have very well-tuned $b$ 
masses except on the very coarse lattices where 
our mistuning amounts to 4\%. 

We also believe that finite volume errors are 
negligible. A study using the heavy quark potential 
derived from a quenched lattice QCD calculation 
in~\cite{Bali:1997am} calculated
wavefunctions for radially excited $\Upsilon$ states. 
None of the wavefunctions for the states being 
considered here extended beyond a 
radius of 1.5~fm and the $2S$ and $1P$ extended 
little beyond 1.0~fm. When sea quarks are included, 
as here, the size of the states will be smaller because 
the Coulomb coefficient in the heavy quark potential is 
larger. Thus 1.5~fm is an overestimate for 
the size of the states. The physical extent of our 
lattices range from 2.3~fm for set 1 to 3.8~fm 
for set 4, so should be large enough to contain 
the $\Upsilon$ states without any finite-volume 
errors from their being 
squeezed.

\begin{table}
\caption{ Complete error budget for the ratios 
of mass splittings, 
$R_P = (1^1P_1 - 1\overline{S})/(2S-1S)_{\Upsilon}$
and $R_S = (3S-1S)_{\Upsilon}/(2S-1S)_{\Upsilon}$. 
Errors are given as a percentage of the ratio. 
Errors which are negligible compared to the 
others are indicated by `0'.
}
\label{tab:ratsysts}
\begin{ruledtabular}
\begin{tabular}{lll}
 & $R_P$ & $R_S$  \\
\hline
stats/fitting & 1.0  &  1.8 \\
$a$-dependence & 0.6 & 1.2 \\
$m_l$-dependence & 0.6 & 0.5 \\
NRQCD $am_b$-dependence & 0.1 & 0.2 \\
NRQCD systematics & 0.5 & 1.0 \\
finite volume & 0 & 0\\
$m_b$ tuning & 0 & 0\\
electromagnetism/ $\eta_b$ annihilation & 0.2 & 0.2 \\
\hline
Total & 1.4 & 2.4  \\
\end{tabular}
\end{ruledtabular}
\end{table}

In Table~\ref{tab:ratsysts} we include 
a 0.2\% error from electromagnetic effects 
and the possibility of $\eta_b$ annihilation, 
neither of which is included in our 
calculation. Electromagnetic effects  
we estimated earlier at 1.6 MeV in the 
$1S$ mass and 0.8 MeV (correlated) in the $2S$ mass. 
If we take the effects on the $1P$ and $3S$ 
masses to be much smaller then we arrive at 
a possible error in $R_S$ and $R_P$ of the 
order of 0.1\% to 0.2\%. $\eta_b$ annihilation 
affects the spin-averaged $1S$ mass, 
shfting it by 
approximately 0.5 MeV. This amounts to a 
possible 0.1\% 
effect in $R_P$, whereas $R_S$ is unaffected.  

Our result for the 
ratio $R_P$
of 0.820(12) is to be compared with the experimental 
result 0.8088(23). Agreement is good within our 1.4\% errors.  
Similarly we obtain 1.625(39) for $R_S$ to be compared
with the experimental result 1.5896(12). 
Again agreement is good, but now with 2.4\% errors, 
dominated by our statistical/fitting error 
because the $3S$ state is a doubly excited state. 
The fact that our central value is slightly higher than 
experiment for both $R_S$ and $R_P$ is consistent with 
the expected effect of missing $v^6$ terms, included 
in our errors, as discussed in section~\ref{subsec:systs}. 

Our result for $R_P$ can 
be converted to a result 
for $M(h_b)-M(1\overline{S})$ = 0.461(7) GeV.  
This can be compared to the result 
$0.440 \pm 17 {+10 \atop -0} \mathrm{GeV}$ 
with over double the error
obtained on configurations including 2+1 flavors of 
sea quarks using the Fermilab heavy quark action~\cite{Burch:2009az}. 
The experimental result for $M(h_b)-M(1\overline{S})$ 
is 0.4553(17) GeV~\cite{pdg, bellehb}. 

Our result for $R_S$ gives 
$M(\Upsilon^{\prime\prime})-M(\Upsilon)$ = 0.914(23) GeV 
compared to an experimental result of 0.8949(6) GeV~\cite{pdg}. 
We have not included in our error budget any effect from 
coupling of the $\Upsilon^{\prime\prime}$ to virtual decay 
channels. The $\Upsilon^{\prime\prime}$ is 200 MeV 
below threshold for real decay to 
a pair of $B$ mesons. This is considered large enough for the 
$\Upsilon^{\prime\prime}$ to be `gold-plated' and 
for the decay channel to have no impact on the mass. 

\subsubsection{Tuned $b$ quark masses}
\label{subsub:tunemb}

\begin{table}
\centerline{
\begin{tabular}{llll}
\hline
\hline
Set  & $am_b(a_{\Upsilon})$  & $am_s(a_{\Upsilon})$ & $am_s(a_{\eta_s})$   \\
\hline
1     &  3.297(11)(35)(7)(16)  & 0.0641(4)(12)(2) & 0.0705(9)(4) \\
2     &  3.263(7)(35)(4)(16)   & 0.0636(3)(12)(2) &  0.0692(5)(4)  \\
3     &  2.696(4)(22)(7)(13)   & 0.0528(2)(8)(2) &  0.0541(6)(3) \\
4     &  2.623(7)(22)(7)(13)  & 0.0512(3)(8)(2) &  0.0531(4)(3) \\
5     & 1.893(6)(12)(5)(9)   & 0.0364(2)(4)(1) &  0.0376(5)(2)  \\
\hline
\hline
\end{tabular}
}
\caption{
Tuned $b$ and $s$ quark masses in lattice 
units on each set of configurations. 
The second column gives $am_b$ and the third $am_s$ 
using the $\Upsilon$ $2S-1S$ splitting to determine 
the lattice spacing. The first two errors in these 
two columns come from statistical errors and 
systematic errors respectively in 
the lattice spacing determination. 
The third and fourth errors in the $am_b$ case are the 
statistical and systematic errors in the determining the kinetic 
mass. Statistical errors in the determination of 
$am_s$ mass are negligible. The third error in the $am_s$ case 
is a correlated 0.3\% error from the square of the physical value of 
the $\eta_s$ mass. 
The fourth column gives 
$am_s$ using the $\eta_s$ decay constant to 
determine the lattice spacing. The first error 
is from statistics/fitting and the second 
is a correlated 0.6\% error from the square of the physical value 
of the $\eta_s$ decay constant. 
}
\label{tab:mtune}
\end{table}

We now return to the tuning of the $b$ quark 
mass. Although not an issue for the mass 
splittings just discussed, it is an important 
source of systematic error for spin-dependent 
mass splittings. We use our determination 
of the lattice spacing from the $\Upsilon$ 
$(2S-1S)$ splitting, given in Table~\ref{tab:aval}
to convert the kinetic mass values given 
in section~\ref{subsec:tune} to physical 
units. 
As described in section~\ref{subsec:tune} 
the appropriate experimental value for comparison 
is 9.445(2) GeV. 

When this is done we see that the masses 
are very well-tuned except on the very coarse 
lattices where they are 4\% high.  
For small changes in the $b$ quark mass the 
change in kinetic mass is approximately twice the 
change in quark mass, as can be seen from 
Table~\ref{tab:kinmassallsets}. 
For the slight changes that we need to 
make this is a sufficiently good approximation. 
We simply adjust the quark mass by 
one half the error in the kinetic mass 
to obtain the tuned quark mass values in lattice units given in 
Table~\ref{tab:mtune}. 

Three errors are given in Table~\ref{tab:mtune}. The first 
is from the statistical error in the kinetic 
mass determined on each ensemble. The second error 
comes from the total error in the determination 
of the lattice spacing in Table~\ref{tab:aval}. 
This includes both statistical and systematic 
errors in determining the $2S-1S$ splitting. 
The third error is a 0.5\% systematic error 
from NRQCD in the kinetic mass obtained from 
analysis of the dispersion relation in 
section~\ref{subsec:tune}. 

%
\subsubsection{The hyperfine splitting}
\label{subsub:hyp}
%

\begin{figure}
\includegraphics[width=0.9\hsize]{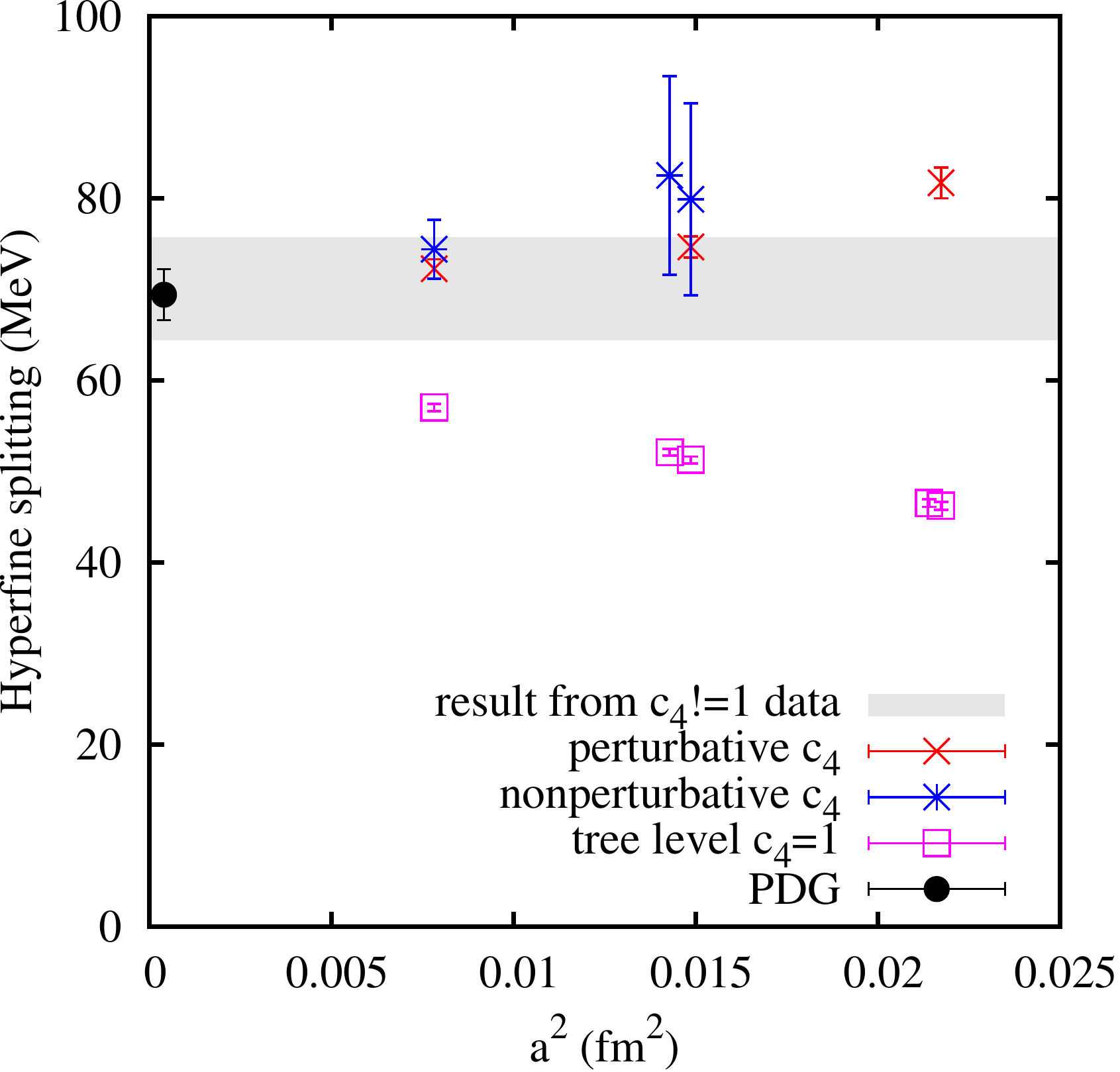}
\caption{ Results for the hyperfine splitting, $M(\Upsilon)-M(\eta_b)$ plotted against the square of the lattice spacing. 
We show results for $c_4=1$ (cyan squares) 
as well as results for 
$c_4$ set equal to its perturbatively improved (red crosses) and 
nonperturbatively improved values (blue stars). The $c_4=1$ results 
include statistical errors only and are shown purely 
for comparison purposes -- they are not included 
in the fit. The results for perturbative and 
nonperturbative $c_4$ include a correction for 
missing 4-quark operators and $m_b$-mistuning. 
The errors on these points are from 
statistics, the lattice spacing 
and the tuning of $m_b$. The results for nonperturbative 
$c_4$ also include statistical errors in the determination 
of $c_4$. 
Our final physical result including our full 
error budget is given by the grey shaded band. 
The full error budget includes errors from 
systematic uncertainties 
in setting $c_4$ and from missing $v^6$ terms 
in the NRQCD action. 
 }
\label{fig:hyp}
\end{figure}

The mass difference between the $^3S_1$ 
and $^1S_0$ states is an important test 
of our calculations because it is statistically 
very precise for the ground-state mesons. 
Controlling systematic errors is the key issue,  
and the main one is
that of radiative corrections to $c_4$, the
coefficient of the ${\bf \sigma}\cdot{\bf B}$
term in the NRQCD action. At leading order 
the hyperfine splitting is proportional to 
$c_4^2$. 
Our previous calculation~\cite{Gray:2005ur}, with $c_4=1$, 
gave a prediction for $M(\Upsilon)-M(\eta_b)$ 
of 61(14) MeV with the error dominated by 
the then-unknown radiative corrections to $c_4$. 

\begin{table}[h]
\caption{ Values for the coefficient of the 
${\bf \sigma}\cdot{\bf B}$ term, $c_4$, for 
different lattice spacing values. The error 
on the perturbative values is $1\times\alpha_s^2$. 
The errors on the nonperturbative values are 
statistics, experiment and NRQCD systematics 
respectively. We did not 
extract a nonperturbative value on the very 
coarse lattices.   
}
\label{tab:c4vals}
\begin{center}
\begin{ruledtabular}
\begin{tabular}{l|c|c}
Sets & $c_4^{\mathrm{pert}}$  & $c_4^{\mathrm{nonpert}}$ \\
\hline
fine & 1.16(5)   & 1.18(2)(1)(5) \\
coarse & 1.20(7) & 1.28(7)(1)(5) \\
very coarse & 1.22(8) & - \\
\end{tabular}
\end{ruledtabular}
\end{center}
\end{table}

\begin{table*}[t]
\centerline{
\begin{tabular}{llllllll}
\hline
\hline
Set  & $am_b$ & $c_{1,5,6}$ & $c_2$ & $c_4$ & $aE_{\eta_b}$ & $aE_{\Upsilon}$ & $aE_{\Upsilon} - aE_{\eta_b}$ \\
\hline
1 & 3.42 & $\alpha_s$ & 1    & 1    & 0.25080(5) & 0.28532(6) &  0.03452(8)  \\
1 & 3.42 & $\alpha_s$ & 1    & 1.22 & 0.21432(5) & 0.26400(6) &  0.04968(7)  \\
1 & 3.5  & $\alpha_s$ & 1    & 1    & 0.25015(6) & 0.28392(9) &  0.03377(10) \\
\hline
2 & 3.39 & $\alpha_s$ & 1    & 1    & 0.25361(3) & 0.28809(3) &  0.03448(4)  \\
2 & 3.42 & $\alpha_s$ & 1    & 1    & 0.25344(5) & 0.28759(5) &  0.03416(6)  \\
\hline
3 & 2.66 & 1          & 1    & 1    & 0.25529(4) & 0.28626(6) & 0.03097(7)  \\
3 & 2.66 & $\alpha_s$ & 1    & 1    & 0.26096(3) & 0.29245(3) & 0.03149(4)  \\
3 & 2.66 & $\alpha_s$ & 1.25 & 1    & 0.25627(24)& 0.28728(33)& 0.03101(41) \\
3 & 2.66 & $\alpha_s$ & 1    & 1.25 & 0.20943(3) & 0.25628(3) & 0.04684(5)  \\ 
3 & 2.66 & $\alpha_s$ & 1    & 1.20 & 0.22040(5) & 0.26394(7) & 0.04354(4)  \\ 
3 & 2.68 & $\alpha_s$ & 1    & 1    & 0.26108(7) & 0.29249(9) & 0.03141(11) \\
3 & 2.7  & 1          & 1    & 1    & 0.24375(8) & 0.27483(11)& 0.03108(13) \\
\hline
4 & 2.62 & $\alpha_s$ & 1    & 1    & 0.26524(2) & 0.29681(2) & 0.03157(3)  \\
4 & 2.62 & $\alpha_s$ & 1    & 1.25 & 0.21289(2) & 0.25978(2) & 0.04689(2)  \\
4 & 2.66 & $\alpha_s$ & 1    & 1    & 0.26546(3) & 0.29662(4) & 0.03116(5)  \\
\hline
5 & 1.91 & 1 & 1    & 1    & 0.24652(3) & 0.27153(5) & 0.02501(6)  \\
5 & 1.91 & $\alpha_s$ & 1    & 1    & 0.25851(2) & 0.28405(2) & 0.02554(3)  \\
5 & 1.91 & $\alpha_s$ & 1    & 1.10 & 0.23204(2) & 0.26206(3) & 0.03003(4)  \\
5 & 1.91 & $\alpha_s$ & 1    & $1.15^*$ & 0.21772(22) & 0.24984(40) & 0.03213(30)  \\
5 & 1.91 & $\alpha_s$ & 1    & 1.16 & 0.21519(2) & 0.24802(4) & 0.03283(2)  \\
5 & 2.0  & $\alpha_s$ & 1    & 1    & 0.25935(3) & 0.28397(4) & 0.02462(5)  \\
\hline
\hline
\end{tabular}
}
\caption{ Fitted energies for ground state $\Upsilon$ and $\eta_b$ mesons on all configuration sets. 
The column $c_{1,5,6}$ denotes whether the $\mathcal{O}(\alpha_s)$ improved coefficients were used in the action. Various values of $c_2$ and $c_4$ have also been used as indicated. $c_3=1$ except for the case indicated by ${}*$ in which $c_3=0.96$. 
Where possible the result from the full $5\times5$ matrix fit was taken. 
Otherwise, the values from the kinetic mass fits were used. In those cases a smaller number 
of configurations and/or time sources was sometimes used and this is reflected in the statistical errors.  
}
\label{tab:hypallsets}
\end{table*}

For this calculation we have results for $c_4$ 
including $\mathcal{O}(\alpha_s)$ corrections 
as well as $c_4$ tuned nonperturbatively.
These determinations of $c_4$ are 
described in Appendix~\ref{appendix:cicalc} 
and Appendix~\ref{appendix:cinonpert} respectively. 
The values obtained by the two methods for $c_4$ 
are given in Table~\ref{tab:c4vals}. 
The nonperturbative values are slightly larger 
than the perturbative ones, but the differences 
are well within the expectations from 
additional $\alpha_s^2$ corrections to the perturbative
values and/or systematic errors in the 
nonperturbative values. Both sets of values get closer to 1 
on the finer lattices, as expected, because they 
are functions of the strong coupling constant 
at a scale related to the inverse of the lattice 
spacing.  

Table~\ref{tab:hypallsets} gives results for 
the energies of the $\Upsilon$ and $\eta_b$ 
for various combinations of values of coefficients 
in the NRQCD action. 
We also give the mass difference between the $\Upsilon$ and $\eta_b$
which is the hyperfine splitting. 
This can be more precise than either mass 
separately because we fit both meson correlators 
together and extract the difference directly 
from the fit taking into account the correlations. 
Where we have 
fits to a $5\times 5$ matrix of correlators, 
as in Tables~\ref{tab:latdata} and~\ref{tab:latdatac4np}, 
we give those results. In other cases we 
calculated only a single local correlator for 
each of the $\Upsilon$ and $\eta_b$ which 
is quite sufficient to extract a splitting 
between the ground state masses. Results 
are not as precise for splittings between 
radially excited states in those cases, however, 
and we do not give them. 

We see from Table~\ref{tab:hypallsets} 
that changing $c_4$ does have a large effect on 
the hyperfine splitting, approximately in line 
with the expectation of variation as $c_4^2$. 
We also see that, on sets 3 and 5 where we have data for 
comparison, changing $c_{1,5,6}$ to their 
$\mathcal{O}(\alpha_s)$ improved values does 
increase the hyperfine splitting slightly. 
It is a small effect, however, of order 2\%.
Changing the coefficient of the Darwin term, $c_2$ 
also has a small effect of order 1(1)\%. 

We conclude from this that we have controlled 
all of the coefficients of $v^4$ terms in our 
NRQCD action at a level required to give few 
percent errors in the hyperfine splitting from 
these sources. We have not, however, included 
4-quark operators in our NRQCD Hamiltonian and 
they can have an impact on the hyperfine splitting 
at an order equivalent to that of $\alpha_s$ 
corrections to $c_4$~\cite{Hammant:2011bt}. 

In Appendix~\ref{appendix:cinonpert}
we give coefficients for the 4-quark operators 
and a formula in equation~\ref{eq:hyp4q} 
for the correction that they would  
induce in the hyperfine splitting.
Table~\ref{tab:hypcorr} gives values
for this correction, to be added to the 
results from Table~\ref{tab:hypallsets}, 
on the very coarse, coarse 
and fine ensembles based on using a spin-averaged 
value of the `wavefunction-at-the-origin', $\psi(0)$, 
for the $\Upsilon$ and $\eta_b$ 
from our fits. This varies only very little 
with $c_4$, falling by at most 2\% from $c_4=1$ to 
our nonperturbative $c_4$ values, so we ignore this 
variation. Our correlators are normalised by dividing 
by 6 after summing over 2 spins and 3 colors. Then 
$\psi(0)$ is given by the 
amplitude of our correlator fits;
$a(l,1)$ for 
the $1S$ and $a(l,2)$ for the $2S$ from 
equation~\ref{eq:fit}. 

\begin{table}[h]
\caption{ Corrections to the $1S$ and $2S$ hyperfine 
splittings from spin-dependent 4-quark 
operators missing from our NRQCD action. 
We use equation~\ref{eq:hyp4q} 
inserting values for $\psi(0)$ from our 
fitted results and values for $\alpha_V(\pi/a)$ 
from Table~\ref{tab:alpha}. We convert 
to physical units using lattice spacing 
values from Table~\ref{tab:aval}.  
}
\label{tab:hypcorr}
\begin{center}
\begin{ruledtabular}
\begin{tabular}{lcc}
Sets & Correction to $1S$  & Correction to $2S$ \\
& hyperfine (MeV) & hyperfine (MeV) \\ 
\hline
fine & -1.7  &  -1.0 \\
coarse & 5.2 & 3.4 \\
very coarse & 12.9 & 8.3 \\
\end{tabular}
\end{ruledtabular}
\end{center}
\end{table}

We see from Table~\ref{tab:hypcorr} that the 
correction is substantial on the very coarse 
lattices and very small on the fine lattices, 
because of the variation of the coefficients 
$d_1$ and $d_2$ with $am_b$.   
The corrected results are shown in Figure~\ref{fig:hyp} 
along with the uncorrected results for $c_4=1$. 
We see that there is a substantial difference between 
the results coming from the change in $c_4$ and, to 
a lesser extent, from the correction for the 4-quark 
operator. The strong dependence on the lattice spacing 
seen in the $c_4=1$ results is reduced, 
in line with the expectation that improving an 
effective theory should reduce the cutoff dependence. 

An additional small factor in Figure~\ref{fig:hyp} is
that we have corrected results for slight mistuning
of the $b$ quark mass and we have included the error 
from the quark mass tuning in the hyperfine 
splitting error. The hyperfine splitting is expected 
to be approximately inversely proportional to the quark 
mass and this is seen in Table~\ref{tab:hypallsets}. We assume 
this relationship to make small adjustments based on the 
tuned $b$ masses from Table~\ref{tab:mtune}. 
The largest effect is a 4\% one on the very coarse lattices. 
The lattice spacing error from the quark mass tuning 
is correlated with the lattice spacing error on the 
hyperfine splitting because of this inverse relationship. 
The lattice spacing error therefore appears with a 
factor of 2 in the hyperfine splitting. 

To obtain a physical result for the hyperfine splitting 
we then combine results with perturbative and nonperturbative 
values of $c_4$ allowing for systematic differences between 
them from uncertainties in the determination of $c_4$. 
We must also allow for uncertainties from higher-order 4-quark operator 
effects and for lattice spacing and sea quark mass dependence. 
The nonperturbative $c_4$ results are given a correlated
systematic error corresponding to the second and third errors
in Table~\ref{tab:c4vals} and remembering that the 
hyperfine splitting is related to $c_4^2$. Similarly 
the results for perturbative 
$c_4$ are given a separate correlated systematic error 
corresponding to the $\alpha_s^2$ errors given in Table~\ref{tab:c4vals}. 
We allow for higher order 4-quark operator effects with 
a correlated systematic error of size $6\alpha_s^3 |\psi(0)|^2/m_b^2$ 
with a coefficient of possible size $\pm 1 \pm \ln(am_b)$. 
This does not assume that the small coefficient seen 
at $\mathcal{O}(\alpha_s^2)$ on the fine lattices 
is repeated at higher order.  

\begin{figure}
\includegraphics[width=0.9\hsize]{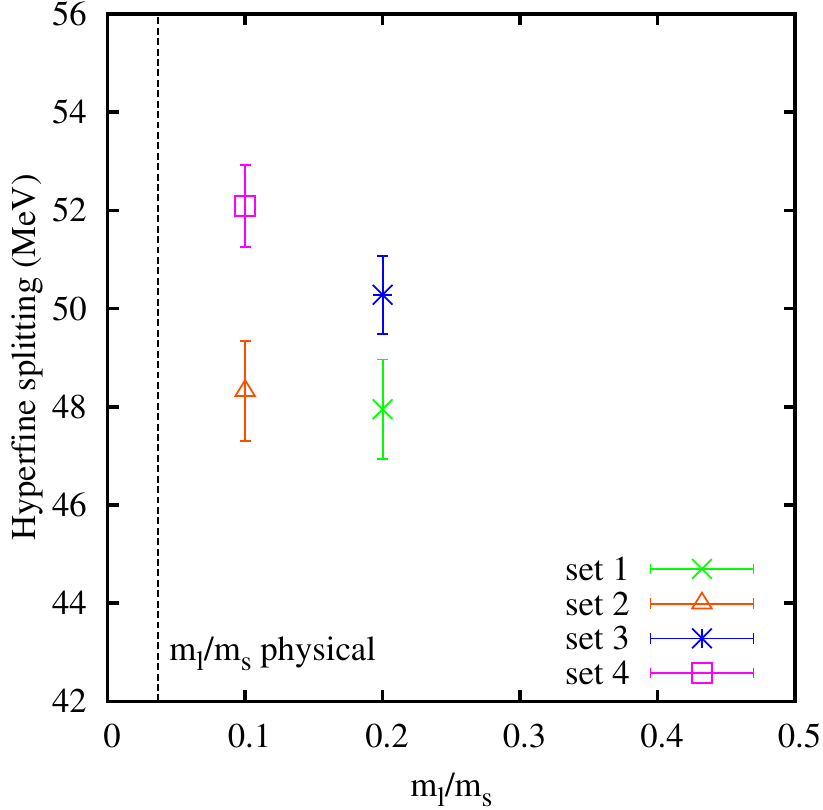}
\caption{ Results for the hyperfine splitting 
obtained for $c_4=1$ (not used in our 
fit for the physical hyperfine splitting) compared as a function 
of sea light quark mass in units of the strange 
quark mass.  Results are given for 
two values of $m_l/m_s$ on very coarse (sets 1 and 2) 
and coarse (sets 3 and 4) lattices. 
The errors on the points include statistical/fitting errors, 
lattice spacing errors and $m_b$ tuning errors. 
 }
\label{fig:hypml}
\end{figure}

We allow for lattice spacing and sea quark mass effects as 
in the fit function of equation~\ref{eq:fitxa}. 
The prior on sea quark mass effects is now taken to allow
15\% effects for $m_l \approx m_s$. This reflects the 
fact that a 40\% difference was seen between quenched and 
dynamical results in~\cite{Gray:2005ur}. 
In fact sea quark mass effects in our data are small. 
Figure~\ref{fig:hypml} shows a comparison of the 
hyperfine splitting as a function of the light sea 
quark mass for the case $c_4=1$ where we have a complete 
set of data. Although these results are {\it not} used 
to determine our final answer for the hyperfine splitting 
they do provide a useful comparison between ensembles 
at the same lattice spacing and different light quark 
mass since the effects of $c_4$ are independent of 
sea quark mass. The results are adjusted for $b$ quark 
mass mistuning and include errors from the $b$ quark 
mass and the lattice spacing. Variation with light quark mass 
is at most ~2 MeV. Note that we are using 
much lighter sea quark masses than in previous 
calculations~\cite{Gray:2005ur, Meinel:2010pv}; indeed 
$m_l$ is within a factor of 3 of its physical value.  

\begin{table}
\caption{ Complete error budget for 
the $1S$ hyperfine splitting and the 
ratio of the $2S$ to the $1S$ hyperfine splittings. 
Errors are given as a percentage of the final result. 
${}^*$ Note that the $m_b$ tuning uncertainty does 
not include the lattice spacing uncertainty in $m_b$. 
Since that is correlated with the $a$ uncertainty on 
converting the hyperfine splitting from lattice to 
physical units, they must be handled together and 
both are included in the $a$-uncertainty.  
}
\label{tab:hyperr}
\begin{ruledtabular}
\begin{tabular}{lll}
 & $M(\Upsilon)-M(\eta_b)$ & $R_H$  \\
\hline
stats/fitting		& 0.1  	& 4  \\
$a$-dependence 		& 1.5 	& 5 \\
$a$-uncertainty 	& 0.5 	& 0 \\
$m_l$-dependence 	& 3 	& 3.5  \\
NRQCD $am_b$-dependence & 2 	& 0.5 \\
NRQCD $v^6$ 		& 10 	& 5 \\
NRQCD $c_4$ uncertainty 	& 7 	& 0 \\
NRQCD 4-quark uncertainty 	& 2 	& 1 \\
$m_b$ tuning$^*$ 		& 0.1 	& 0 \\
$\eta_b$ annihilation 	& 1 	& 0.5 \\
\hline
Total & 13 &  9 \\
\end{tabular}
\end{ruledtabular}
\end{table}

We obtain a physical value for the hyperfine splitting 
from the above fit of 70(6) MeV. Fitting the results for 
perturbative $c_4$ on their own gives a consistent 67(7) MeV and 
the results for nonperturbative $c_4$ alone gives 75(9) MeV. 
Our best result therefore comes from combining the two. 
An additional 10\% error must 
be allowed for higher order ($v^6$) spin-dependent terms 
in the NRQCD action, giving a final result of:
\begin{equation}
 M(\Upsilon) - M(\eta_b) = 70(9) \mathrm{MeV}. 
\end{equation}
Our complete error budget is given in Table~\ref{tab:hyperr}. 
The shift for the effect of $\eta_b$ annihilation 
is included in our 4-quark operator correction (as 
discussed in Appendix~\ref{appendix:cinonpert}) but 
we separate out an error for that from the rest of the 4-quark 
operator error. 

\begin{figure}
\includegraphics[width=0.9\hsize]{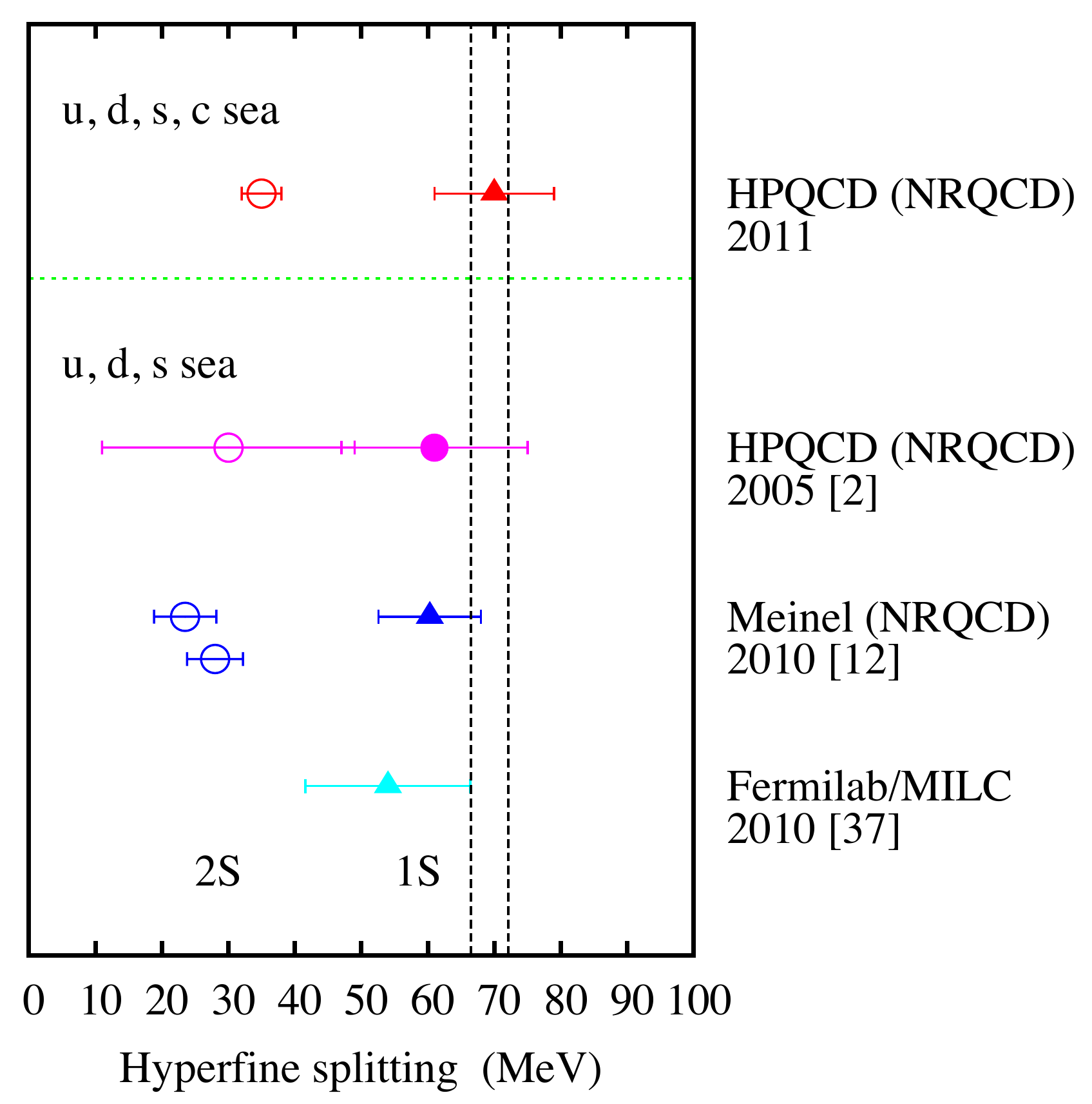}
\caption{ Comparison of results 
for the hyperfine splittings, $M(\Upsilon)-M(\eta_b)$
and $M(\Upsilon^{\prime})-M(\eta_b^{\prime})$,
from different full lattice QCD calculations. 
Filled symbols indicate the $1S$ hyperfine and 
open symbols, the $2S$ hyperfine. 
Circles indicate predictions and triangles postdictions. 
The top (red) points are the new results from this paper, 
and the points below that (pink) are from~\cite{Gray:2005ur}, 
when the $1S$ hyperfine was a prediction. 
The third line gives results from~\cite{Meinel:2010pv}. Two 
results are given for the $2S$ hyperfine; that from a ratio
to the $1S$ hyperfine, as here, and that from a ratio to 
the combination of $P$-wave spin splittings sensitive to 
$c_4$ (see Appendix~\ref{appendix:cinonpert}). 
The top three results use the NRQCD formalism 
for $b$ quarks; the bottom (cyan) result uses the 
Fermilab heavy quark action~\cite{Burch:2009az}. 
The black dashed lines mark the current experimental 
average~\cite{pdg}. 
 }
\label{fig:hypcomp}
\end{figure}

In Figure~\ref{fig:hypcomp} we compare our new result for the hyperfine 
splitting to earlier full lattice QCD results and to experiment~\cite{pdg}. 
Earlier results using NRQCD are: 61(14) MeV from~\cite{Gray:2005ur}
with a treelevel $v^4$ NRQCD action and 60.3(7.7) MeV from 
~\cite{Meinel:2010pv} using an NRQCD with $v^6$ spin-dependent terms 
and $c_4$ determined from $P$-wave splittings but with no 4-quark operator 
corrections, which are potentially more significant than 
$v^6$ terms, or errors from them. The result obtained from the 
Fermilab heavy quark action~\cite{Burch:2009az} is 54.0(12.4) MeV.   
For this action the hyperfine splitting is sensitive to the 
coefficient of the $\mathcal{O}(a)$ improvement term known as 
the clover term. In principle this coefficient does not have 
to be tuned but the approach to the continuum limit is slow. 
Here it was taken to have its tree-level 
value after tadpole-improvement 
and the quark mass was tuned using the spin-averaged kinetic mass of 
the $B_s$ and $B_s^*$ mesons.  

All four lattice QCD results agree well with the current 
experimental average of 69.3(2.8) MeV~\cite{pdg} obtained 
from averaging results from experiments on radiative transitions 
to $\eta_b$ from $\Upsilon^{\prime}$ 
and $\Upsilon^{\prime\prime}$~\cite{exptetab, Aubert:2009pz, Bonvicini:2009hs}. Preliminary experimental results using radiative transitions from 
the $h_b$ indicate a somewhat lower value~\cite{mizukqwg11}. 

Our new result above contains the most complete analysis at 
$\mathcal{O}(v^4)$ in the NRQCD action. To improve it would 
require the inclusion of spin-dependent operators at $\mathcal{O}(v^6)$.  
The effect of spin-dependent $v^6$ operators was 
studied in~\cite{Meinel:2010pv}, taking ratios of the hyperfine 
splitting to $P$-wave spin 
splittings to cancel the effect of $c_4$. Those results indicate 
that spin-dependent $v^6$ terms tend to reduce the hyperfine 
splitting by about 10\%. We have included a (symmetric) 10\% error in our 
results to account for missing $v^6$ terms. 

The hyperfine splitting has also been calculated 
using continuum QCD perturbation 
theory~\cite{Kniehl:2003ap}. A 
considerably smaller result is obtained of 41(14) MeV. 
This is not in disagreement with the nonperturbative lattice 
QCD results 
given the size of the errors. It has been suggested, however, 
that the inclusion of radiative corrections to $c_4$ 
in the lattice NRQCD calculation would reduce the value 
of the hyperfine splitting obtained~\cite{Penin:2009wf}. 
That expectation was based on an incorrect analysis of the 
form of $c_4$ in the lattice NRQCD calculation and 
we see indeed from the 
results given here that 
the inclusion of radiative corrections to 
$c_4$ has had the opposite effect and 
increased our value for the hyperfine splitting.  

\begin{figure}
\includegraphics[width=0.9\hsize]{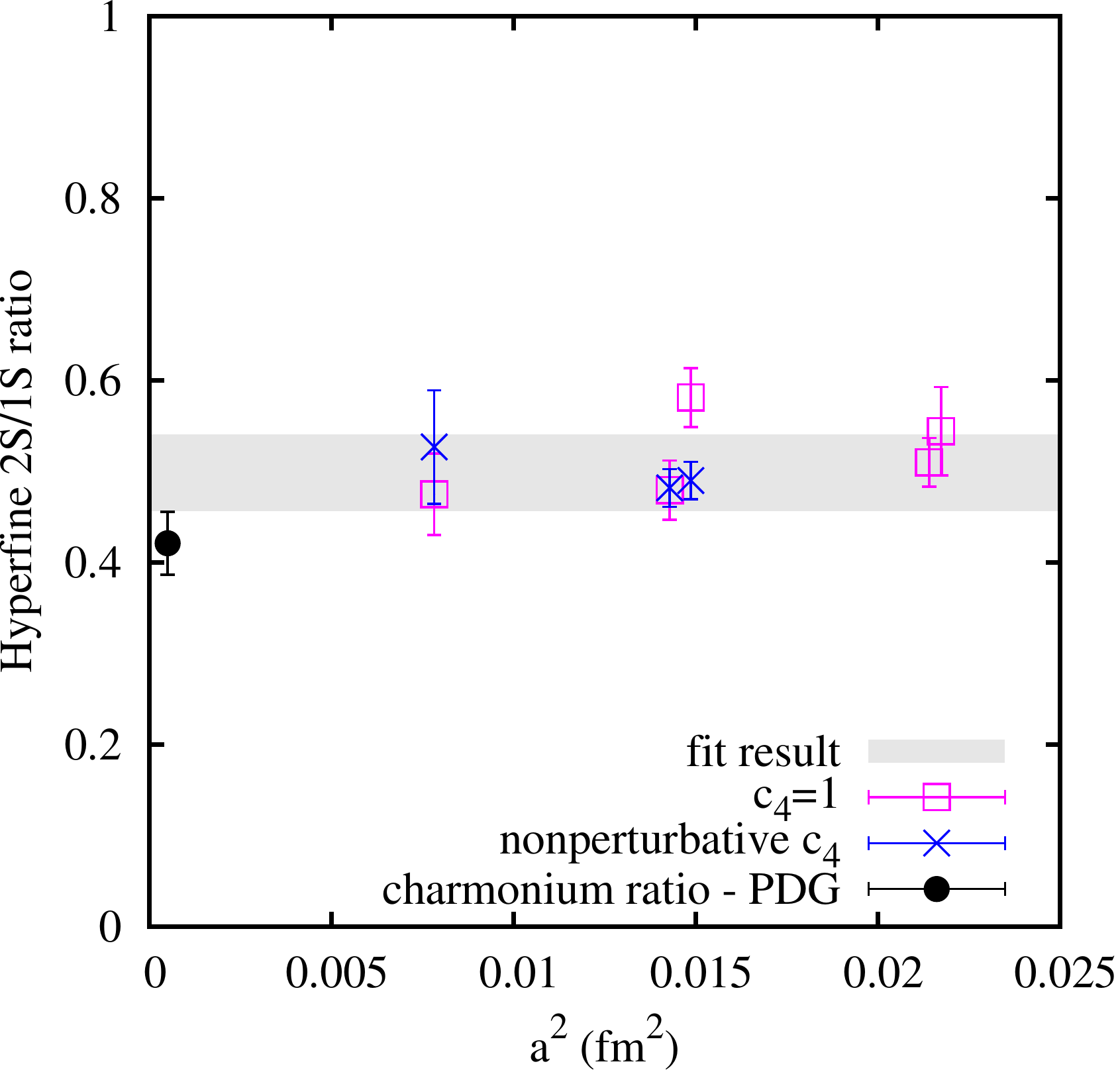}
\caption{ The ratio of hyperfine splittings for $2S$ and 
$1S$ states, $(M(\Upsilon^{\prime})-M(\eta_b^{\prime}))/(M(\Upsilon)-M(\eta_b))$, plotted against the square of the 
lattice spacing. Points have been corrected for missing 
4-quark operator effects. The shaded band shows our final 
physical result including our full error budget.   
 }
\label{fig:hyprat}
\end{figure}

The best way to study the $2S$ hyperfine splitting, $M(\Upsilon^{\prime})-M(\eta_b^{\prime})$ is through the ratio to the $1S$ hyperfine splitting. 
We define:
\begin{equation}
R_H = \frac{M(\Upsilon^{\prime})-M(\eta_b^{\prime})}{M(\Upsilon)-M(\eta_b)}.
\label{eq:rh}
\end{equation}
Then $c_4$ effects cancel as can be seen in the numbers in 
Tables~\ref{tab:latdata} and~\ref{tab:latdatac4np} and plotted 
in Figure~\ref{fig:hyprat}. In the Figure we have corrected the 
results for missing 4-quark operator effects which are slightly 
different in the $1S$ and $2S$ states and so have an effect 
on the ratio. This is at most 4\%, on the very coarse lattices. 
We have made no correction for the slight mistunings of $m_b$ 
since they should largely cancel in this ratio. 

Again we extract a physical result for $R_H$ by allowing for $m_l$ and 
$a$ dependence  as in equation~\ref{eq:fitxa}. Here we relaxed
the priors on the $a$-dependence so that they all had the form 
$0.0 \pm 1.0$. The prior on $m_l$-dependence (in units of $m_s$) 
was taken as $0.0 \pm 0.15$ as for the $1S$ hyperfine, allowing 
a 15\% change from $m_l=m_s$ down to the physical point. 

Our final physical value is: 
\begin{equation}
R_H = 0.499(42). 
\end{equation}
The full error budget is given in Table~\ref{tab:hyperr} where 
we allow a 5\% error for missing spin-dependent $v^6$ terms, 
allowing for some cancellation of $v^6$ effects between the 
$1S$ and $2S$ hyperfine splittings. 

Combining our result for $R_H$ with the current experimental 
average for the $1S$ hyperfine splitting gives a result 
for the $2S$ hyperfine splitting of 35(3)(1) MeV, where the 
second error comes from the experimental $1S$ splitting. 
We predict the mass for the $\eta_b^{\prime}$ to be 9988(3) GeV. 

Figure~\ref{fig:hypcomp} compares our result for 
the $2S$ hyperfine splitting to earlier predictions. 
Meinel~\cite{Meinel:2010pv} gives 23.5(4.7) MeV from 
a ratio to $P$-wave spin splittings and 
28.0(4.7) from a ratio to the $1S$ hyperfine splitting, 
as used here.  He includes the effect of spin-dependent 
$v^6$ operators but without such a complete analysis as 
we have done here of $v^4$ operators.  
The conclusion from Figure~\ref{fig:hypcomp} is that 
lattice QCD results give a fairly clear prediction of 
the $2S$ hyperfine splitting around 30 MeV, half the 
result for the $1S$ hyperfine splitting. 

Figure~\ref{fig:hyprat} compares our result for 
the bottomonium ratio of $2S$ to $1S$ hyperfine splittings 
to the experimental charmonium ratio of 0.421(35). 
Our bottomonium result is somewhat higher, but not in 
disagreement with this value. This indicates that the 
heavy quark mass dependence in the $1S$ and $2S$ hyperfine 
splittings
is very similar over the wide range of quark masses 
from $c$ to $b$ so that the ratio remains the same. 

The $S$-wave hyperfine splittings can be compared to 
the much smaller result for the $P$-wave states. 
In Appendix~\ref{appendix:cinonpert} we determine the 
$P$-wave hyperfine splitting to be 2(2) MeV, 
consistent with zero. 

%
\section{The $\eta_s$ mass and decay constant}
\label{sec:etas}
%

To complement the computation in the previous section, 
the lattice spacing was also determined using the decay 
constant of the fictitious $\eta_s$ meson, $f_{\eta_s}$. 
This is a pseudoscalar particle consisting of an 
$s \bar{s}$ pair whose properties can easily be computed 
in lattice QCD. It is particularly suitable for fixing the 
lattice spacing since there are no $u/d$ valence quarks 
meaning that the error coming from the chiral extrapolation to 
physical $u/d$ masses is small. 
The ``physical'' values of the $M_{\eta_s}$ and $f_{\eta_s}$ 
have to be fixed by comparison to $M_{\pi}$, $M_K$, $f_{\pi}$ and 
$f_K$ as in~\cite{Davies:2009tsa}. This requires a simultaneous 
chiral and continuum extrapolation for the masses and decay 
constants of the $\pi$, $K$ and $\eta_s$. Previously we 
found, on ensembles including 2+1 flavors of sea quarks, that 
properties of the $\eta_s$ were very close to those expected 
from leading order chiral perturbation theory, i.e.  
$M^2_{\eta_s} \approx 2M_K^2 - M_{\pi}^2$ and 
$f_{\eta_s} \approx 2f_K - f_{\pi}$. We re-examine that issue 
here on these ensembles containing 2+1+1 sea quarks. 

%
\subsection{Simulation details and Fitting}
\label{subsec:etasfits}
%

$s$ and $u/d$ valence quark propagators were calculated on the ensembles 
given in Table~\ref{tab:params} using the same HISQ action as used 
for the sea quarks. The HISQ action~\cite{Follana:2006rc} is 
a further improved version 
of the improved staggered (asqtad) action that reduces discretisation 
errors coming from staggered taste effects for that action by 
about a factor of 3. The improved staggered action smears 
the gluon fields that appear in the quark action in a very 
specific way to reduce high-momentum taste-exchange interactions 
but without increasing discretisation errors~\cite{Lepage:1998vj}. The HISQ 
action takes this one step further by performing two smearing 
steps. The original version of the HISQ action used an SU(3) 
projection of the smeared links between the two smearing steps. 
However, this caused difficulties for the updating algorithm 
when these quarks were included as sea quarks~\cite{Bazavov:2010ru}. 
Instead the sea quarks here use the HISQ action but with only 
a U(3) projection between the smearing steps. Whether U(3) 
or SU(3) projection it makes very little difference to 
the spectrum of mesons from HISQ quarks and so all the 
good features of the HISQ action demonstrated in~\cite{Follana:2006rc} remain 
essentially unaltered. Note that the HISQ action does not 
use tadpole-improvement -- the U(3) (or SU(3)) projection 
effectively takes care of large tadpole contributions in 
the same way that the use of the $u_0$ parameter does in 
the NRQCD and gluon actions.      

\begin{table}
\caption{ Valence light and strange quark mass parameters 
on each ensemble. The valence light quark masses are the 
same as in the sea (given in Table~\ref{tab:params}), 
except for a slight difference on set 3. 	
The valence strange quark masses have been retuned slightly to be 
closer to the physical values. 
Columns 4 and 5 give the number of configurations used 
from each ensemble and the number of time sources for 
propagators per 
configuration. 
\label{tab:hisqparams}
}
\begin{ruledtabular}
\begin{tabular}{lllll}
Set & $am_l^{\rm val}$ & $am_s^{\rm val}$ & $n_{cfg}$ & $n_t$ \\
\hline
1 & 0.013   & 0.0688 &  1021 & 16  \\
2 & 0.0064  & 0.0679 &  1000 & 16  \\
\hline
3 & 0.01044 & 0.0522 &  1053 & 16  \\ 
4 & 0.00507 & 0.0505 &  1000 & 16  \\ 
\hline
5 & 0.0074  & 0.0364 & 1008 & 16 \\ 
\end{tabular}
\end{ruledtabular}
\end{table}

The parameters of the valence quark propagators are listed 
in Table~\ref{tab:hisqparams}. We took the light quark 
mass to be the same as that in the sea (except for a small 
difference on set 3), but we retuned the 
valence strange quark masses slightly to allow for 
mistuning of the strange sea quark mass (of course the final 
well-tuned $s$ quark mass values cannot be decided until 
a value for the lattice spacing is determined so we 
will revisit this issue at the end of this section). 
We used delta function random wall sources as for the 
l-smeared $b$ quark propagators discussed in section~\ref{sec:ups}. 
We also used 16 evenly-spaced time sources per configuration 
to increase statistics. The starting position of these 
time sources was shifted from configuration to configuration 
in the ensemble. 
 
The light meson pseudoscalar correlators are calculated 
by combining the light and strange quark propagators. 
Here we use the goldstone mesons, 
made with the local $\gamma_5$ operator. 
Then the correlators are simply given by 
the squared modulus of the propagators summed over a time-slice 
to project onto zero momentum. The correlators were binned over 
all time sources on a configuration. 

\begin{figure*}
\includegraphics[width=6.0cm]{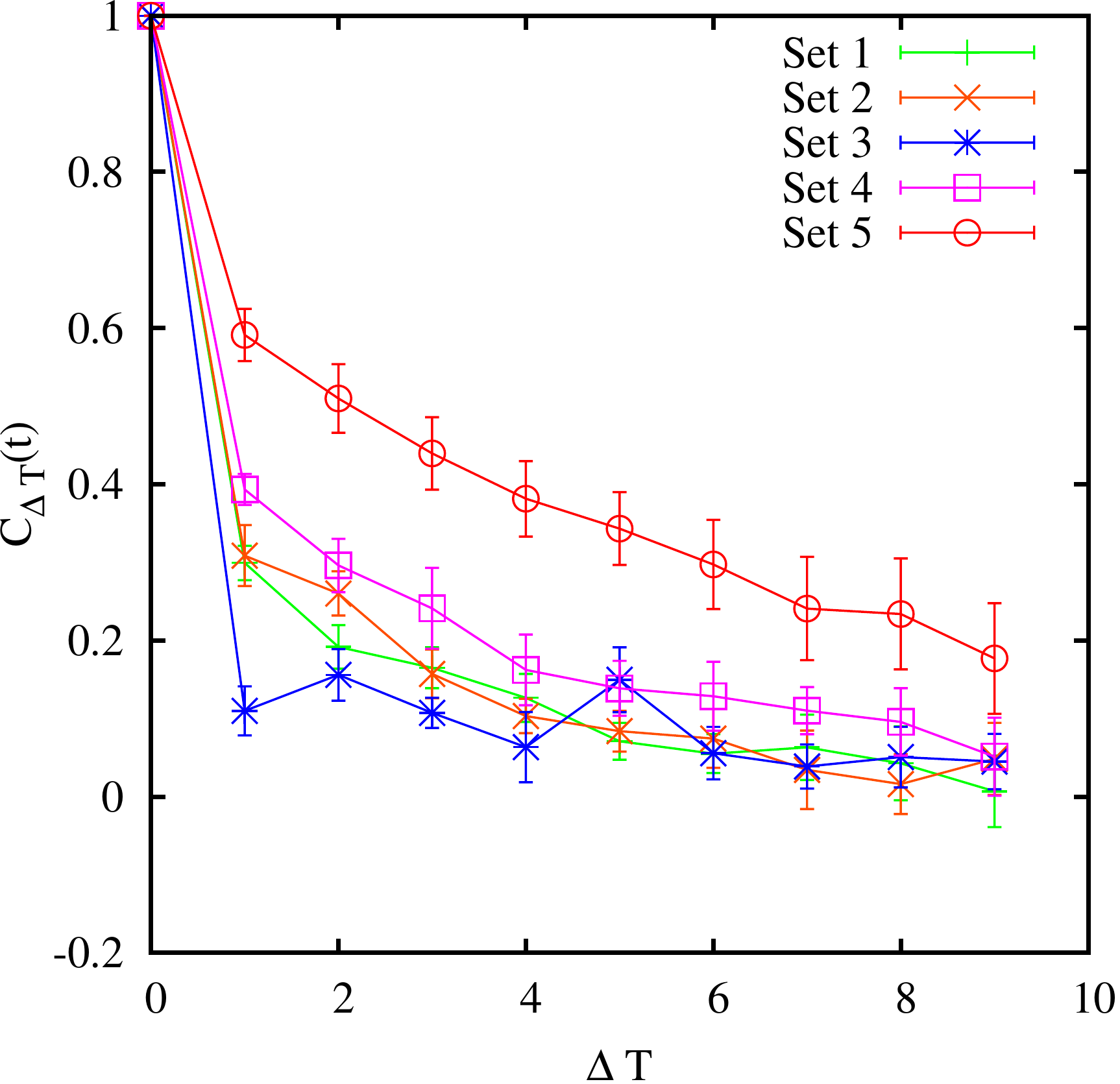}
\includegraphics[width=6.0cm]{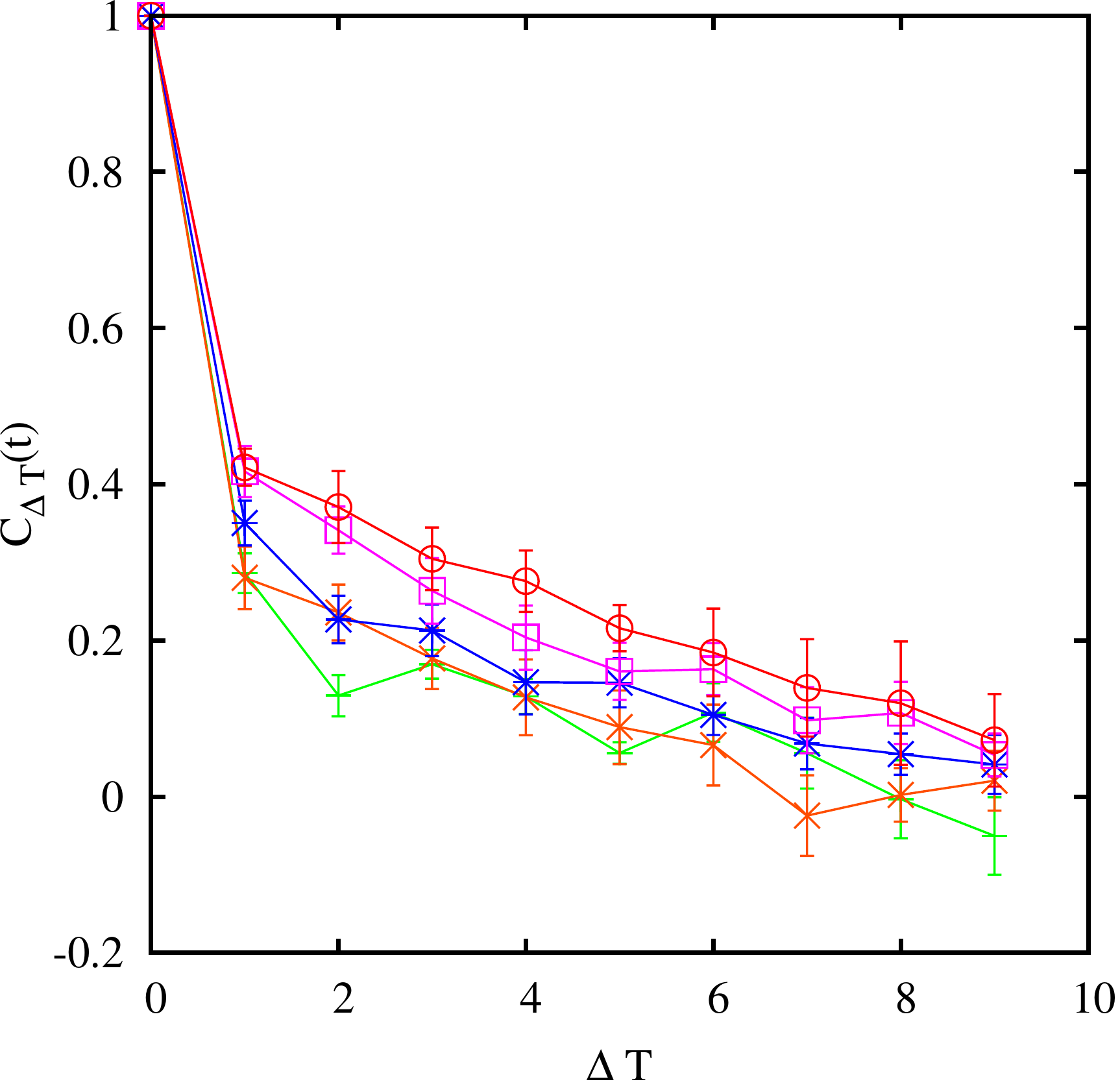}
\caption{Autocorrelation function $C_{\Delta T}$ for $\pi$ (left) 
and $\eta_s$ (right) correlators. These are made with a 
$\delta$ function random wall as described in the text.  
The key to results from different ensembles is (color online): 
set 1, green plus; set 2, orange cross; set 3 blue star; set 4 pink open 
square; set 5, red open circle. 
The correlators are evaluated at lattice time separation $t/a=6$ on 
very coarse lattices (sets 1 and 2), $t/a=8$ on coarse lattices (sets 3 and 4) 
and $t/a=10$ on fine lattices (set 5). This corresponds to a $t$ value, 
approximately constant in physical units across the lattice spacing 
values, 
where the $\pi$ and $\eta_s$ correlators have 
reached the ground-state plateau. $\Delta T$ is given in units of 
configuration number in the ordered list for each ensemble. 
 }
\label{fig:pion-autocorr}
\end{figure*}

To study the autocorrelations between configurations we proceed 
as in subsection~\ref{subsec:smear} to calculate the autocorrelation 
function $C_{\Delta T}$. Figure~\ref{fig:pion-autocorr} shows 
$C_{\Delta T}$ against $\Delta T$ for both the $\pi$ and $\eta_s$ 
correlators at a source-sink lattice time separation appropriate 
to our fits. This time separation is increased as the lattice spacing 
decreases to remain approximately physical. We see from the Figure 
that the $\eta_s$ shows little autocorrelation, although more than 
was visible for the $\Upsilon$ in Figure~\ref{fig:ups-autocorr}. We expect 
longer autocorrelation times for lighter mesons because they have a larger 
spatial extent and therefore decorrelation of the gluon field configuration 
on relevant spatial scales takes longer in Monte Carlo time. The $\pi$  
correlators on the coarse and very coarse lattices show a similar 
autocorrelation function to the $\eta_s$. However, on the 
fine lattices where autocorrelations might be expected to 
be worst, there are clear signs that neighbouring configurations 
in time are correlated. Note the difference between our result and 
that of~\cite{Bazavov:2010ru}. There very little autocorrelation 
was seen in the $\pi$ meson correlator, but fewer time sources were 
used per configuration (typically 4). 
Both here and in~\cite{Bazavov:2010ru} 
the time sources are moved 
randomly from one configuration to the next. 
However, with 16 time sources there is not much scope 
for a large shift in time between configurations. 
To reduce autocorrelations we bin all of our 
fine light meson correlators 
by a factor of 8 in configuration number before fitting.
From Figure~\ref{fig:pion-autocorr} this 
can be seen to reduce the autocorrelation function 
well below $e^{-1}$. 

The fitting method used for the correlators was the same as 
for the Upsilon correlators but with only a single source 
and sink smearing. 
For meson correlators made from relativistic quarks the 
fit function takes a `cosh' form rather than simple 
exponentials because of propagation in both time directions. 
For staggered quarks in 
general we have to include an additional oscillating term from 
opposite parity mesons that couple through the time-doubler 
quark. 
The fit function then becomes: 
\begin{eqnarray}
\label{eq:fitHISQ}
G_{\mathrm{meson}}(t)
 &=& \sum_{k=0}^{n_{\mathrm{exp}}} a_k( e^{-E_kt} + e^{-E_k(T-t)} )
\\
&-&
(-1)^{t/a}
\sum_{ko=0}^{n_{exp}} a_{ko} ( e^{-E_{ko}t} + e^{-E_{ko}(T-t)}  ). \nonumber
\end{eqnarray}
The oscillating piece is absent for the $\pi$ and $\eta_s$ because 
the valence quark and antiquark have equal mass and the oscillation 
cancels. It is necessary to include it for the $K$ meson. 
For each ensemble a simultaneous fit to all three correlators was 
performed using the appropriate form for each. This allowed us to 
take into account the correlations between the fit results for each meson 
in our subsequent chiral extrapolations.  
We use the full range of $t$ values in the fit apart from 
the first 3--5 time-slices. Priors for energies and amplitudes 
are chosen as for the $\Upsilon$ fits described in 
section~\ref{subsec:smear}. We take results from 
4 exponential fits, where ground-state masses 
and their errors have clearly stabilised.  

The results that we use are the ground-state meson masses and 
amplitudes, i.e. $k=0$ in equation~\ref{eq:fitHISQ}. 
The meson masses are given by 
the parameter $E_0$ from each fit, since there is no 
energy offset for staggered quarks as there is for 
NRQCD. 
The decay constants are extracted from the fit using
\be
f_{ab} = (m_a + m_b) \sqrt{ \frac{2a_0}{E_0^3}   }
\label{eq:decayconstant}
\ee
for a meson containing quarks $a$ and $b$ with 
masses $m_a,m_b$, ground state mass $E_0$ and ground 
state amplitude, $a_0$, 
from the fit form above.
Equation~\ref{eq:decayconstant} uses the PCAC relation, 
valid for staggered quarks, to relate the matrix 
element of the pseudoscalar 
density to that of the temporal axial current and therefore 
the decay constant. The existence of the PCAC relation 
means that the temporal axial current is absolutely 
normalised and there is no uncertainty from lattice 
to continuum current matching factors as there can 
be in some other quark formalisms. 

%
\subsection{Results and chiral extrapolations}
\label{subsec:etasres}
%

Our results for the $\pi$, $K$, and $\eta_s$  
meson masses and decay constants are 
listed in tables \ref{tab:stagmassdata} and 
\ref{tab:stagdecaydata}. We also give various ratios that are 
useful indicators of the sensitivity of the $\eta_s$ parameters 
to the chiral extrapolation in the $u/d$ sea quark mass, and 
to the lattice spacing.

%
%
\begin{table*}
\begin{tabular}{lcccccccccc}
\hline
\hline
Set & 1 & 2 & 3 & 4 & 5  \\
\hline
$aM_\pi$ 	& 0.23637(15) & 0.16615(7) & 0.19153(9) & 0.13413(5) & 0.14070(9)   \\
$aM_K$	 	& 0.41195(17) & 0.39082(9) & 0.32781(10)& 0.30757(7) & 0.23933(11)  \\
$aM_{\eta_s}$ 	& 0.53361(14) & 0.52797(8) & 0.42351(9) & 0.41476(6) & 0.30884(11)  \\
\hline
 $M_{\eta_s}^2/(2M_{K}^2 - M_\pi^2 )$ & 1.00426(43) & 1.00317(28) & 1.00636(34) & 1.00474(26) & 1.00660(27)   \\
\hline
\hline
\end{tabular}
\caption{Values for the ground state masses in lattice 
units ($E_0$ from eq.~\ref{eq:fitHISQ}) 
for $\pi$, $K$ and $\eta_s$ mesons. The fourth row gives the ratio 
of the square of the $\eta_s$ mass to a combination of $\pi$ and $K$ 
masses that would be 1 in leading order chiral perturbation theory. 
}
\label{tab:stagmassdata}
\end{table*}

\begin{table*}
\begin{tabular}{lcccccccccc}
\hline
\hline
Set & 1 & 2 & 3 & 4 & 5  \\
\hline
$af_\pi$ 	& 0.11183(9) & 0.10511(5) & 0.09075(5) & 0.08451(4) & 0.06621(5) \\
$af_K$	 	& 0.12689(8) & 0.12268(4) & 0.10185(5) & 0.09788(3) & 0.07427(4) \\
$af_{\eta_s}$ 	& 0.14199(6) & 0.14026(3) & 0.11312(4) & 0.11119(2) & 0.08238(4) \\
\hline
$f_{K}/f_{\pi}$ 	& 1.13467(58) & 1.16717(38) & 1.12231(38) & 1.15819(35) & 1.12170(39) \\
$f_{\eta_s}/f_{\pi}$ 	& 1.26974(80) & 1.33442(59) & 1.24653(68) & 1.31568(53) & 1.24416(69) \\
$f_{\eta_s}/(2f_{K} - f_\pi )$ 	& 1.00031(62) & 1.00007(27) & 1.00154(45) & 0.99948(26)& 1.00061(32)  \\
$f_{\eta_s}/M_{\eta_s}$ 	& 0.26609(11) & 0.26566(6)  & 0.26711(10) & 0.26809(6) & 0.26674(12) \\
\hline
\hline
\end{tabular}
\caption{Values for the ground state decay constants in lattice 
units (derived from $a_0$ in eq.~\ref{eq:fitHISQ} as described in 
the text) 
for $\pi$, $K$ and $\eta_s$ mesons. We also give various ratios 
of decay constants obtained from the simultaneous fit. 
The sixth row gives the ratio 
of the $\eta_s$ decay constant to a combination of $\pi$ and $K$ 
decay constants that would be 1 in leading order chiral perturbation theory. 
}
\label{tab:stagdecaydata}
\end{table*}

\begin{figure}
\includegraphics[width=0.9\hsize]{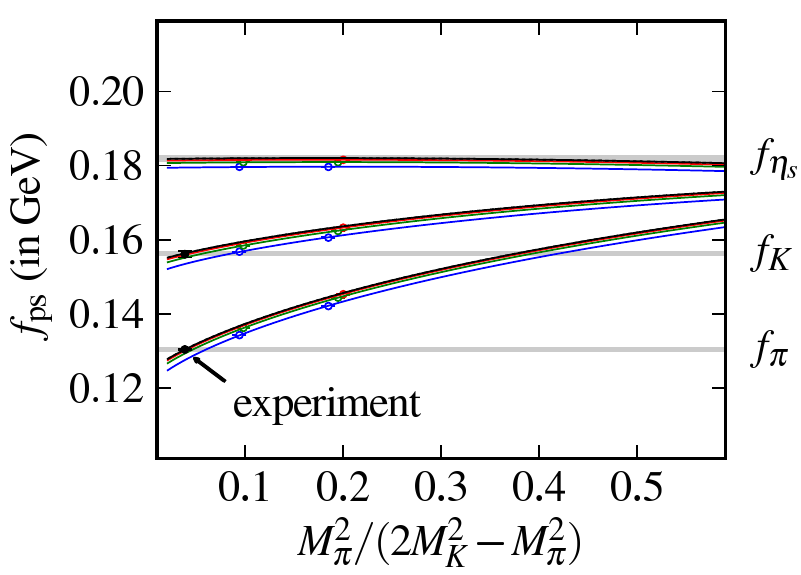}
\caption{ The pseudoscalar decay constants plotted 
against the ratio of squared pseudoscalar masses that 
is approximately equal to $m_l/m_s$. 
The points have been adjusted for finite volume 
effects and for mistuning of the strange quark mass. 
The lines are from the tuned fit function at each 
lattice spacing, with results increasing in 
value from very coarse (blue) to fine (red) (color online). 
The top (black) line is
the $a=0$ curve and the black leftmost data points 
give the experimental value for $f_{\pi}$ and $f_K$ 
given values for $V_{ud}$ and $V_{us}$~\cite{pdg}. 
 }
\label{fig:fpi}
\end{figure}

\begin{figure}
\includegraphics[width=0.9\hsize]{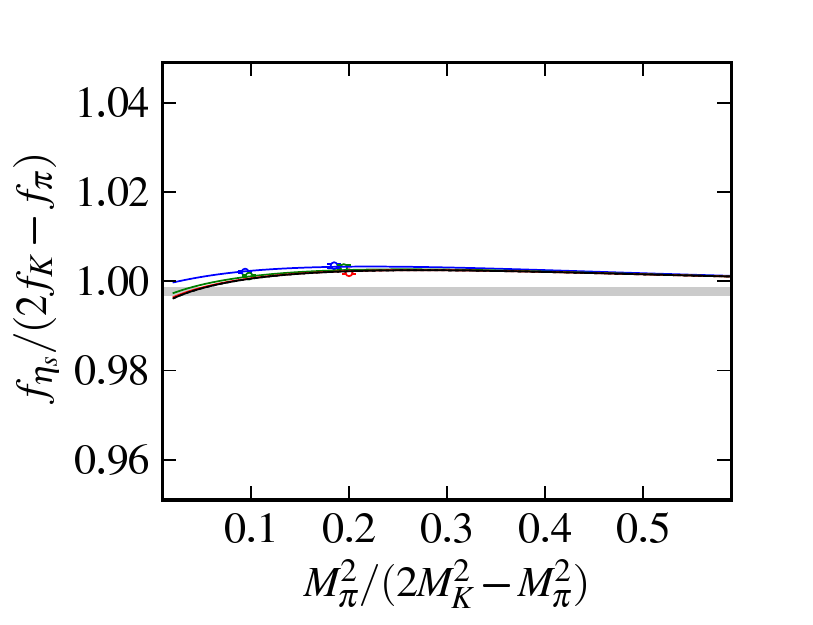}
\caption{ The ratio of $f_{\eta_s}$ to 
$2f_K-f_{\pi}$, which would be  1 in leading order chiral perturbation 
theory. The ratio of squared meson masses on the 
$x$-axis corresponds approximately to $m_l/m_s$. 
The blue, green and red points and fit curves 
correspond to very coarse, coarse and fine lattices
respectively (color online). The black line is 
the continuum, $a=0$, fit curve. 
 }
\label{fig:fetas}
\end{figure}

\begin{figure}
\includegraphics[width=0.9\hsize]{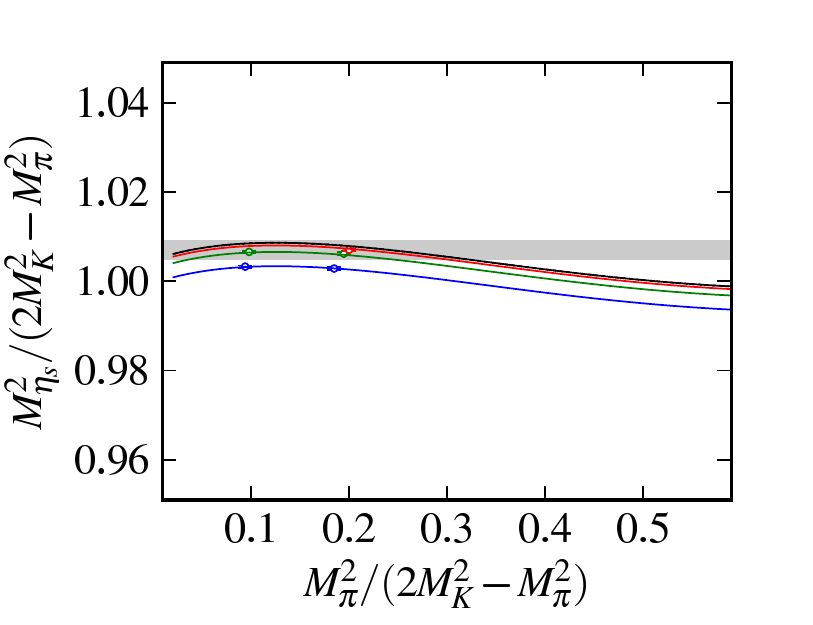}
\caption{ The ratio of $M_{\eta_s}^2$ to 
$2M_K^2-M_{\pi}^2$, which would be  1 in leading order chiral perturbation 
theory. The ratio of squared meson masses on the 
$x$-axis corresponds approximately to $m_l/m_s$. 
The blue, green and red points and fit curves 
correspond to very coarse, coarse and fine lattices
respectively (color online). The black line is 
the continuum, $a=0$, fit curve. 
 }
\label{fig:metas}
\end{figure}

We fit the three decay constants and meson masses simultaneously 
using SU(3) chiral perturbation theory, adapted to 
include discretisation effects. 
The usual approach is to use values for the decay constant 
and meson mass 
in GeV, having chosen a value of the 
lattice spacing on each ensemble. Extrapolation to 
the point where $M_{\pi}$ and $M_K$ 
take their physical values then allows comparison to 
experiment of the resulting values for $f_{\pi}$ 
and $f_K$.  
Here instead we use values for $r_1/a$ to fix 
the relative lattice spacing between ensembles
and keep the physical value of $r_1$ as a parameter 
to be obtained from the fit. The value for $r_1$
is determined by the requirement to match 
$f_{\pi}$ and $f_K$ from experiment in the chiral limit
where the experimental values are included as 
extra pieces of `data' for the fit. 
The experimental values for $f_{\pi}$ and 
$f_K$ come from experimental measurement 
of the leptonic decay rate and values of 
$V_{ud}$ and $V_{us}$ taken from elsewhere. 
We use~\cite{pdg}
\begin{eqnarray}
f_{\pi} &=& 0.1304(2) \mathrm{GeV} \nonumber \\
f_K &=& 0.1561(9) \mathrm{GeV}. 
\end{eqnarray}
The meson mass values that go with these decay constants 
in a world appropriate to lattice QCD 
without electromagnetism and in which 
$m_u=m_d$ are~\cite{Aubin:2004fs}:
\begin{eqnarray}
M_{\pi}^2 &=& M_{\pi^0}^2  \\
M_K^2 &=& \frac{1}{2}\left(M_{K^0}^2 + M_{K^+}^2 - (1+\Delta_E)(M_{\pi^+}^2-M_{\pi^0}^2) \right). \nonumber 
\label{eq:pikvals}
\end{eqnarray}
We take $\Delta_E$, which parameterizes the violations of 
Dashen's theorem, to have the value $1 \pm 1$. 
The decay constants are already defined to be 
results in pure QCD, provided electromagnetic effects 
have been removed from the experimental leptonic 
decay rates~\cite{pdg}. The residual error in 
$f_K$ from the fact that it is the decay constant 
of the $K^+$ whereas the $K$ mass in equation~\ref{eq:pikvals}
is the isospin average is less than 
0.1\%~\cite{Aubin:2004fs}, so we ignore it 
here. 

Our fit also returns 
a physical value for $f_{\eta_s}$ and $M_{\eta_s}$. 
This can be used in subsequent analyses to tune 
the $s$ quark mass and to fix the lattice spacing. 
It is a good meson to use for this purpose because 
its parameters are very insensitive to the 
sea quark masses, as we see in Tables~\ref{tab:stagmassdata} 
and~\ref{tab:stagdecaydata}. 

The analysis is 
the same as that used in~\cite{Davies:2009tsa} except 
for two improvements. 
The first is that the fit is simplified because HISQ quarks 
are used in both the valence and sea sectors. 
The second is that we include correlations between all of 
the decay constants and meson masses on each 
ensemble by feeding into the fit the covariance 
matrix that resulted 
from the simultaneous fit to all three 
meson correlation functions. 
Below we provide a brief description of the chiral/continuum 
extrapolations following~\cite{Davies:2009tsa}. 

On each ensemble 
the decay constants and meson masses
are a function of the masses of the valence quarks for that 
meson and of the masses of the sea quarks for that ensemble,
with the coefficients of
the mass dependence constrained to be the same for 
the $\pi$, $K$ and $\eta_s$ mesons. 
Because the quark masses run with energy scale 
it simplifies the fits to use a dependent variable 
related to the square of 
appropriate goldstone meson masses instead the  
quark mass. Thus we write
\begin{equation}
x_l = \frac{M_{\pi}^2/2}{\Lambda_{\chi}^2}, 
\label{eq:xdef}
\end{equation}
where $\Lambda_{\chi}$ provides the cut-off 
scale of the chiral expansion, 
\begin{equation}
\Lambda_{\chi}=4\pi f_{\pi}/\sqrt{2}.
\label{eq:chidef}
\end{equation}
Similarly
\begin{equation}
x_s = \frac{M_K^2 - M_{\pi}^2/2}{\Lambda_{\chi}^2}. 
\label{eq:xsdef}
\end{equation}
In the cases where the sea and valence quark masses
differ, the $x$ parameters for the sea quark 
masses are obtained from those of the valence 
masses by rescaling in proportion to the quark mass.  
Then the decay constant made from valence 
quarks $a$ and $b$ takes the functional form
\begin{equation}
f(x_a,x_b,x_l^{\rm sea},x_s^{\rm sea},a) = f^{\rm NLO} + \delta f_{\chi} + \delta f_{\rm lat},
\label{eq:chifdef}
\end{equation}
where $f^{\rm NLO}$ is the full partially quenched 
chiral perturbation theory formula 
at next-to-leading order~\cite{Sharpe:2000bc} 
and $\delta f_{\chi}$ and $\delta f_{\rm lat}$ 
include possible correction terms coming from higher order terms 
in the quark mass and finite lattice spacing corrections. 
Each of the terms contains a set of unknown coefficients which are given 
prior constraints in our fit allowing us to test their effect 
on our final result. 

$f^{\rm NLO}$~\cite{Sharpe:2000bc} includes 
terms proportional to the squares of 
the appropriate meson masses as well as logarithmic terms 
that appear in combinations such as, for example, 
$(x_a+x_l^{\rm sea})\log(x_a+x_l^{\rm sea})$. 
The logarithmic terms are corrected for finite volume 
effects through the use of finite volume chiral perturbation 
theory. The finite volume correction is significant for 
$f_{\pi}$, particularly on set 1 where $M_{\pi}L=3.8$ and 
the finite volume correction is 1.8\%. For the other sets, 
with $M_{\pi}L > 4$, the correction ranges from 0.4\% 
to 0.7\%. For $f_K$ and $f_{\eta_s}$ the correction 
is much smaller. It is at its largest on set 1 with 0.7\% 
for $f_K$ and 0.2\% for $f_{\eta_s}$. 

$\delta f_{\chi}$ includes polynomial dependence on various combinations 
of the $x_i$ up to and including $x_i^4$ terms~\cite{Davies:2009tsa}. 
Most of these terms only matter for the $s$ quark and they allow 
for differences between the $s$ and $l$ sectors within 
SU(3) chiral perturbation theory. 
Since $x_s =0.17$ including $x_i^4$ terms means that missing terms at $x_s^5$ 
are $\mathcal{O}(10^{-4})$, smaller than our statistical errors. 
It is sufficient to include polynomials because we cannot 
distinguish high-order logarithms from polynomials 
over this range in $x_i$. 

$\delta f_{\rm lat}$ allows for dependence on powers
of the square of the lattice spacing, since this is 
the form that discretisation errors take for 
staggered quarks. We include terms up 
to $(a\Lambda_{\rm QCD})^8$ where $\Lambda_{\rm QCD}$ is 
taken to be $\mathcal{O}(0.6 \mathrm{GeV})$. 
The coefficients of the $a$-dependence 
are also allowed to have dependence on valence and 
sea mass dependence. This includes dependence on $\log(x_l)$ to model 
discretisation errors coming from staggered taste-changing 
effects~\cite{Davies:2009tsa}. 

The terms in the chiral expansion are generally written 
so that the coefficients are expected to be $\mathcal{O}(1)$. 
For these coefficients we take the prior in our fit to be $0\pm 1$. 
This is true for the higher order terms in the chiral expansion 
that relate to $a$-dependence and mass-dependence, except where 
the masses involved are sea-quark masses and then the prior is 
taken as $0 \pm 0.3$, simply because sea-quark effects are typically suppressed
over valence quark effects by this amount. The prior 
on the bare decay constant parameter in chiral perturbation theory, 
$f_0$, is taken as $0.11 \pm 0.02$. In fact the parameter that is 
tuned by the fit is $\log(f_0)$ in order to keep $f_0$ positive.
The prior on $\log(f_0)$ is then taken as $-2.2 \pm 0.18$. 
The priors for the coefficients $L_{4,5,6,8}$ 
that multiply analytic terms at NLO in chiral perturbation 
theory are taken as $0\pm 0.01$. 

The meson masses are fitted simultaneously 
with the decay constants feeding in the 
$6\times 6$ covariance matrix on each ensemble. 
The leading behaviour in chiral perturbation theory 
for the meson masses is 
now trivial. 
However the chiral fit, which shares some of the same coefficients 
as that of the decay constants
~\cite{Sharpe:2000bc}, allows us to fix the higher order 
behaviour as a function of sea and valence masses. In 
particular it allows us to fix the behaviour of the 
$\eta_s$ mass as the $\pi$ and $K$ masses vary, so that
we can obtain its value at the physical point. 
The priors in the chiral fit for the meson masses 
take the same form as described above for the decay 
constant. 

The fitting forms above were extensively tested for 
robustness against both
real and fake data in~\cite{oldfds} and~\cite{Davies:2009tsa}.

\begin{table}
\caption{ Complete error budget for 
$r_1$, $f_{\eta_s}$, $M_{\eta_s}$ and $f_{\eta_s}/M_{\eta_s}$.
Errors are given as a percentage of the physical 
value. 
Errors which are negligible compared to the 
others are indicated by `0'.
}
\label{tab:etaserror}
\begin{ruledtabular}
\begin{tabular}{lllll}
 & $r_1$ & $f_{\eta_s}$ & $M_{\eta_s}$ & $f_{\eta_s}/M_{\eta_s}$ \\
\hline
stats/fitting & 0.24 & 0.16 & 0.07 & 0.18 \\
$a$-extrapolation & 0.46 & 0.14 & 0.03 & 0.16 \\
$m_l$-extrapolation & 0.09 & 0.12 & 0.04 & 0.11 \\
finite volume & 0.04 & 0 & 0 & 0\\
$r_1/a$ & 0.73 & 0.12 & 0.02 & 0.12 \\
initial $r_1$ uncertainty & 0.26 & 0.02 & 0 & 0.02\\
$M_{\pi}$,$M_K$ & 0 & 0.05 & 0.14 & 0.09\\
Total & 0.90 & 0.28 & 0.17 & 0.30 \\
\end{tabular}
\end{ruledtabular}
\end{table}

The results of our fit are shown in Figure~\ref{fig:fpi}.
The data points, adjusted for finite volume effects and 
for the slight mistuning of the valence and sea strange quark masses,
are plotted as a function of $x_l/x_s$. 
The fit lines at each value of the lattice spacing are shown
along with the $a=0$ line. At the physical value for 
$x_l/x_s$ we give the experimental values for $f_{\pi}$ 
and $f_K$. 
This plot should be compared with 
Figure 4 in~\cite{Davies:2009tsa}. It is evident that these
`second generation' configurations have significantly 
smaller discretisation errors~\cite{Bazavov:2010ru}.  

The fit has a $\chi^2/\mathrm{dof}$ value of 0.3 for 36 degrees of 
freedom. The fitted value of $f_0$ is $\exp(-2.174 \pm 0.028)$,   
in agreement with SU(3) chiral fits using asqtad improved staggered 
quarks~\cite{Bazavov:2009tw}. The resulting physical values for $f_{\eta_s}$
and $M_{\eta_s}$ are
\begin{eqnarray}
f_{\eta_s} &=& 0.1819(5) \mathrm{GeV} \nonumber \\
M_{\eta_s} &=& 0.6893(12) \mathrm{GeV} \nonumber \\
f_{\eta_s}/M_{\eta_s} &=& 0.2638(8). 
\label{eq:etasvals}
\end{eqnarray}
These are in agreement with the results obtained on 
$n_f=2+1$ dynamical asqtad configurations~\cite{Davies:2009tsa} 
but considerably more accurate because our statistical precision 
is improved, and we have smaller continuum and chiral extrapolation 
errors. These last two are reduced because of the improvements
in the gluon field configurations and because we are working 
closer to the chiral limit.  Complete error budgets for 
$f_{\eta_s}$, $M_{\eta_s}$ and their ratio are given in 
Table~\ref{tab:etaserror}. 

The results in equation~\ref{eq:etasvals} are very close, but 
in fact differ significantly from the 
expected result from leading order chiral perturbation theory. 
This is illustrated in Figures~\ref{fig:fetas} and~\ref{fig:metas} 
in which the ratios $f_{\eta_s}/(2f_K-f_{\pi})$ and 
$M_{\eta_s}^2/(2M_K^2-M_{\pi}^2)$ are plotted against $x_l/x_s$. 
Both ratios are very flat in $x_l/x_s$, never differing by 
as much as 1\% from 1. We determine the physical values for 
the ratios to differ significantly from 1, however, with 
results:
\begin{eqnarray}
f_{\eta_s}/(2f_K-f_{\pi}) &=& 0.9977(6) \nonumber \\
M_{\eta_s}^2/(2M_K^2-M_{\pi}^2) &=& 1.0070(18). 
\label{eq:etasrats}
\end{eqnarray}

Since our fit uses $r_1/a$ to set the relative lattice 
spacing we can determine a value for $r_1$ from the 
final match with experiment for $f_{\pi}$ and $f_K$.   
We obtain 
\begin{equation}
r_1(f_{\eta_s}) = 0.3209(29) \mathrm{fm}.
\end{equation}
The error budget for $r_1$ is given in Table~\ref{tab:etaserror}. 
This physical result for $r_1$ agrees with the value 
obtained from the same analysis on $n_f = 2+1$ 
dynamical asqtad lattices~\cite{Davies:2009tsa}, but 
is almost twice as accurate. 

We can use $f_{\eta_s}$ and $M_{\eta_s}$ to determine the 
$s$ quark mass and lattice spacing 
on each ensemble. This is done by tuning 
the $s$ quark mass so that 
$f_{\eta_s}/M_{\eta_s}$ takes the value in equation~\ref{eq:etasvals} 
and then the lattice spacing is read from the 
value of $f_{\eta_s}$. We do this here retrospectively 
by using our chiral fits to tune the $s$ quark 
mass and work out the corresponding changes in 
$f_{\eta_s}$ and $M_{\eta_s}$. For simplicity the 
sea light quark masses were also retuned to the 
physical value. The values of $a$ obtained are 
given in Table~\ref{tab:aval}.  

In Table~\ref{tab:mtune} we give tuned values 
of $am_s$ on each ensemble as a result of 
tuning the $\eta_s$ to the physical value 
given in equation~\ref{eq:etasvals}. 
Over the 
short range needed for the retuning the relationship 
$m_s \propto M_{\eta_s}^2$ works very well. 
We give results 
for both the case of using the $\eta_s$ decay 
constant to fix the lattice spacing and of 
using the $\Upsilon$ $2S-1S$ splitting. 
The values of $am_s$ obtained from the 
two methods differ substantially on 
the very coarse lattices but come into 
agreement on the fine lattices as expected. 

The errors in the tuned values of $am_s$ 
are dominated by the errors in the lattice 
spacing. The relative error in $a$ is 
doubled in $am_s$ because the quark mass 
is proportional to the square of the meson 
mass. When the quark mass is converted to 
physical units one factor of the lattice 
spacing error disappears. 

%
\section{$r_1$}
\label{sec:r1}
%

\begin{figure}
\includegraphics[width=0.9\hsize]{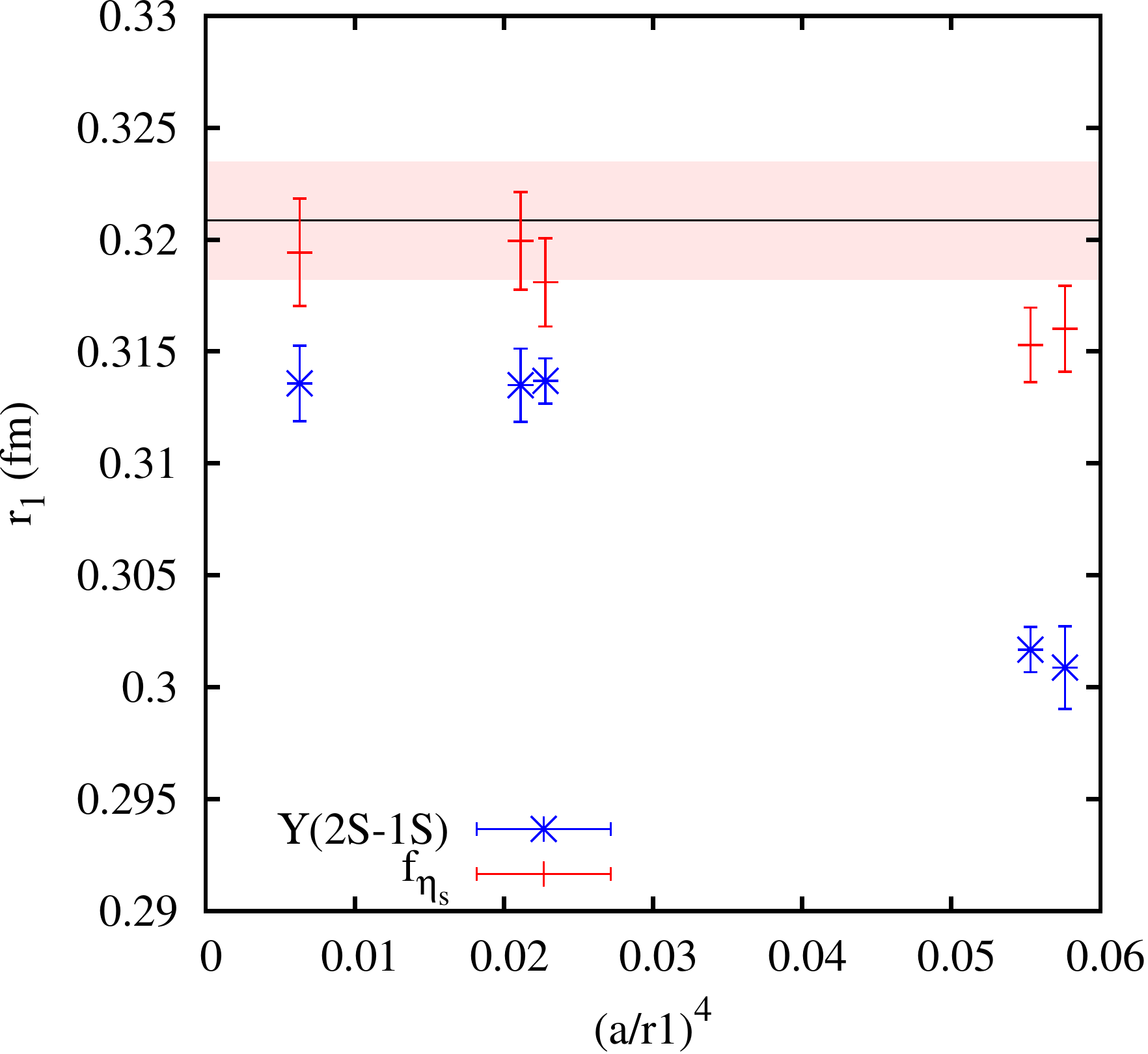}
\caption{ Values for the heavy quark potential 
parameter $r_1$ obtained by combining values 
for $r_1/a$ from MILC with either of our 
two methods for determining the lattice spacing. 
The red plus symbols correspond to using the 
$\eta_s$ and the blue stars to using the 
$\Upsilon$ $2S-1S$ splitting (these points 
do not include the NRQCD systematic error which 
is correlated between the points). The black line 
with light red error band corresponds to 
the final value for $r_1$ from the combined fit 
to the results from both methods and includes the 
total error. 
 }
\label{fig:r1}
\end{figure}

The values of the heavy quark potential parameter, $r_1$, 
can be determined by combining the values for $r_1/a$ from 
MILC given in Table~\ref{tab:params} with the values for 
the lattice spacing given from our two different methods in 
Table~\ref{tab:aval}. 
We use `unsmoothed' values of $r_1/a$ which 
are the results of an independent fit to the heavy quark potential 
on each ensemble. 
Figure~\ref{fig:r1} shows 
the results for $r_1$ from each method as a function 
of the lattice spacing. Differences are evident on 
the very coarse lattices as a result of discretisation errors 
but there is clear convergence as $a \rightarrow 0$. 
The results are plotted against $(a/r_1)^4$ since the 
leading tree-level discretisation errors are at $a^4$. 
Note that the behaviour of this plot is rather different 
from that obtained previously on the 2+1 flavor configurations 
(Figure 3 of~\cite{Davies:2009tsa}). There is a little less variation 
with $a$, to be expected because of the various improvements 
to the discretisation of QCD. The main difference 
however is the direction of approach 
to $a=0$. The results for $r_1$ from 
$f_{\eta_s}$ are now very flat and the results from 
the $\Upsilon$ approach $a=0$ from below. 
This reflects a change in the relative discretisation errors 
of the quantities involved. 

As discussed above, the chiral fits involving 
the $\eta_s$ give a result for $r_1$ of 0.3209(29) fm. 
We can fit the results from using the $\Upsilon$ method 
to test if they are consistent with this. Using 
the $\eta_s$ result as a prior for the $\Upsilon$ fit 
then enables us to extract an improved result for 
$r_1$ which combines both methods. 

 To extract a physical value for $r_1$ using 
the $\Upsilon$ results we use the same functional form for the 
fit as was used earlier for $R_S$ and $R_P$, equation~\ref{eq:fitxa}. 
This includes an allowance for variations as a function 
of the sea light quark masses, although it is clear from 
the results that any such dependence is very 
small. Indeed the quantities being used were chosen for their 
insensitivity to such effects. We also include an 
allowance for discretisation errors, both of the 
standard type, varying as $(a\Lambda)^n$ and the  from various 
$am_b$-dependent type coming from radiative corrections 
in NRQCD. 
We also allow for the NRQCD systematic error from 
Tables~\ref{tab:systrelrad} and~\ref{tab:systdisc} 
as a correlated error for all 5 ensembles.  

The fit to the $\Upsilon$ values using a large width 
prior for the physical value (0.32(10)fm) 
gives $\chi^2/{\mathrm{dof}} = 0.79$ 
for 5 degrees of freedom and $r_1(\Upsilon)=0.310(6)$ fm. 
This shows the required consistency in the determination 
of the lattice spacing from the two methods 
as $a\rightarrow 0$. 
The fit including the prior value from the $\eta_s$ 
analysis gives $\chi^2/{\mathrm{dof}} = 0.76$ and 
result:
\begin{equation}
r_1 = 0.3209(26) \mathrm{fm}. 
\label{eq:finalr1}
\end{equation}
This is slightly improved over the $\eta_s$ value 
on its own.

This final value for $r_1$ can now be used to determine 
$a$ on other ensembles if values
of $r_1/a$ are available. We include in Table~\ref{tab:aval} 
the lattice spacing values on sets 1 to 5 from using $r_1$. 

Our result for $r_1$ can be compared to our previous 
result of 0.3133(23) fm on the MILC 2+1 flavor dynamical 
asqtad lattices~\cite{Davies:2009tsa}. 
This is 2\% lower than our current result 
with a combined uncertainty of 1\% and so is not significant. 
In principle the two 
results do not have to agree because we are now including 
$c$ quarks in the sea. However we expect this 
to have a small effect and then only in short-distance 
quantities~\cite{Gregory:2010gm}. 
We can obtain estimates for the effect on the $\Upsilon$ $2S-1S$
splitting from the fact that it is proportional 
to the hyperfine splitting. Missing
$c$ quarks in the sea increases the $\Upsilon$ mass 
by approximately 5 MeV with a smaller amount for 
excited states. It therefore 
reduces the $2S-1S$ splitting by approximately 2.5 MeV or 
0.4\%. This could have led previously to a 0.4\% 
underestimate of $r_1$ from the $\Upsilon$ $2S-1S$ splitting 
if $r_1/a$ itself was not affected. This effect is 
no larger than other sources of systematic error 
in the earlier calculation~\cite{Gray:2005ur} coming from radiative 
corrections to $v^4$ terms that are also now included.   
Thus we cannot claim to see any strong evidence of 
an effect from $c$ quarks in the sea. 
Indeed if we compare the $r_1$ values coming from 
the $\Upsilon$ analysis alone there is a change 
of 2(2)\%, in which a 0.4\% effect from $c$ in the 
sea would be invisible. 
Any allowance for an effect on $r_1$ itself, 
also a fairly short distance quantity, would reduce this 
expected variation further.  
The $\eta_s$ analysis would be expected to be 
very insensitive to sea charm because of the low internal 
momenta inside these light hadrons. For that case 
we see only a 0.5\% change in the value of $r_1$ obtained, 
again with a 2\% error. 

%
\section{$m_b/m_s$}
\label{sec:mbms}
%

\begin{figure}
\includegraphics[width=0.9\hsize]{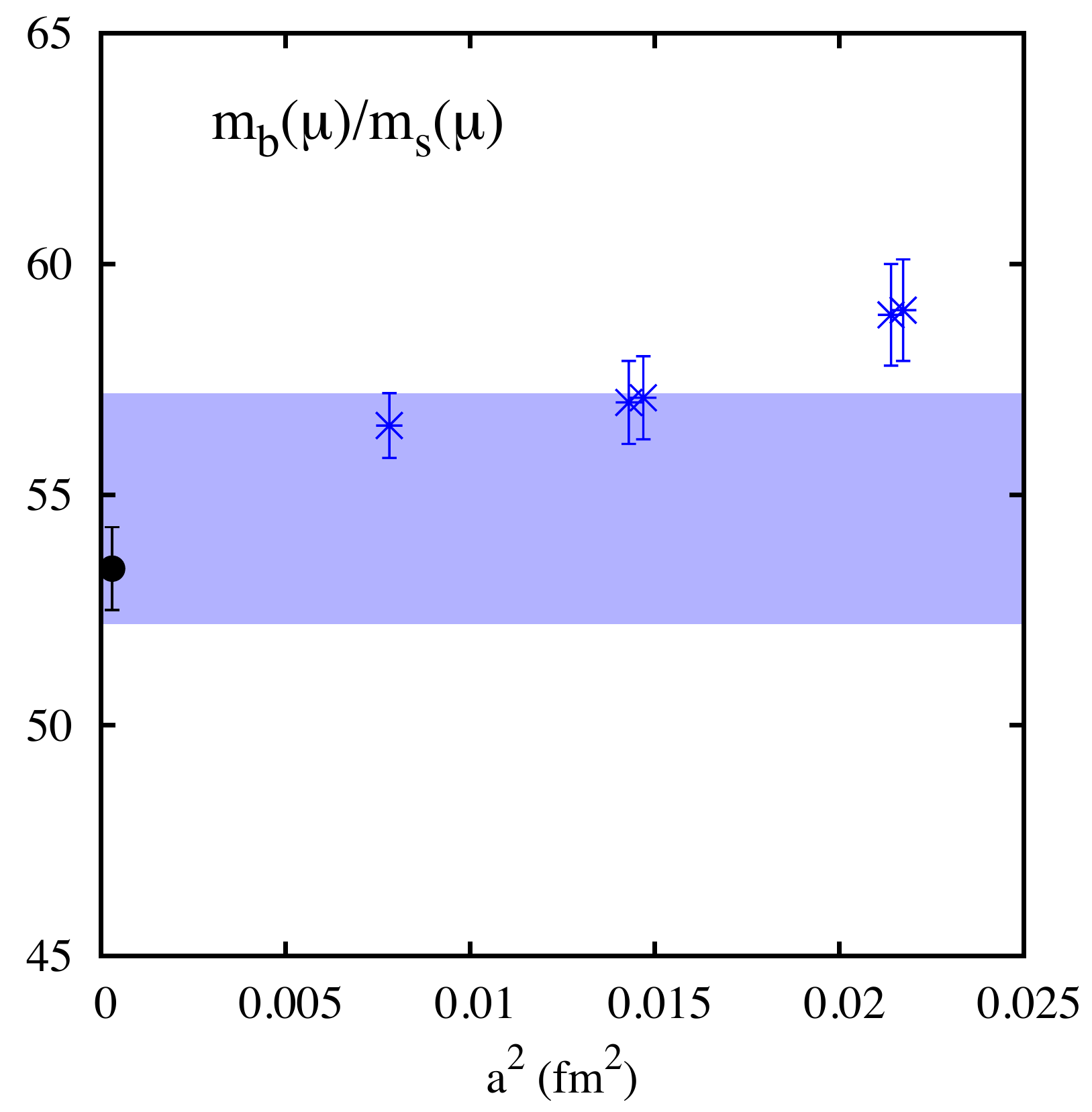}
\caption{ Values for the ratio of 
the $b$ quark mass to the $s$ quark mass 
in the $\overline{MS}$ scheme at a given scale plotted 
against the square of the lattice spacing. 
Results are obtained
from combining NRQCD $b$ quark masses and 
HISQ $s$ quark masses with an $\mathcal{O}(\alpha_s)$ perturbative 
renormalisation.  
The errors on the points include statistical/fitting 
errors, lattice spacing errors and NRQCD systematic errors. 
The final result, including the $\mathcal{O}(\alpha_s^2)$ 
perturbative 
error is plotted as the shaded blue band. 
The result from our previous fully nonperturbative 
calculation on ensembles including 
2+1 flavors of sea quarks~\cite{McNeile:2010ji, Davies:2009ih}  
is given by the black filled circle 
at $a=0$. 
 }
\label{fig:mbms}
\end{figure}

From Table~\ref{tab:mtune} we can determine the 
ratio of the bare NRQCD $b$ quark mass to the bare 
HISQ $s$ quark mass on each ensemble. To do this 
we must use the same determination of the lattice 
spacing for the tuning of each mass, and so we 
use the lattice spacing determined from the $\Upsilon$ $2S-1S$ 
splitting (columns 2 and 3). The lattice spacing error 
appears doubled in $m_s$ and once in $m_b$ because of 
their different dependence on the meson masses 
used to fix them. These errors are correlated 
in the ratio $m_b/m_s$ so one factor of the lattice 
spacing error cancels between numerator and 
denominator. 

The ratio of masses in different 
schemes (NRQCD and HISQ) is not
particularly useful. However, we can convert this 
using perturbation theory 
to a ratio of masses in the same mass-independent 
scheme, such as $\overline{MS}$, at the same scale, $\mu$. 
The ratio then becomes scale-independent and the 
same in any scheme related to $\overline{MS}$ by a 
simple renormalisation. 
For both the NRQCD and the HISQ actions the mass 
renormalisation is known
to $\mathcal{O}(\alpha_s)$.  

The lattice to $\overline{MS}$ mass renormalisation constant 
is calculated by multiplying the lattice bare mass to pole mass 
renormalisation by the continuum pole mass to $\overline{MS}$ 
renormalisation. This latter renormalisation is given by:
\begin{equation}
m_q^{\overline{MS}}(\mu) = m_{\mathrm{q,pole}}\left(1 + \alpha_s[-\frac{4}{3\pi} -\frac{2}{\pi}\ln{\frac{\mu}{m_{\mathrm{q,pole}}}}] + \ldots \right) .
\label{eq:msbar}
\end{equation}
The lattice bare mass to pole mass
renormalisation for HISQ quarks is given for small 
quark masses by~\cite{Mason:2005bj, quentin}~\footnote{Note that
there is a typographical error in~\cite{Mason:2005bj} so 
that the lattice logarithm is given the wrong sign in 
equation 11.}: 
\begin{equation}
m_{\mathrm{s,pole}} = \frac{am_s}{a}\left(1 + \alpha_s[-\frac{2}{\pi}\ln{am_s} + 0.5387] \ldots \right) ,
\label{eq:renhisq}
\end{equation}
where we have written the equation explicity for the strange quark mass. 
When equations~\ref{eq:msbar} and~\ref{eq:renhisq} are combined to obtain 
the conversion factor from the lattice bare mass to the 
$\overline{MS}$ mass at scale $\mu$ and $\mathcal{O}(\alpha_s)$ the logarithm multiplying 
$\alpha_s$ becomes $\ln(a\mu)$, and there is a constant 
given by 0.5387-$4/(3\pi)$.   

We can also write the NRQCD mass renormalisation in the form 
\begin{equation}
m_{\mathrm{b,pole}} = \frac{am_b}{a}\left(1 + \alpha_s[-\frac{2}{\pi}\ln{am_b} + A^{\mathrm NRQCD}] \ldots \right) .
\label{eq:rennrqcd}
\end{equation}
although no $\ln(am)$ term is explicit 
in that calculation. On doing this we find that the remainder 
term, $A^{\mathrm NRQCD}$ given in Table~\ref{tab:c4zm}, 
has very little $am_b$ dependence. 

Combining equations~\ref{eq:msbar},~\ref{eq:renhisq} and~\ref{eq:rennrqcd} 
it is then clear that the
ratio of $\overline{MS}$ masses for $b$ and $s$ is given to $\mathcal{O}(\alpha_s)$ by:
\begin{equation}
\frac{m_b^{\overline{MS}}(\mu)}{m_s^{\overline{MS}}(\mu)} = \frac{am_b}{am_s} \left[1+ \alpha_s(A^{\mathrm{NRQCD}}-0.5387) + \ldots \right]
\label{eq:massrat}
\end{equation}
where the $\mu$ dependence cancels out. 
The ratio of bare lattice masses from columns 2 and 3 of 
Table~\ref{tab:mtune} varies very little with lattice 
spacing with values between 51 and 52. The renormalisation 
in equation~\ref{eq:massrat} is a relatively mild one, 
with $\alpha_s$ coefficient varying between 0.31 and 0.39 with $am_b$ value. 
We apply this one-loop renormalization with $\alpha_s$ values 
taken as $\alpha_V(1.8/a)$ from Table~\ref{tab:alpha}. 
The energy scale for $\alpha_s$ is then 
in agreement with the Brodsky-Lepage-Mackenzie 
scale calculated for the light quark (asqtad) mass 
renormalisation in~\cite{Mason:2005bj}. This gives 
the values for the $\overline{MS}$ $m_b/m_s$ ratio 
plotted in Figure~\ref{fig:mbms}. 

The results in Figure~\ref{fig:mbms} show very little 
dependence on lattice spacing or sea quark mass within 
the ~1\% statistical and systematic errors from the 
lattice calculation. A much larger error is that
from missing higher order powers of $\alpha_s$ in 
equation~\ref{eq:massrat}. 
We take account of this error by allowing a correlated
error between the points of $1\times \alpha_V(1.8/a)^2$ 
along with a possible variation with $am_b$ of the form 
$\alpha_V(1.8/a)^2 \times \delta x_m/4 $ (see equation~\ref{eq:fitxa} 
for a definition of $\delta x_m$). This allows 
the $\alpha_s^2$ term to have both a coefficient and a mass 
dependence which is three times that of the known 
$\alpha_s$ term. 
We allow for possible dependence on sea quark masses and 
the lattice spacing by using a fit of 
the same form as that in equation~\ref{eq:fitxa}. 
The final fit result is then:
\begin{equation}
\frac{m_b^{\overline{MS}}(\mu)}{m_s^{\overline{MS}}(\mu)} = 54.7(2.5),
\end{equation}
plotted as the shaded blue band in Figure~\ref{fig:mbms}. 
The error is dominated, not surprisingly, by the error from 
the unknown $\alpha_s^2$ term.

We can compare this new result to a combination of our earlier 
results for $m_b/m_c$ (4.51(4)) from~\cite{McNeile:2010ji} and 
$m_c/m_s$ (11.85(16)) from~\cite{Davies:2009ih}. These results were 
obtained entirely nonperturbatively by using 
the HISQ action for all the quarks. Then 
the ratio of lattice bare quark 
masses in the continuum limit is the
ratio of $\overline{MS}$ masses at a given scale - 
the renormalisation factor cancels completely. From the numbers above 
we have $m_b/m_s$ = 53.4(9) which is plotted 
as the black point at $a=0$ on Figure~\ref{fig:mbms}. 
Our new, completely independent, result agrees well 
with this earlier value although it is much less accurate.

%
\section{Conclusions}
\label{sec:conclude}
%

In this paper we have determined the $\Upsilon$ spectrum 
using the NRQCD formalism for the $b$ quarks in lattice 
QCD. We include several improvements over our earlier work.  
The key improvements are:
\begin{itemize}
\item  we use gluon field configurations with 
a fully $\mathcal{O}(\alpha_s a^2)$ improved gluon action and 
HISQ quarks in the sea, provided by the MILC collaboration; 
\item $c$ quarks are now included in the sea; 
\item we take the NRQCD action to a new level of 
accuracy by including radiative corrections 
to the terms at next-to-leading relativistic order ($v^4$); 
\item we improve the method for tuning the $b$ quark mass so 
that systematic errors are reduced to 0.5\%. 
\end{itemize}

With significantly improved systematic 
errors from NRQCD we are then able to determine the lattice spacing 
to better than 1\% from the $2S-1S$ splitting. 
Using this we obtain $M(h_b)-M(1\overline{S})$ to 1.4\% 
and $M(\Upsilon^{\prime\prime})-M(\Upsilon)$ to 2.4\% which 
is a strong test of NRQCD. 
This gives $M(h_b)$ = 9905(7) MeV to be compared with the 
experimental result of 9898.3(1.5) MeV~\cite{bellehb} and 
$M(\Upsilon^{\prime\prime})$ = 10375(22) MeV to be compared 
to the experimental result of 10355.2(5) MeV~\cite{pdg}. 

We have examined the 
$\Upsilon$ and $\eta_b$ dispersion relations in 
much more detail than before, so that we can quantify the effect of 
the radiative corrections to the $v^4$ kinetic 
terms in the action. We are also able to show 
how small are the deviations from continuum rotational invariance.
This enables us to tune the $b$ quark mass 
to 0.5\%. 

Our result for the hyperfine splitting between the 
$\Upsilon$ and $\eta_b$ states is much more accurate 
than in our earlier work because we have included 
the critical renormalisation of $c_4$ (the coefficient 
of the ${\bf \sigma}\cdot{\bf B}$ term ) in our analysis. 
We obtain $M(\Upsilon)-M(\eta_b)$ = 70(9) MeV now with a 
13\% error.  This gives $M(\eta_b)$ of 9390(9) MeV to 
be compared to the experimental result of 9390.9(2.8) MeV.  

Our result for $M(\Upsilon^{\prime})-M(\eta_b^{\prime})$ 
is also much more accurate largely because of a huge improvement 
in the statistical error. We find a $2S$ hyperfine splitting
that is half as big as the $1S$ hyperfine splitting at 35(3) MeV, 
predicting $M(\eta_b^{\prime})$ = 9988(3) MeV.  

\begin{figure}
\includegraphics[width=0.98\hsize]{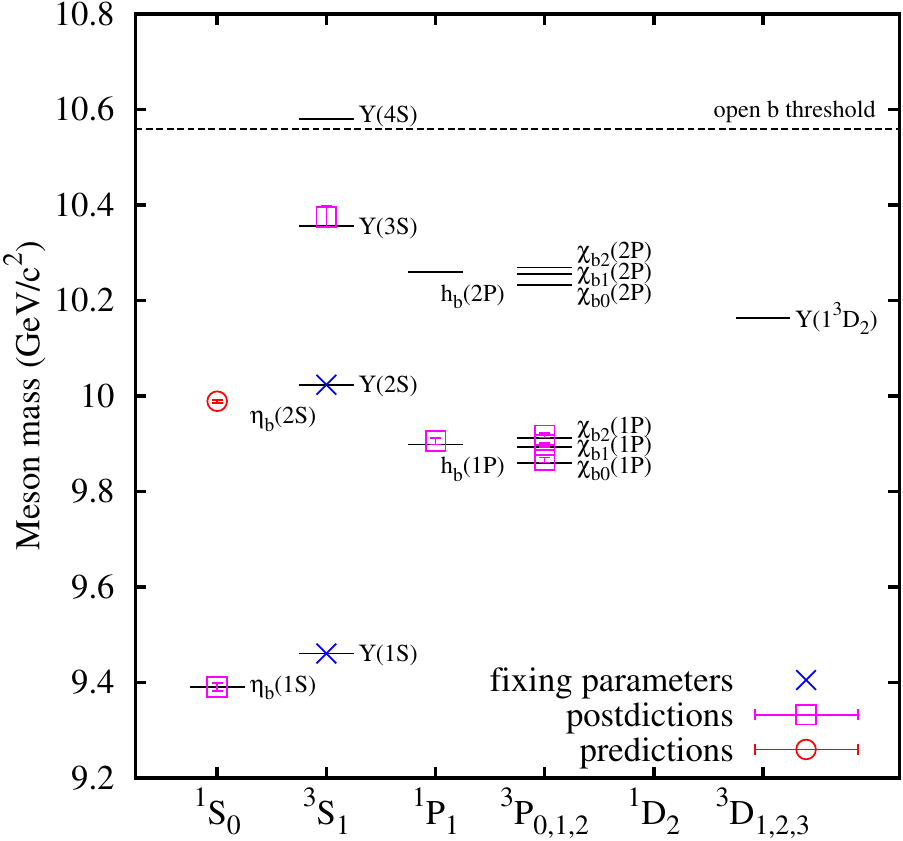}
\caption[xyz]{ The spectrum of bottomonium states 
from lattice NRQCD (colored symbols with error 
bars) compared to experiment (black lines). 
Blue crosses denote results used to tune 
parameters, 
pink open squares results to be 
compared to experiment and red circles predictions 
ahead of experiment~\footnote{ For 
simplicity we mark the $\Upsilon$ with a blue 
cross although in fact we use the spin-average of 
$\Upsilon$ and $\eta_b$ to tune the $b$ quark mass.} 
 }
\label{fig:spectrum}
\end{figure}

These new results are collected together in 
a plot of the $\Upsilon$ spectrum from improved 
lattice NRQCD in Figure~\ref{fig:spectrum}. We mark with different 
symbols those results used to tune parameters, those 
which correspond to masses already known from experiment, 
and those (the $\eta_b^{\prime}$) which are predictions.  
We include the $P$-wave fine structure from our 
results for $c_4=1.15$ on the fine lattices, set 5, 
since this $c_4$ is close to the perturbative value
on those lattices. We include an additional 10\% error 
for missing $v^6$ terms in our NRQCD action. 
$D$-wave $\Upsilon$ masses from our calculation will be 
reported elsewhere. 

Light meson ($\pi$, $K$ and $\eta_s$) 
masses and decay constants are also given here 
that enable us to determine the properties of the $\eta_s$ 
meson and give a complementary determination of the lattice 
spacing to better than 1\%. The calculation shows significantly 
improved discretisation errors over our earlier results on ensembles 
including 2+1 flavors of asqtad quarks~\cite{Davies:2009tsa}. 
The results on the properties of the $\eta_s$ are in 
agreement with our earlier work. However, our earlier result
was not able to distinguish the mass and decay constant 
of the $\eta_s$ from that that would be obtained in 
leading order chiral perturbation theory. We now obtain 
$M(\eta_s)$ = 0.6893(12) GeV and $f_{\eta_s}$=0.1819(5) GeV. 
In both cases these values disagree significantly, 
but by less than 1\%, 
from the leading order expectation. 
The $\eta_s$ particle is relatively insensitive 
to sea $u/d$ quark masses and so it is very useful 
to have accurate results for its properties for 
tuning the $s$ quark mass and determining the
lattice spacing on other lattice ensembles. 

The $\Upsilon$ $2S-1S$ and $\eta_s$ 
determinations of the lattice spacing can be compared through a third 
parameter, $r_1$, from the heavy quark potential. 
We show that both determinations agree in the continuum 
and chiral limits and give a physical value for $r_1$ of 
0.3209(26) fm. This can also be used to determine 
the lattice spacing on other lattice ensembles.   

We also combine $\Upsilon$ and $\eta_s$ calculations 
through a determination of the ratio of $\overline{MS}$ $b$ quark 
to $s$ quark masses of $m_b/m_s$ = 54.7(2.5), in 
agreement with our earlier result from HISQ quarks alone 
of 53.4(9)~\cite{Davies:2009ih, McNeile:2010ji}. 

Finally we comment on the effect of including 
$c$ quarks in the sea. We have seen no significant 
effect on any of the observables that we have 
calculated compared to results obtained with 
2+1 flavors of sea quarks. The results cannot 
be compared lattice spacing by lattice spacing 
because of changes to the lattice QCD action that 
reduce the size of discretisation errors 
in our new results. 
Final physical results
can be compared, however, with and without sea $c$ quarks 
to see if there is a difference. 
In our earlier 2+1 flavor calculations~\cite{Gregory:2010gm} 
we estimated that the presence of sea $c$ quarks would 
shift the $\Upsilon$ and $\eta_b$ masses downwards by 5 MeV
through an induced additional local potential which 
was proportional to $\alpha_s^2\delta^3(r)/m_c^2$. 
This would have a smaller effect on radial excitations 
of the $\Upsilon$ than on the ground state masses and very little effect 
on $P$-wave states. We then estimate the effect 
on, for example the $\Upsilon$ $1P-1S$ splitting to be $\mathcal{O}(1\%)$.  
This would barely be visible above the errors in our 
current calculation and the errors in the earlier calculation 
were somewhat larger, so any comparison certainly has 
an error of greater than 1\%. However, it is clear from 
our results that no unexpectedly large effect has 
appeared. For light hadrons we expect even smaller effects and 
there we can limit any differences in $M_{\eta_s}$ 
and $f_{\eta_s}$ to smaller than 1\%, with the main 
error coming from our earlier calculation~\cite{Davies:2009tsa}.  

We are now combining $b$ quark propagators from our 
improved NRQCD action with $l$, $s$ and $c$ propagators 
on these ensembles
to study $B$, $B_s$ and $B_c$ meson masses and matrix elements. 
Significantly improved systematic errors should be possible 
both from the NRQCD action and because we are working much 
closer to physical light sea quark masses than before with an improved 
gluon and sea quark action. 

%
\section*{Acknowledgements}
%

We are grateful to the MILC collaboration for the use of their 
gauge configurations and particularly to Doug Toussaint for help 
in reading them and in providing values for $r_1/a$. 
The results described here were obtained using the Darwin Supercomputer 
of the University of Cambridge High Performance 
Computing Service as part of the DiRAC facility jointly
funded by STFC, the Large Facilities Capital Fund of BIS 
and the Universities of Cambridge and Glasgow. This work was funded 
by STFC, MICINN, DGIID-DGA and NSF with support from the Scottish Universities 
Physics Alliance and the EU under ITN-STRONGnet. 

\appendix

%
\section{Gauge Action}
\label{appendix:gauge_action}
%

For clarity, the gauge action $S_G$ used in the generation of the MILC ensembles will be summarised in this section. See~\cite{Bazavov:2010ru}. 
The action is a tadpole and one-loop improved L\"{u}scher-Weisz action, 
\bea
S_G &=&
\beta \left[ 
 c_P \sum_P 
 \left( 1 -\frac{1}{3} {\rm Re }\Tr(P)     \right)
\right.
\nn \\ 
&&+
  c_R \sum_R 
  \left( 1 -\frac{1}{3} {\rm Re }\Tr(R)     \right)
\nn \\
&&+ 
\left.
  c_T \sum_T 
  \left( 1 -\frac{1}{3} {\rm Re }\Tr(T)     \right)
\right]
\eea
where the sums are over plaquettes $P$, rectangles $R$ and twisted loops (or parallelograms) $T$.
The coefficients are calculated perturbatively through $\mathcal{O}(\alpha_s)$ 
including both gluonic loops \cite{Alford:1995hw} and contributions from 
HISQ sea quarks \cite{Hart:2008sq}. The tadpole improvement parameter was chosen to be the fourth root of the plaquette 
$u_{0P}=( \frac{1}{3}   {\rm Re \ Tr}  \langle P  \rangle)^{1/4}$ and, via a perturbative calculation of the plaquette, gives an 
expression for the strong coupling constant 
$\alpha_s = - 1.303615 \log u_{0P}$. $u_{0P}$ also 
appears in the gauge coupling as  $\beta = 10/(g^2 u_{0P}^4)$. The coefficients used are
\bea
C_P &=&
1.0
\nn \\
C_R &=&
\frac{-1}{20 u_{0P}^2}(1-(0.6264 - 1.1746 N_f)\log(u_{0P}^2))
\nn \\
C_T &=&
\frac{1}{u_{0P}^2}(0.0433 - 0.0156 N_f) \log(u_{0P}^2)
\eea
The inclusion of these terms mean that the gauge action is improved completely through order ${\cal O} (\alpha_s a^2)$. As mentioned in the text, sea quarks are included using the HISQ action~\cite{Follana:2006rc} with 
a $U(3)$ projection (only) for the intermediate re-unitarization step.

%
\section{Perturbative determination of radiative corrections to $c_i$ coefficients in the NRQCD action and the mass renormalization}
\label{appendix:cicalc}
%

{\it Spin-independent coefficients}. The $c_i$ coefficients appearing in the NRQCD action, equation~\ref{eq:deltaH}, 
have expansion $1+c^{(1)}_i\alpha_s + \mathcal{O}(\alpha_s^2)$. 
The $c^{(1)}_i$ for the kinetic terms, $i=1,5,6$, are determined 
following the method of~\cite{Morningstar:1993de, Morningstar:1994qe}. 
The NRQCD quark self-energy is calculated through $\mathcal{O}(\alpha_s)$ 
and the $c^{(1)}_i$ are given by the requirement that the correct 
energy-momentum relationship be obtained through $\mathcal{O}(\alpha_sv^4)$. 
The terms proportional to $(\Delta^{(2)})^2$ in equation~\ref{eq:deltaH} 
can be merged together so that this term in $\delta H$ appears as: 
\begin{equation}
\tilde{c}_1(1+\frac{am_b}{2n})\frac{(\Delta^{(2)})^2}{8(am_b)^3}.
\label{eq:c1tilde}
\end{equation}
Thus only two radiative corrections need to be calculated for the complete 
set of kinetic terms at $\mathcal{O}(v^4)$, i.e. for $\tilde{c}_1$ and $c_5$. 
The radiative correction for $\tilde{c}_1$ then applies equally to $c_1$ and 
$c_6$.

The full inverse NRQCD quark propagator at $\mathcal{O}(\alpha_s)$ is 
\begin{equation}
aG^{-1}(p) = Q^{-1}(p) - \alpha_s a\Sigma(p)
\label{eq:ginv}
\end{equation}
where $ap=(a{\bf p},ap_4)$ is a 4-vector in lattice Euclidean space. 
The pole in the propagator is identified 
as $a\omega({\bf p}) = ip_4a$. The expansion of $\omega({\bf p})$ in 
powers of the spatial momentum can be used to identify the quark mass 
renormalisation factor $Z_m$ and wavefunction renormalisation factor 
$Z_2$ but also to tune $\tilde{c}_1^{(1)}$ and 
$c_5^{(1)}$ to appropriate values. 
$Q^{-1}(p)$ is the quark propagator obtained at tree level from 
the NRQCD action, including the (as yet unknown) radiative corrections 
to $\tilde{c_1}$ and $c_5$. Its pole is then given by:  
\begin{eqnarray}
a\omega_0(\bf{p}) &=& \frac{a^2{\bf p}^2}{2am_b} - \frac{(a^2{\bf p}^2)^2}{8(am_b)^3} 
+ \alpha_s \left\{ c_5^{(1)} \frac{a^4{\bf p}^4}{24am_b} \right.\nonumber \\
&-& \left. \tilde{c_1}^{(1)}\left( \frac{1}{2n} + \frac{1}{am_b}\right) \frac{(a^2{\bf p}^2)^2}{8(am_b)^3}\right\} 
\label{eq:omega0}
\end{eqnarray}

\begin{figure}[t]
\includegraphics[width=5cm]{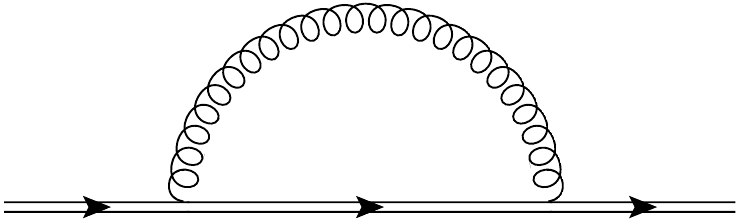}
\vspace{5mm}

\includegraphics[width=5cm]{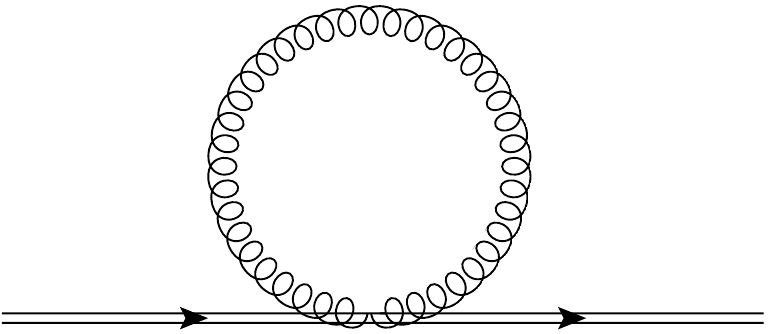}
\vspace{10mm}

\includegraphics[width=5cm]{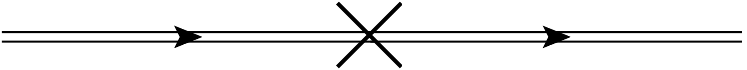}
\caption{The Feynman diagrams needed for the calculation of 
$\mathcal{O}(\alpha_s)$ corrections to the heavy quark 
self energy. From top to bottom, rainbow, tadpole and 
$u_0$ counterterm diagrams. }
\label{fig:feyndiags}
\end{figure}

$\Sigma(p)$ is the one-loop self-energy and consists, as shown 
in Figure~\ref{fig:feyndiags}, of rainbow and tadpole 
diagrams as well as diagrams containing insertions of the one-loop piece 
of the tadpole-improvement factor, $u_0$. Writing $u_0 = 1 + \alpha_s u_0^{(1)}$, 
we have $u_0^{(1)}=0.750$ for the Landau link tadpole 
parameter, $u_{0L}$~\cite{Nobes:2001tf}. 
We have 
\begin{equation}
\omega({\bf p}) = \omega_0({\bf p}) - \alpha_s \Sigma(\omega_0({\bf p}),{\bf p})
\label{eq:omega}
\end{equation}
and can expand $\Sigma$ to $v^4$ as: 
\begin{eqnarray}
a\Sigma(p) &=& \Sigma_0(\omega) + \Sigma_1(\omega)\frac{a^2{\bf p}^2}{2am_b}  \\
&+& \Sigma_2(\omega)\frac{(a^2{\bf p}^2)^2}{8(am_b)^3} + \Sigma_3(\omega)a^4{\bf p}^4. \nonumber
\label{eq:sigma}
\end{eqnarray}
The $\Sigma_i$ are extracted from suitable combinations of partial derivatives of 
$\Sigma$: 
\begin{eqnarray}
\Sigma_0 &=& a\Sigma({\bf p}=0) \\
\Sigma_1 &=& \left. am_b\frac{\partial^2 a\Sigma}{\partial a^2p_z^2}\right|_{{\bf p}=0} \nonumber \\
\Sigma_2 &=& \left. (am_b)^3\frac{\partial^4 a\Sigma}{\partial a^2p_z^2 \partial a^2p_y^2}\right|_{{\bf p}=0} \nonumber \\
\Sigma_3 &=& \frac{1}{24}\left( \frac{\partial^4 a\Sigma}{\partial a^4p_z^4} -3\frac{\partial^4 \Sigma}{\partial a^2p_y^2\partial a^2p_z^2}\right)_{\bf{p}=0}. \nonumber 
\end{eqnarray}
Each of the $\Sigma_i$ also has an expansion in powers of $\omega$ as 
$\Sigma_i = \Sigma_{l=0}^{\infty} \Sigma_i^{(l)}\omega^l$.
Then 
\begin{equation}
a\omega({\bf p}) = \frac{a^2{\bf p}^2}{2am_{b,r}} - \frac{(a^2{\bf p}^2)^2}{8(am_{b,r})^3} - \alpha_s a \delta \omega({\bf p})
\label{eq:omega2}
\end{equation}
where $m_{b,r}=Z_m m_b$ and 
\begin{equation}
Z_m = 1 + \alpha_sZ_m^{(1)} = 1 + \alpha_s (\Sigma_0^{(1)} + \Sigma_1^{(0)} ) 
\label{eq:zm}
\end{equation}
to this order. 
The correction term $\delta \omega$ is given by: 
\begin{eqnarray}
a\delta \omega &=& W_0 + \\
&+& \left( W_1 + \tilde{c}_1^{(1)}\left( \frac{1}{2n}+\frac{1}{am_b}\right)\right) \frac{(a^2{\bf p}^2)^2}{8(am_b)^3} \nonumber \\
&+& \left( W_2 - \frac{c_5^{(1)}}{24am_b} \right) a^4{\bf p}^4 \nonumber
\label{eq:omega3}
\end{eqnarray}
with 
\begin{eqnarray}
W_0 &=& \Sigma_0^{(0)} \nonumber \\
W_1 &=& \frac{2\Sigma_0^{(1)}}{am_b} + 2 \Sigma_0^{(2)} + \frac{3\Sigma_1^{(0)}}{am_b} + 2\Sigma_1^{(1)} + \Sigma_2^{(0)} \nonumber \\
W_2 &=& \Sigma_3^{(0)}.
\label{eq:wdef}
\end{eqnarray}
The requirement that lattice NRQCD reproduce the low-energy 
physics of full QCD means that $\delta \omega$ can only be 
a pure energy shift independent of spatial momentum,  i.e. the coefficients
of $({\bf p}^2)^2$ and ${\bf p}^4$ in equation~\ref{eq:omega3} must be 
zero. Thus
\begin{eqnarray}
\tilde{c}_1^{(1)} &=& - \left(\frac{1}{2n}+\frac{1}{am_b} \right)^{-1} W_1 \nonumber \\
c_5^{(1)} &=& 24am_bW_2 
\label{eq:ciw}
\end{eqnarray}  

The Feynman rules were generated automatically using the HiPPy package and the Feynman 
lattice integrals for $\Sigma$ and its derivatives were constructed and evaluated 
numerically using the HPsrc package and VEGAS contained therein~\cite{Hart:2004bd,Hart:2009nr}.
We use analytic differentiation using the TaylUR package in the HPsrc Fortran code
together with numerical differentiation which, for sufficiently smooth functions,
can be up to an order of magnitude faster than analytic differentiation.

Because the kinetic $(a^2{\bf p}^2)^2$ term is included at tree level both
$W_1$ and $W_2$ are Infra-Red (IR) finite and so no IR regulation is needed although
a gluon mass was used to regularize intermediate divergences. However,
the integrals arising from the rainbow diagram still have large peaks in
the IR region. These peaks arise because the differentiation generates extra powers
of the heavy quark NRQCD propagator in the integrand. In this case to use
numerical differentiation alone proves to be unstable and it is imperative  
to use a mixture of analytic and numerical approaches and also to introduce
a suitable subtraction function to remove the most severe behaviour of the integrand.
In contrast, the integrals arising from the tadpole diagram are well behaved because 
they contain no quark propagators but they are expensive to evaluate since the two-gluon vertex
contains a large number of terms. In this case, numerical differentiation proved to be the 
most efficient for the higher order mixed derivatives without compromising accuracy. 
In all cases the temporal derivatives were done using the analytic method.

We checked that the results agree well with those of Morningstar~\cite{Morningstar:1993de, 
Morningstar:1994qe} for his gluon and NRQCD actions. For the simplest gluon and NRQCD
actions the results agree with the analytic calculation of Monahan~\cite{Monahan:2011}.

The contribution to the $c_i^{(1)}$ from the $u_0^{(1)}$ insertions 
of Figure~\ref{fig:feyndiags} can be calculated analytically. 
This gives: 
\begin{eqnarray}
\frac{\tilde{c}_1^{(1)}}{u_0^{(1)}} &=& -\frac{1}{8}\left( 1 + \frac{am_b}{2n} \right)^{-1} \left[ \frac{12}{n^2} - \frac{1}{n} +\frac{1}{2am_b}\left(\frac{3}{n^2}-4 \right) \right.  \nonumber \\
&+& \left. \frac{6}{(am_b)^2}\left(\frac{1}{n} - 12 \right) + \frac{6}{(am_b)^3} \right] \\
\frac{c_5^{(1)}}{u_0^{(1)}} &=& -\frac{4}{3} + \frac{1}{4am_b} + \frac{3}{(am_b)^2} - \frac{3}{8n(am_b)^2} -\frac{3}{4(am_b)^3}.  \nonumber 
\end{eqnarray}
These contributions are sizeable and act to cancel contributions coming 
from the other diagrams, 
as part of the `tadpole-improvement' 
mechanism~\cite{Lepage:1992xa, Morningstar:1994qe}. 
This is particularly true for $c_5^{(1)}$; less so for 
$\tilde{c}_1^{(1)}$, as in~\cite{Morningstar:1994qe}. 
The $c_i^{(1)}$ values will then change depending on the tadpole-improvement 
parameter chosen, for example $u_{0P}$ or $u_{0L}$, because 
the $c_i$ must compensate perturbatively for changes in $u_0$. 
Here we use $u_{0L}$ in the NRQCD action and this is the only 
$u_0$ 
that affects the $c_i$ to $\mathcal{O}(\alpha_s)$. $u_{0P}$ 
is used in the gluon action and counterterms from this will 
appear in the $c_i$ at higher order. 

\begin{table}
\caption{ Coefficients $\tilde{c}_1^{(1)}$ and $c_5^{(1)}$ 
that multiply $\alpha_s$ in the one-loop correction 
to the kinetic terms in the NRQCD action used here in conjunction 
with the improved gluon action described in 
Appendix~\ref{appendix:gauge_action}. 
}
\label{tab:ci}
\begin{ruledtabular}
\begin{tabular}{llll}
$am_b$ & n & $\tilde{c}_1^{(1)}$ & $c_5^{(1)}$ \\
\hline
1.95 & 4 & 0.774(21) & 0.392(17) \\
2.8 & 4 & 0.951(26) & 0.406(11) \\
3.4 & 4 & 0.952(30) & 0.445(10) \\
\end{tabular}
\end{ruledtabular}
\end{table}

Table~\ref{tab:ci} gives the results for $\tilde{c}_1^{(1)}$ 
and $c_5^{(1)}$ for 3 different values of $am_b$ and stability 
parameter $n=4$. The mass values are 
not exactly the $am_b$ values used for the numerical work 
in this paper but the $c_i^{(1)}$ show very 
mild dependence on $am_b$, so we can simply interpolate to the 
$am_b$ values we are using. The results are 
different from those of~\cite{Morningstar:1994qe} because 
both the gluon action and the NRQCD action have changed. 
However, qualitative features are the same in that the 
values are not large and only mild dependence on $am_b$ 
is seen for $am_b$ larger than 1. 
We note that the $c_i^{(1)}$ coefficients will change if 
higher order terms are added to the NRQCD action. For example~\cite{eikephd} 
tests were done with an NRQCD action which included 
a term in $\delta H$ of $-\Delta^{(6)}/(180am_b)$, removing 
$\mathcal{O}(a^6)$ discretisation errors from $H_0$. This 
changes $c_5^{(1)}$ to 0.017(4) for $am_b$ = 1.95 and
$n=4$ to be compared with the result of 0.392(17) in Table~\ref{tab:ci}. 

\begin{table}
\caption{ Values for $\alpha_V$ used in calculating 
the $\mathcal{O}(\alpha_s)$-corrected coefficients 
$c_1$, $c_5$, $c_6$ and $c_4$.  
}
\label{tab:alpha}
\begin{ruledtabular}
\begin{tabular}{lllll}
Sets & $1/a$  & $\alpha_V^{(4)}(1.4/a)$ & $\alpha_V^{(4)}(1.8/a)$ & $\alpha_V^{(4)}(\pi/a)$ \\
 & GeV & & & \\
\hline
fine & 2.2 & 0.32 &  0.28 & 0.225\\
coarse & 1.6 & 0.39 & 0.33 & 0.255\\
very coarse & 1.3 & 0.46 & 0.38 & 0.275\\
\end{tabular}
\end{ruledtabular}
\end{table}

The coefficients in Table~\ref{tab:ci} need to be 
combined with a value for $\alpha_s$ to give final results 
for $c_1$, $c_5$ and $c_6$. The scale, $q^*$, used for $\alpha_s$ 
was taken as that calculated for the Brodsky-Lepage-Mackenzie 
scheme in Figure 10 of~\cite{Morningstar:1994qe}, assuming that this 
does not change significantly with the changes in the 
action used. This gives 
$q^*\approx 1.4/a$ for $c_5$ and $q^*\approx 1.8/a$ for $\tilde{c_1}$. 
We take $\alpha_s$ from~\cite{McNeile:2010ji}, specifically the value 
$\alpha_{\overline{MS}}(M_z,n_f=5) = 0.1183$. We convert 
this to $\alpha_V$~\cite{Davies:2008sw} and run perturbatively to 
values using $n_f=4$ and appropriate 
scales $q^*$. The $q^*$ values are 
calculated for the very coarse, coarse and fine ensembles using 
$a^{-1} \approx$ 1.3, 1.6 and 2.2 GeV respectively from Table~\ref{tab:aval}.  
The $\alpha_V$ values obtained are listed in Table~\ref{tab:alpha}. 
These are combined with the coefficients in Table~\ref{tab:ci} 
to give the values used in Table~\ref{tab:wilsonparams}. 
The coefficient $\tilde{c}_1^{(1)}$ was reduced slightly on 
the fine lattices (to 0.766) to account for the fact that 
the value of $am_b$ used was slightly smaller than that 
for which the coefficient was calculated. 

The remaining errors in the kinetic $c_i$ coefficients 
after this one-loop correction has been made 
will be $\mathcal{O}(\alpha_s^2)$. From Table~\ref{tab:alpha}
we can see that $0.5\alpha_V^2$ ranges from 0.1 
on the very coarse ensembles to 0.05 on fine. 
The impact of these errors can then be estimated from 
the size of the effect that we see on physical observables from 
the $\mathcal{O}(\alpha_s)$ corrections. 

We have not corrected the coefficient of the Darwin 
term, $c_2$, in our NRQCD Hamiltonian. We have however 
assessed the effect of taking $c_2$ to be as large 
as 1.25 on the meson kinetic mass and on the hyperfine 
splitting and we find the effects to be small. 

\begin{table}
\caption{ Coefficients $Z_m^{(1)}$ and $c_4^{(1)}$ 
that multiply $\alpha_s$ in the one-loop correction 
to the mass renormalization and the 
${\bf \sigma}\cdot{\bf B}$ term 
in the NRQCD action 
respectively. These are calculated with the 
NRQCD action used here and 
the improved gluon action described in 
Appendix~\ref{appendix:gauge_action}. 
$A^{\mathrm{NRQCD}}$ in column 4 
is $Z_m^{(1)} + 2\ln(am_b)/\pi$, as described in 
the text.
}
\label{tab:c4zm}
\begin{ruledtabular}
\begin{tabular}{lllll}
$am_b$ & n & $Z_m^{(1)}$ & $A^{\mathrm{NRQCD}}$ & $c_4^{(1)}$ \\
\hline
1.9 & 4 & 0.439(3) & 0.848(3) &  0.691(7) \\
2.65 & 4 & 0.263(5) & 0.883(5) & 0.775(8)  \\
3.4 & 4 & 0.150(3) & 0.929(3) &  0.818(4)\\
\end{tabular}
\end{ruledtabular}
\end{table}

\begin{figure}[t]
\includegraphics[width=7.5cm]{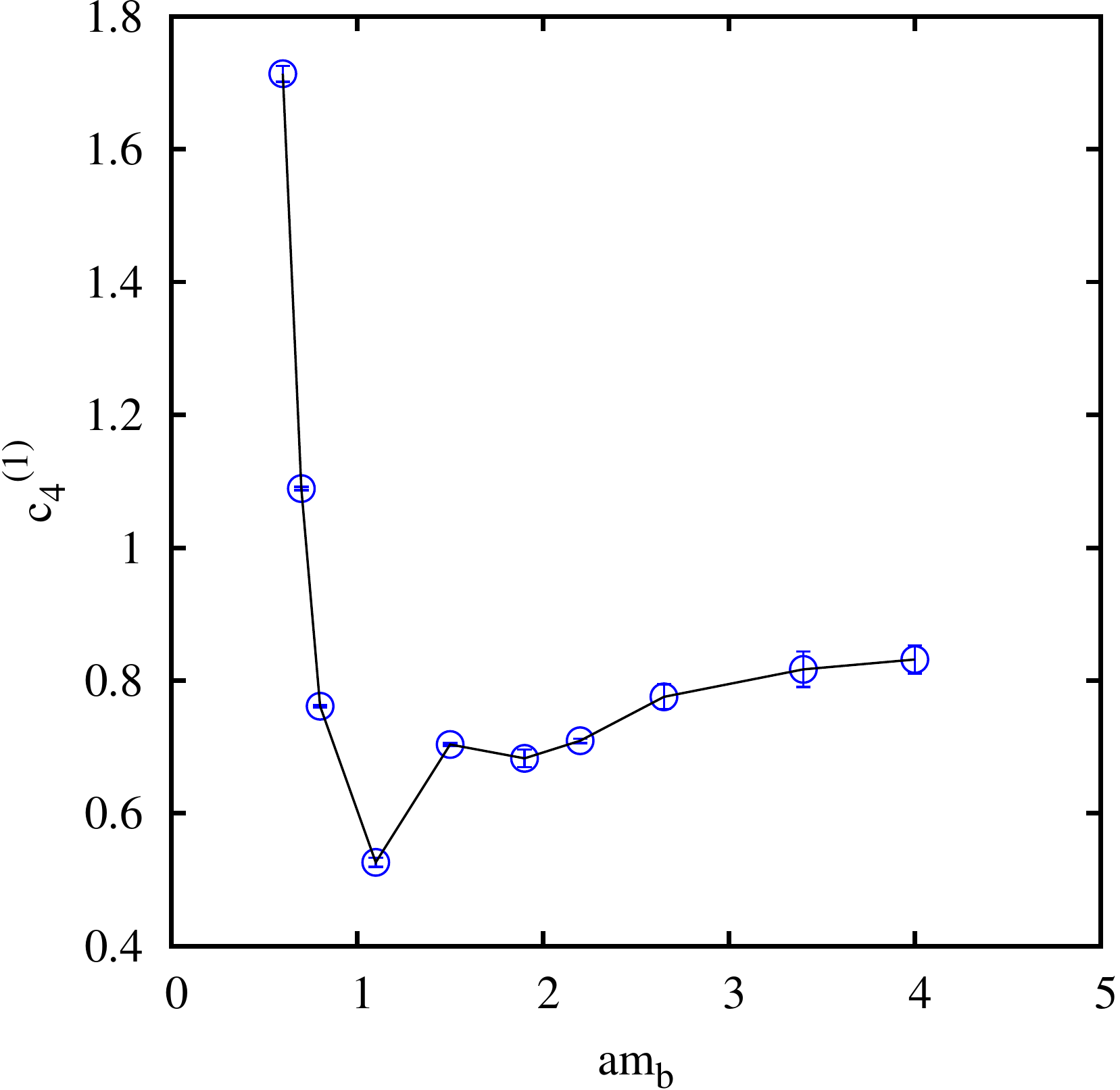}
\caption{The $\mathcal{O}(\alpha_s)$ coefficient in 
the perturbative expansion of $c_4$, coefficient 
of the ${\bf \sigma}\cdot{\bf B}$ term, plotted 
against the bare $b$ quark mass.  
}
\label{fig:c4n4}
\end{figure}

{\it Mass renormalization}. 
In Table~\ref{tab:c4zm} we give values 
for $Z_m^{(1)}$, the coefficient of $\alpha_s$ in 
the mass renormalization of equation~\ref{eq:zm}.  
$Z_m$ was calculated previously for different parameter values 
in~\cite{Dalgic:2003uf}. We also show the 
result of adding $2\ln(am_b)/\pi$ to $Z_m^{(1)}$ 
as $A^{\mathrm{NRQCD}}$. 
$A^{\mathrm{NRQCD}}$ 
will be used in section~\ref{sec:mbms} to derive 
the ratio $m_b/m_s$ in the $\overline{MS}$ scheme. 
Only mild dependence on $am_b$ is seen in $A^{\mathrm{NRQCD}}$. 

{\it Spin-dependent terms}. The perturbative 
renormalisation of field-dependent terms 
has to be done in a different way and this includes 
all the spin-dependent terms. Recently the radiative 
corrections to $c_4$ have become available~\cite{Hammant:2011bt}. They 
were calculated by matching the effective action in NRQCD 
to continuum QCD using the background field 
method.  We give 
values for $c_4^{(1)}$ in Table~\ref{tab:c4zm} appropriate 
to the $am_b$ and $n$ values we are using here. 

A number of pieces go in to the calculation of $c_4^{(1)}$. 
These include renormalisation of the chromomagnetic moment, 
renormalisation of the wavefunction and, because 
$c_4$ multiplies a term in the bare lattice 
NRQCD Hamiltonian which 
includes the bare lattice quark mass, the mass 
renormalisation. $c_4^{(1)}$ is the sum of two pieces,  
one of which has polynomial dependence on the 
bare quark mass $am_b$ and the other is proportional 
to $\log(am_b)$. The logarithmic term has 
coefficient $-3/(2\pi)$~\cite{Hammant:2011bt}. 
Both terms are included in the total result 
given in Table~\ref{tab:c4zm} - the logarithmic 
term is of similar size to the polynomial term 
over the range of $am_b$ that we are using. 
The $c_4^{(1)}$ values are combined 
with values for $\alpha_V^{(4)}(\pi/a)$ (given 
in Table~\ref{tab:alpha}) to give 
the results for the 1-loop corrected $c_4$ coefficients given 
in Table~\ref{tab:c4vals}.   

Figure~\ref{fig:c4n4} gives a more 
complete picture of $c_4^{(1)}$ 
by plotting 
values as a function of $am_b$.
We see relatively little $am_b$ dependence until 
$am_b$ becomes smaller than 1 when 
$c_4^{(1)}$ starts to diverge. This is 
typical of the behaviour of radiative corrections 
to coefficients in the NRQCD action. 

\begin{table}
\caption{ Coefficients $d_1$ and $d_2$ 
of spin-dependent 4-quark operators that 
give rise to a correction to the hyperfine 
splitting. 
These are calculated with the 
NRQCD action used here with the parameters 
given in columns 1 and 2 and 
the improved gluon action described in 
Appendix~\ref{appendix:gauge_action}. 
}
\label{tab:d1d2}
\begin{ruledtabular}
\begin{tabular}{llll}
$am_b$ & n & $d_1$ & $d_2$ \\
\hline
1.9 & 4 & $-\ln(1.9)$ + 0.796(4) & $\ln(1.9)/3$ -0.311(1)  \\
2.65 & 4 & $-\ln(2.65)$ + 0.448(6) & $\ln(2.65)/3$ -0.195(2)  \\
3.4 & 4 & $-\ln(3.4)$ + 0.038(8) & $\ln(3.4)/3$ -0.058(2) \\
\end{tabular}
\end{ruledtabular}
\end{table}

Finally we give results for the coefficients of 
the spin-dependent 4-quark operators that contribute 
to the hyperfine splitting. These terms have coefficients 
$d_1$ and $d_2$ that multiply terms that would appear 
in the NRQCD action at order $\mathcal{O}(\alpha_s^2v^3)$: 
\begin{eqnarray}
S_{4q}&=&
d_1\frac{\alpha_s^2}{(am_b)^2}(\psi^{\dag}\chi^*)(\chi^{T}\psi) \nn \\
&&+ d_2\frac{\alpha_s^2}{(am_b)^2}(\psi^{\dag}{\bf \sigma}\chi^*)\cdot(\chi^{T}{\bf \sigma}\psi).
\label{eq:4qops}
\end{eqnarray}
These terms are subleading compared to treelevel $v^4$ 
operators but contribute to the hyperfine splitting at 
the same order as $\alpha_s$ corrections to 
$c_4$~\cite{Hammant:2011bt}. We do not include these 
4-quark terms in our NRQCD action but we can estimate 
their effect because they give rise a shift 
in the relative energies of the $\Upsilon$ and $\eta_b$ 
which is proportional to the `wavefunction-at-the-origin', 
and given by:
\begin{equation}
\Delta E_{\mathrm hyp} = \frac{6\alpha_s^2(d_2-d_1)}{m_b^2}|\psi(0)|^2.
\label{eq:hyp4q}
\end{equation}

The relevant coefficients, $d_1$ and $d_2$ are given 
in Table~\ref{tab:d1d2}. They were calculated previously 
for a slightly different NRQCD action in~\cite{Hammant:2011bt}. 
The coefficients are related by: 
\begin{equation}
d_1 = -3d_2 -\frac{4}{9}(1-\ln(2))
\label{eq:d1d2}
\end{equation}
where the term proportional to $(\ln 2 - 1)$ is from 
$\eta_b$ annihilation to 2 gluons. This term increases 
the hyperfine splitting from equation~\ref{eq:hyp4q} 
since it pushes the $\eta_b$ mass down.  
When equation~\ref{eq:hyp4q} is applied for just this 
piece  we obtain an estimate of its size of 
about 1 MeV. This is smaller, but in agreement with, 
the earlier estimate of 2.4(2.4) MeV applied in 
section~\ref{subsec:tune} in deriving an 
appropriate spin-averaged $1S$ meson mass to tune 
the $b$ quark mass against. For that purpose the 
shift is completely negligible, representing a 
tiny fraction of the $\eta_b$ mass. For the hyperfine 
splitting it is a more important issue. For this we take 
the results from equation~\ref{eq:hyp4q} because  
that provides a consistent treatment of all the 4-quark 
operator effects. 

$d_1$ and $d_2$ are separated into logarithmic and 
nonlogarithmic pieces in Table~\ref{tab:d1d2}. 
The nonlogarithmic piece has significant mass dependence
here, becoming small at large $ma$. The logarithmic and 
nonlogarithmic terms in fact cancel for $ma$ around 1.9. 
In assessing the error in the hyperfine splitting 
from missing higher order terms multiplying  
the 4-quark operators we are careful not to assume 
that this is generic behaviour. 

We combine the $d_1$ and $d_2$ coefficients 
with $\alpha_V(\pi/a)$ values 
from Table~\ref{tab:alpha} and values for $|\psi(0)|^2$ from 
our fits to obtain corrections to the hyperfine splitting 
that are applied in section~\ref{subsub:hyp}. 

%
\section{Nonperturbative determination of radiative corrections 
to $c_3$ and $c_4$ coefficients in the NRQCD action}
\label{appendix:cinonpert}
%

\begin{figure}[t]
\includegraphics[width=0.9\hsize]{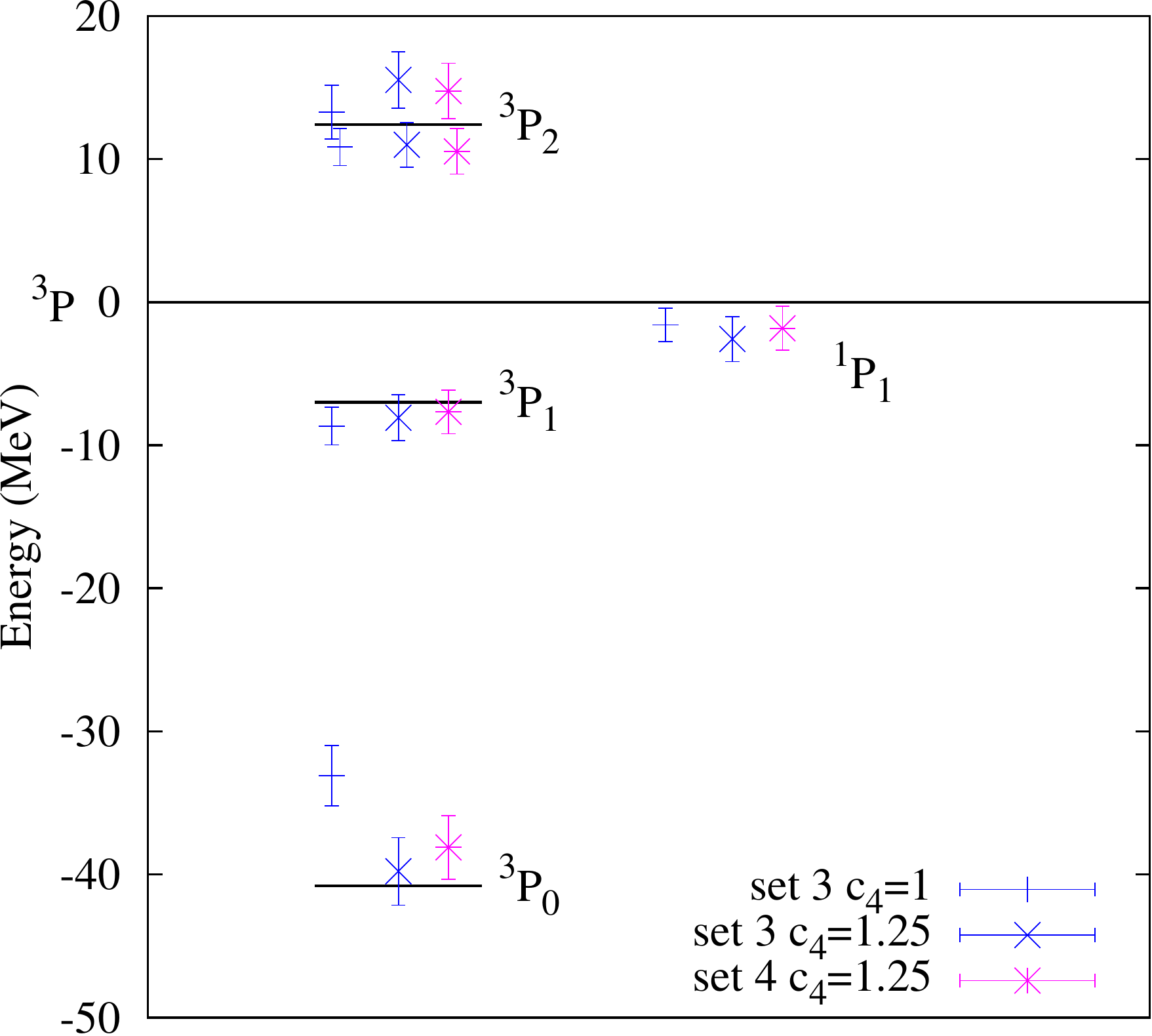}
\includegraphics[width=0.9\hsize]{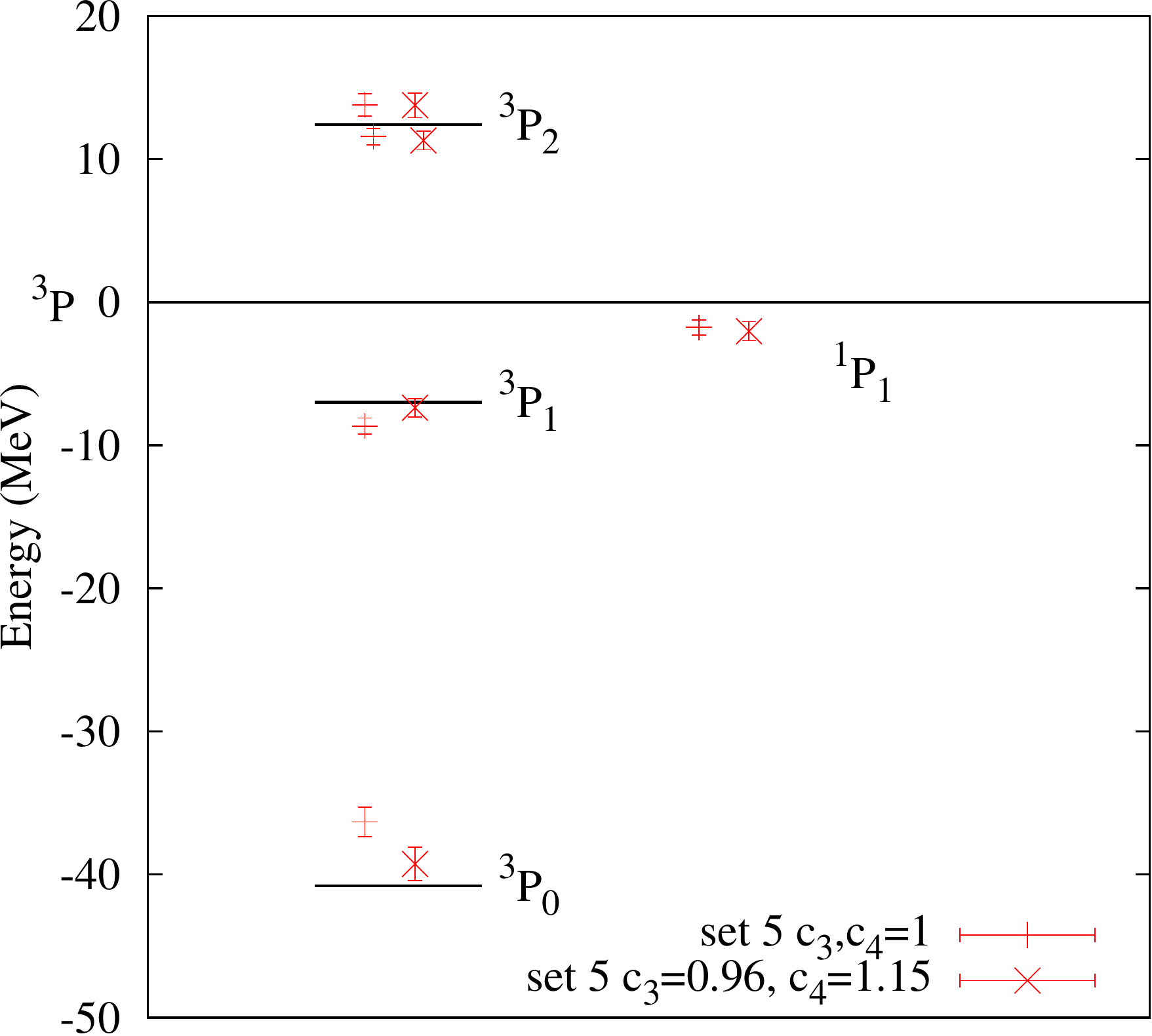}
\caption{The masses of the lowest-lying $P$-wave states in the $\Upsilon$
spectrum plotted relative to the spin-average of the $^3P$ states for the coarse 
lattices, sets 3 and 4 (top plot) and the fine lattices, set 5 (lower plot). 
In each plot we compare mass splittings for $c_4=1$ with a nonperturbatively 
tuned value for $c_4$ chosen to match a combination of mass splittings 
to experiment (see text). $c_3=1$ in all cases. For the ${}^3P_2$ states 
the ${}^3P_{E}$ is plotted to the left of the ${}^3P_{T_2}$.
}
\label{fig:pwaves}
\end{figure}

\begin{table*}[t]
\centerline{
\begin{tabular}{llllll}
\hline
    & Set 3 	& Set 3 	     & Set 4 	  & Set 5 		& Set 5 		\\
    & $am_b=2.66$   & $am_b=2.66$ & $am_b=2.62$ & $am_b=1.91$ 	& $am_b=1.91$	\\
    & $c_4=1.0$   & $c_4=1.25$ & $c_4=1.25$ & $c_4=1.0$ 	& $c_4=1.15$	\\
    & $c_3=1.0$   & $c_3=1.0$ & $c_3=1.0$ & $c_3=1.0$ 	& $c_3=0.96$	\\
\hline \hline
$aE(1^1P_1 )$  	   & 0.5655(23) & 0.5247(22) & 0.5253(20) & 0.4833(10) & 0.4478(11) \\
$aE(1^3P_0 )$  	   & 0.5460(20) & 0.5017(20) & 0.5034(18) & 0.4678(9)  & 0.4312(9)  \\
$aE(1^3P_1 )$      & 0.5611(24) & 0.5213(26) & 0.5218(24) & 0.4802(10) & 0.4454(11) \\
$aE(1^3P_{E} )$    & 0.5747(30) & 0.5359(28) & 0.5354(25) & 0.4903(11) & 0.4549(12) \\
$aE(1^3P_{T_2} )$  & 0.5732(28) & 0.5331(28) & 0.5328(27) & 0.4893(11) & 0.4538(12) \\
\hline
$aE(1^1P_1 )-aE(^3\overline{P}) $ 	& -0.0010(7)  & -0.0016(10) & -0.0011(9) & -0.0008(2) & -0.0009(2) \\
$aE(1^3P_0 )-aE(^3\overline{P}) $ 	& -0.0204(13) & -0.0246(14) & -0.0231(13)& -0.0163(4) & -0.0176(5) \\
$aE(1^3P_1 )-aE(^3\overline{P}) $ 	& -0.0053(8)  & -0.0050(10) & -0.0046(9) & -0.0039(2) & -0.0033(3) \\
$aE(1^3P_{E}   )-aE(^3\overline{P}) $ 	& 0.0082(12)  &  0.0096(12) & 0.0089(11) & 0.0062(3)  & 0.0062(4)  \\
$aE(1^3P_{T_2} )-aE(^3\overline{P}) $ 	& 0.0067(8)   &  0.0068(10) & 0.0064(10) & 0.0052(3)  & 0.0051(3)  \\
\hline
$2aE(1^3P_2E ) + 3aE(1^3P_2T_2)$ 
& 0.093(8)  & 0.104(9)  & 0.097(8)  & 0.072(3)   & 0.073(3)   \\
$- 3aE(1^3P_1 ) -2aE(1^3P_0 )$ 
& & & & & \\
$0.4aE(1^3P_2{E} ) + 0.6aE(1^3P_2{T_2})$  
& -0.018(2) & -0.026(4) & -0.025(3) & -0.0153(7) & -0.0198(9) \\
$- 3aE(1^3P_1 ) +2aE(1^3P_0 )$  
& & & & & \\
\hline
\hline
\end{tabular}
}
\caption{Fitted energies for $P$-wave states on sets 3, 4 and 5 
for the NRQCD parameters given. The last two rows give the 
combination of energies used to determine $c_3$ and $c_4$. 
Errors are from statistics/fitting only. }
\label{tab:pval}
\end{table*}

\begin{table*}[t]
\centerline{
\begin{tabular}{llllll}
\hline
    & 3 	& 3 	     & 4 	  & 5 		& 5 		\\
    & $c_4=1$   & $c_4=1.25$ & $c_4=1.25$ & $c_4=1$ 	& $c_4=1.15$   \\ 
    & $c_3=1.0$ & $c_3=1.0$ & $c_3=1.0$ & $c_3=1.0$ &    $c_3=0.96$	\\
\hline \hline
$5E(1^3P_2 ) - 3E(1^3P_2 ) -2E(1^3P_2 )$ 
& 0.151(13) & 0.168(14) & 0.160(13) & 0.161(6)  & 0.162(7)  \\
$E(1^3P_2 ) - 3E(1^3P_2 ) +2E(1^3P_2 )$  
& -0.028(4) & -0.042(6) & -0.041(5) & -0.0342(16) & -0.0442(20) \\
\hline
\hline
\end{tabular}
}
\caption{Combinations of $P$-wave energies needed to fix 
$c_3$ and $c_4$ in GeV. Lattice spacing values are taken 
from Table~\ref{tab:aval}. Errors are statistical/fitting only.}
\label{tab:pcomb}
\end{table*}

An alternative approach to tuning the spin-dependent coefficients 
$c_3$ and $c_4$ is from matching fine structure in the spectrum 
to experiment. Here we use the $P$-wave fine structure in the 
$\Upsilon$ spectrum to do this~\cite{Gray:2005ur, Meinel:2010pv}. 
The $P$-wave masses are shifted from the spin-average by 
amounts that depend on spin-spin coupling terms proportional 
to  ${\bf S}\cdot{\bf S}$ 
or $S_{ij}$ and spin-orbit terms proportional to ${\bf L}\cdot{\bf S}$. 
The spin-spin terms are proportional to $c_4^2$ and the 
spin-orbit terms to $c_3$. We can make a combinations of 
the $^3P$ state masses in which the eigenvalues of $S_{ij}$ 
and ${\bf S}\cdot{\bf S}$ cancel, and a separate 
combination in which the eigenvalues of 
${\bf S}\cdot{\bf S}$ and ${\bf L}\cdot{\bf S}$ cancel. 
Thus each of these combinations gives a result which should 
depend on only one of the spin-dependent couplings in 
our current NRQCD action. Comparing these combinations 
to experiment allows us to tune $c_3$ and $c_4$.  
Note that 4-quark operators, discussed in Appendix~\ref{appendix:cicalc}
with reference to their effect on $S$-wave hyperfine splittings, 
have negligible effect on $P$-wave states, and they therefore 
give a very clean determination of $c_3$ and $c_4$. 

The eigenvalues for ${\bf L} \cdot {\bf S}$, $S_{ij}$ and 
${\bf S}\cdot{\bf S}$ for $^{2S+1}L_J$ states 
$\{^3P_0, ^3P_1, ^3P_2, ^1P_1\}$ (i.e. $\{\chi_{b0},\chi_{b1},\chi_{b2},h_b\}$) 
are:
\begin{eqnarray}
{\bf L} \cdot {\bf S} &:& \{-2,-1,1,0\}; \nonumber \\ 
S_{ij}&:& \{-4,2,-2/5,0\}; \nonumber \\
{\bf S}\cdot{\bf S} &:& 
\{1/4,1/4,1/4,-3/4\}. 
\label{eq:eigs}
\end{eqnarray}
We see that the ${\bf S}\cdot{\bf S}$ terms affect the 
splitting between the $^1P_1$ and the spin-average of 
the $^3P$ states. We expect this splitting to be small 
because, in a potential model approach, 
the accompanying spin-dependent potential 
would be a delta function at the origin with very little 
overlap for $P$-wave states. 
Both $S_{ij}$ and ${\bf L}\cdot{\bf S}$ terms 
affect the splittings within the $^3P$ sector but 
not the splitting between $^1P_1$ and $^3\overline{P}$. 
The ${\bf L}\cdot{\bf S}$ terms give the 
conventional ordering of $^3P_0$, $^3P_1$ and $^3P_2$, 
but with the $^3P_2$ splitting from the $^3P_1$ 
larger than that between the $^3P_1$ and $^3P_0$. 
The $S_{ij}$ terms will push down the $^3P_2$ 
relative to the others. Thus the final splittings 
depend on the relative strength of the accompanying 
potentials for these terms and, in NRQCD language, the coefficients 
$c_3$ and $c_4$.  

The combination of spin-splittings that depends 
on $c_4^2$ (through $S_{ij}$) and is independent of $c_3$ is~\cite{Gray:2005ur} 
\begin{equation}
M(\chi_{b2})-3M(\chi_{b1})+2M(\chi_{b0})
\label{eq:pcombc4}
\end{equation}
with experimental value: -47.4(1.3) MeV~\cite{pdg}, determining 
the errors on mass differences by adding the errors 
on the masses in quadrature. 
Likewise the combination that depends on $c_3$ only
\begin{equation}
5M(\chi_{b2})-3M(\chi_{b1})-2M(\chi_{b0})
\label{eq:pcombc3}
\end{equation}
with experimental value: 163.8(3.2) MeV~\cite{pdg}. 
Comparison of our results with experiment for these 
combinations then allows us to fix $c_3$ and $c_4$. 

Table~\ref{tab:pval} gives the results for the energies in 
lattice units of the lowest-lying $P$-wave states for 
the $am_b$ values given in Table~\ref{tab:upsparams} on 
the coarse (sets 3 and 4) and fine (set 5) lattices. 
The results were obtained from 5 exponential fits 
of the form given in equation~\ref{eq:fit} to the $2\times 2$ 
matrix of correlators for each $P$-wave meson done as 
a single simultaneous fit. This enables us to extract 
mass differences more precisely from the fit than the 
individual masses and the splitting between each state 
and the spin-average of all the ${}^3P$ states is also given.  
Note that we give separate values for ${}^3P_2E$ and ${}^3P_2T_2$ 
lattice representations of the $J=2$ state. 
Differences between the values obtained for $T_2$ and 
$E$ would be a sign 
of discretisation errors. We do not have a significant signal 
for this but the $E$ state is higher than the $T_2$ in all 
of our fits. The difference is about 3(2) MeV on the coarse 
lattices and 2(1) MeV on the fine lattices. 

Results are given for the case $c_4=1$ and a nonzero 
value of $c_4$ chosen to give reasonable agreement 
with the experimental value of the combination 
in equation~\ref{eq:pcombc4}. Results in GeV for the two 
combinations tested are given in Table~\ref{tab:pcomb} 
using values of the lattice spacing from the $(2S-1S)$ 
splitting in 
Table~\ref{tab:aval}, and combining the $E$ and $T_2$ 
representations for the ${}^3P_2$ state with 
the appropriate number of spin states. We see that $c_3=1$ within 
our errors, but $c_4$ needs to be larger than 1, more 
so on the coarse lattices than the fine. We take 
the same value of $c_4$ on both coarse sets since 
the tuning should not depend on the sea quark 
masses and indeed our results demonstrate that it does 
not. 

Figure~\ref{fig:pwaves} shows the spectrum 
of $P$-wave states relative to the $^3P$ spin-average 
($5M(\chi_{b2})+3M(\chi_{b1})+M(\chi_{b0})$). 
The results with $c_4>1$ clearly agree better with 
experiment than for $c_4=1$. The main effect 
of increasing $c_4$ is to push the 
$\chi_{b0}$ state down relative to the spin-average. 
Very little else changes. In particular 
we see that the splitting between the 
$^1P_1$ and the $^3P$ spin-average is very small 
and negative in all cases. It increases with 
increasing $c_4$ but, because the splitting itself
is so small, this is not significant.  
We obtain a $P$-wave hyperfine splitting of 
2(2) MeV on both coarse and fine lattices, 
where the experimental result is $1.6 \pm 1.5$ MeV~\cite{bellehb}. 

On the fine lattices
we took $c_4=1.15$, based on initial calculations.
From Table~\ref{tab:pcomb} and Figure~\ref{fig:pwaves} 
this appears to be an underestimate and  
an improved value would be 1.18.       
We also used $c_3=0.96$ but that value is
indistinguishable within our errors from 
$c_3=1.0$. 
On the coarse lattices, $c_4=1.25$ is also 
a slight underestimate, although it agrees 
within statistical errors with the correct answer. 

The final nonperturbatively tuned values 
for $c_3$ and $c_4$ that we obtain are then: 
$c_3 = 1.00(9)(2)(10)$ on coarse lattices and 1.00(4)(2)(10) 
on fine lattices. Our best estimates for $c_4$ are:
\begin{eqnarray}
c_4 (\mathrm{coarse}) &=& 1.28(7)(1)(5) \nonumber \\
c_4 (\mathrm{fine}) &=& 1.18(2)(1)(5). 
\end{eqnarray}
The first error is from the statistical/fitting error 
on the $P$-wave masses along with the lattice spacing 
error. The second error 
is from experiment. The third error is a systematic error 
from $v^6$ terms in NRQCD that are missing from our 
calculation but that have effectively been absorbed into 
the value of $c_3/c_4$ from matching to experiment.  
The spin splittings could change by $\mathcal{O}(10\%)$ 
from $v^6$ terms. 
Note that $c_4$ gets closer to 1 as the lattice 
gets finer as we expect. Note also that the relationship between 
$c_3$ and $c_4$ that holds in potential models or NRQCD 
in the continuum 
because of Lorentz covariance~\cite{Bali:1997am} is not applicable 
to lattice NRQCD in this formulation. 

The agreement between the $c_4$ 
coefficients obtained nonperturbatively and 
the $c_4$ coefficients obtained at $\mathcal{O}(\alpha_s)$ 
in Appendix~\ref{appendix:cicalc} is good, and 
certainly within possible $\alpha_s^2$ variation 
of the perturbative coefficients (see Table~\ref{tab:c4vals}). 

Our nonperturbative results for $c_4$ and $c_3$ agree well 
with those derived for the same NRQCD action on 
different gluon configurations at similar 
lattice spacing in~\cite{Meinel:2010pv}. There 
spin-dependent terms at $v_b^6$ are also 
included in a separate calculation and then 
the values derived for the $v^4$ coefficients $c_3$ and 
$c_4$ change (both increasing). Note that in 
that paper the calculations were done with 
tree-level coefficients for all $c_i$ and the results rescaled 
for the derived values of $c_4$. 

%
\section{ Results for the Kinetic mass}
\label{appendix:mkin}
%
We give here the tables of results for the ground-state energies
and kinetic masses (as defined in equation~\ref{eq:mkin}) of the $\Upsilon$ and $\eta_b$ mesons 
for different lattice meson momenta in 
units of $2\pi a/L$. We also give the spin-averaged $1S$ kinetic mass.  
Results are taken from simultaneous fits to local correlators with 
the given momentum and with momentum zero using the fit form in equation~\ref{eq:fit}. 
The energies for each momentum and the energy difference which 
yields the kinetic mass are given directly by the fit. Correlations 
between the correlators mean that the error on the energy difference 
is typically smaller than the combined errors from the separate energies. 
This is particularly true of the `on-axis' momenta which have only 
one non-zero component. So, for example, the kinetic mass for momentum 
(2,0,0) is more precise than that for (1,1,1). 

The results are given for coarse ensemble set 3 and fine ensemble set 
5 with separate results for the case where $c_{1,5,6}$ are taken to 
be 1 and the case where $c_{1,5,6}$ are $\alpha_s$-improved. 
We take 9 exponential fits on set 3 and 7 exponential fits on set 5. 

\begin{table*}[h]
\centerline{
\begin{tabular}{lllllll}
\hline
\hline
  & (0,0,0) & (1,0,0) & (1,1,1) & (2,0,0) & (2,2,1) & (3,0,0) \\
\hline
$aE(^1S_0,{\bf P})$  			& 0.25529(4) & 0.26119(4) & 0.27309(4) & 0.27890(4) & 0.30830(8) & 0.30814(6) \\
$aE(^3S_1,{\bf P})$  			& 0.28626(6) & 0.29220(7) & 0.30426(7) & 0.31007(9) & 0.33977(17) & 0.33957(14) \\
$aM_{\mbox{\tiny Kin}}(\eta_b)$    	& - & 5.773(10) & 5.767(7) & 5.788(2) & 5.787(7) & 5.805(3) \\
$aM_{\mbox{\tiny Kin}}(\Upsilon)$  	& - & 5.716(25) & 5.703(17) & 5.739(7) & 5.732(15) & 5.753(11) \\
$\overline{aM}_{\mbox{\tiny Kin}}(1S)$ 	& - & 5.730(20) & 5.719(14) & 5.751(6) & 5.746(12) & 5.766(8) \\
\hline
\hline
\end{tabular}
}
\caption{$\Upsilon$ and $\eta_b$ energies and kinetic masses in 
lattice units for various lattice momenta on coarse set 3 for $b$ quark 
mass $am_b=2.66$ and $c_{1,5,6}$ set to 1.}
\label{tab:kinmass-coarse-ci-eq1}
\end{table*}

\begin{table*}[h]
\centerline{
\begin{tabular}{lllllllll}
\hline
\hline
  & (0,0,0) & (1,0,0) & (1,1,0) & (1,1,1) & (2,0,0) & (2,1,1) & (2,2,1) & (3,0,0) \\
\hline
$aE(^1S_0,{\bf P})$  			& 0.26096(4) & 0.26684(4) & 0.27273(4)  & 0.27860(4) & 0.28438(4) & 0.29610(4)  & 0.31348(6) & 0.31335(6) \\
$aE(^3S_1,{\bf P})$  			& 0.29243(6) & 0.29838(6) & 0.30434(6)  & 0.31030(7) & 0.31611(8) & 0.32799(8)  & 0.34555(14) & 0.34536(14) \\
$aM_{\mbox{\tiny Kin}}(\eta_b)$    	& - & 5.818(7) & 5.819(7)  & 5.817(7) & 5.839(3) & 5.834(4)  & 5.844(7) & 5.859(4) \\
$aM_{\mbox{\tiny Kin}}(\Upsilon)$  	& - & 5.747(18) & 5.748(17)  & 5.742(17) & 5.778(7) & 5.764(9)  & 5.778(13) & 5.798(12)\\
$\overline{aM}_{\mbox{\tiny Kin}}(1S)$ 	& - & 5.764(15) & 5.766(14) & 5.761(14) & 5.793(5) & 5.782(8)  & 5.795(11) & 5.813(10) \\
\hline
\hline
\end{tabular}
}
\caption{$\Upsilon$ and $\eta_b$ energies and kinetic masses in 
lattice units for various lattice momenta on coarse set 3 for $b$ quark 
mass $am_b=2.66$ and $c_{1,5,6}$ set to 
their $\mathcal{O}(\alpha_s)$ improved values. Slight differences 
with Table~\ref{tab:latdata} for zero momentum energies arise because 
we fit a single zero momentum correlator rather than a $5\times 5$ matrix. }
\label{tab:kinmass-coarse-ci-neq1}
\end{table*}

\begin{table*}[t]
\centerline{
\begin{tabular}{lllllll}
\hline
\hline
  & (0,0,0) & (1,0,0) & (1,1,1) & (2,0,0) & (2,2,1) & (3,0,0) \\
\hline
$aE(^1S_0,{\bf P})$  			& 0.24652(3) & 0.25107(3) & 0.26010(3) & 0.26461(3) & 0.28713(4) & 0.28712(4) \\
$aE(^3S_1,{\bf P})$  			& 0.27153(5) & 0.27610(4) & 0.28518(5) & 0.28974(5) & 0.31244(6) & 0.31246(7) \\
$aM_{\mbox{\tiny Kin}}(\eta_b)$    	& - & 4.244(6) & 4.252(6) & 4.2548(14) & 4.2516(23) & 4.251(3) \\
$aM_{\mbox{\tiny Kin}}(\Upsilon)$  	& - & 4.222(14) & 4.230(13) & 4.225(3) & 4.223(4) & 4.215(6) \\
$\overline{aM}_{\mbox{\tiny Kin}}(1S)$ 	& - & 4.228(12) & 4.236(11) & 4.2327(24) & 4.230(4) & 4.224(5) \\
\hline
\hline
\end{tabular}
}
\caption{$\Upsilon$ and $\eta_{b}$ energies and kinetic masses in lattice units for various lattice momenta on fine set 5 for $b$ quark mass $am_b=1.91$ and $c_{1,5,6}$ set to $1$}
\label{tab:kinmass-fine-ci-eq1}
\end{table*}

\begin{table*}[t]
\centerline{
\begin{tabular}{llllllll}
\hline
\hline
  & (0,0,0) & (1,0,0) & (1,1,0) & (1,1,1) & (2,0,0) & (2,2,1) & (3,0,0) \\
\hline
$aE(^1S_0,{\bf P})$  			& 0.25827(3) & 0.26278(3) & 0.26727(4) & 0.27173(3) & 0.27620(3) & 0.29850(4) & 0.29847(3) \\
$aE(^3S_1,{\bf P})$  			& 0.28390(5) & 0.28844(4) & 0.29299(6) & 0.29747(5) & 0.30199(5) & 0.32451(6) & 0.32447(6) \\
$aM_{\mbox{\tiny Kin}}(\eta_b)$    	& - & 4.278(7) & 4.286(9) & 4.287(6) & 4.2914(14) & 4.2920(23) & 4.2951(19) \\
$aM_{\mbox{\tiny Kin}}(\Upsilon)$  	& - & 4.245(15) & 4.251(17) & 4.256(14) & 4.2515(29) & 4.2523(44) & 4.2538(41) \\
$\overline{aM}_{\mbox{\tiny Kin}}(1S)$ 	& - & 4.253(12) & 4.260(15) & 4.264(11) & 4.2615(24) & 4.2622(37) & 4.2641(35) \\
\hline
\hline
\end{tabular}
}
\caption{$\Upsilon$ and $\eta_{b}$ energies and kinetic masses in lattice units for various lattice momenta on fine set 5 for $b$ quark mass $am_b=1.91$ and $c_{1,5,6}$ set to their $\mathcal{O}(\alpha_s)$ improved values. Slight 
differences in zero momentum energies are seen compared to Table~\ref{tab:latdata} because we used $u_{0L}$=0.85246 rather than 0.8525 and are fitting to single correlators rather than 
a $5\times 5$ matrix. }
\label{tab:kinmass-fine-ci-neq1}
\end{table*}

\bibliography{ups}

\end{document}